\documentclass[twocolumn,showkeys,preprintnumbers,longbibliography,amsmath,amssymb,epsfig,floatfix,prx]{revtex4-2}

\usepackage[table]{xcolor} 
\definecolor{LightCyan}{rgb}{0.88,1,1}
\usepackage{graphicx}      
\usepackage{dcolumn}       
\usepackage{ifthen}

\usepackage{amsmath}
\usepackage{mathtools}
\usepackage{amsfonts}
\usepackage{amssymb}
\usepackage{physics}       
\usepackage{bm}            
\usepackage{calrsfs}
\DeclareMathAlphabet{\altmathcal}{OMS}{cmsy}{m}{n}
\DeclareMathAlphabet{\mathcalligra}{T1}{calligra}{l}{m}

\usepackage[utf8]{inputenc}
\usepackage[english]{babel}
\usepackage{ulem}
\usepackage{siunitx}
\usepackage{multirow}
\usepackage{booktabs}
\usepackage[version=3]{mhchem}

\usepackage{natbib}  
\usepackage{hyperref}
\hypersetup{colorlinks=true, citecolor=blue, urlcolor=blue, linkcolor=blue}

\newcolumntype{C}{>{$}c<{$}}
\AtBeginDocument{
\heavyrulewidth=.08em
\lightrulewidth=.05em
\cmidrulewidth=.03em
\belowrulesep=.65ex
\belowbottomsep=0pt
\aboverulesep=.4ex
\abovetopsep=0pt
\cmidrulesep=\doublerulesep
\cmidrulekern=.5em
\defaultaddspace=.5em
}

\newcolumntype{L}[1]{>{\raggedright\arraybackslash}p{#1}}
\newcolumntype{C}[1]{>{\centering\arraybackslash}p{#1}}
\newcolumntype{R}[1]{>{\raggedleft\arraybackslash}p{#1}}

\definecolor{green}{RGB}{0,100,0}
\definecolor{violet}{RGB}{142, 68, 173} 
\definecolor{edit}{RGB}{0, 128, 255}

\begin{document}
\title{High-temperature ferromagnetism and antiferromagnetism in monolayer \ce{CrTe2}: Roles of strong spin-lattice coupling and charge doping }
\author{Anupama S}
\author{Mukul Kabir}
\email{mukul.kabir@iiserpune.ac.in}
\address{Department of Physics, Indian Institute of Science Education and Research, Pune-411008, India}

\begin{abstract}
The interplay of structural, electronic, and magnetic degrees of freedom governs phase stability and critical temperatures in two-dimensional magnets. Controlling this coupling is essential for advancing fundamental understanding and spintronic applications. Combining first-principles calculations with Heisenberg Monte Carlo simulations, we reveal a rich magnetic phase diagram governed by the interplay of lattice strain and carrier density. These results provide a unified framework that reconciles diverse experimental reports on epitaxial layers and predicts a novel double-stripe antiferromagnetic phase, further stabilized by electron doping. Moreover, structural and electronic perturbations enable room-temperature ferromagnetism and antiferromagnetism. This magnetic evolution arises from competing, highly tunable direct and ligand-mediated exchange interactions in the presence of Ruderman-Kittel-Kasuya-Yosida coupling. By disentangling their individual contributions, we elucidate the underlying microscopic mechanisms, which transcends the conventional conduction electron picture. Finally, we quantify the colossal magnetoelastic response and identify zone-folded Raman modes that serve as unique experimental fingerprints for phase identification. Together, these results establish \ce{CrTe2} as a versatile platform for two-dimensional spintronics, where magnetic order and transition temperatures are tailorable via structural and electrical engineering. 
\end{abstract}
\maketitle

\section{Introduction}
The Hohenberg-Mermin-Wagner theorem establishes that continuous symmetry cannot be spontaneously broken at finite temperatures in two-dimensional  (2D) systems characterized by isotropic, short-range interactions~\citep{PhysRev.158.383,PhysRevLett.17.1133}.  Consequently,  long-range magnetic order is fundamentally prohibited by the emergence of gapless Goldstone modes associated with low-energy spin-orientation fluctuations. This restriction can be circumvented by explicitly breaking the rotational symmetry through spin-orbit coupling, which introduces magnetocrystalline anisotropy necessary to stabilize magnetic order~\citep{nanolett.6b03052,nature22060,nature22391,Gibertini2019,acsnano.1c09150}.  However, most experimentally realized 2D magnets exhibit ordering only at cryogenic temperatures, hindering their integration into practical quantum technologies. It is therefore essential to identify and control novel material platforms capable of sustaining robust, high-temperature magnetic order~\citep{s41565-018-0121-3,s41565-018-0135-x,acs.nanolett.8b02806, s41563-018-0040-6,s41586-018-0626-9, acs.nanolett.9b03316,  PhysRevB.103.214411, PhysRevMaterials.6.084407}. 

Chromium-based materials have emerged as a central platform for investigating 2D magnetism, with \ce{Cr}-trihalides serving as the most prominent examples, exhibiting intrinsic Curie temperatures up to 45 K~\citep{nature22060,nature22391,science.abd5146,Kim11131,acs.nanolett.9b00553,s41565-019-0565-0,s41567-019-0651-0}. While carrier doping effectively tunes exchange interactions and can drive ordering temperatures toward the room-temperature regime in these systems~\citep{s41586-018-0626-9,PhysRevB.103.214411, PhysRevMaterials.6.084407}, monolayer $1T$-\ce{CrTe2} is distinguished by a substantially higher intrinsic $T_{\rm C}$ of approximately 200 K~\citep{s41467-021-22777-x,s41467-022-30738-1}. This robust magnetic behavior is consistent with its bulk counterpart, which is a room-temperature ferromagnet with $T_{\rm C}$ of 310 K~\citep{Freitas_2015,s12274-020-3021-4}.  Notably, exfoliated \ce{CrTe2} flakes with thicknesses down to $\sim$10 nm retain bulk-like magnetic properties, sustaining magnetic ordering well above room temperature~\citep{s12274-020-3021-4,PhysRevMaterials.5.034008}.

Experimental characterizations of monolayer \ce{CrTe2} have yielded seemingly contradictory results, with reported magnetic ground states ranging from ferromagnetic (FM) to zigzag antiferromagnetic (Z-AFM) order~\citep{s41467-021-22777-x, s41467-022-30738-1, s41535-025-00772-5,aelm.202400720,s41563-026-02537-2,s41467-021-27834-z,h4h8-j473}.  Such diverse observations are rooted in the sensitivity of the in-plane lattice to substrate-mediated strain and growth kinetics, where the resulting variation in magnetism highlights a remarkably strong magnetoelastic coupling. Specifically, a Z-AFM ground state is stabilized at smaller lattice constants~\citep{s41467-021-27834-z,aelm.202400720,s41563-026-02537-2,h4h8-j473,s41535-025-00772-5}, whereas an FM state emerges as the lattice expands~\citep{s41467-021-22777-x,s41467-022-30738-1,aelm.202400720,s41563-026-02537-2}. The nature of magnetic anisotropy in the ultrathin limit further remains unsettled. While bulk \ce{CrTe2} and relatively thick flakes exhibit in-plane magnetization~\citep{Freitas_2015,s12274-020-3021-4,PhysRevMaterials.5.034008,acsami.0c07017}, ultrathin samples have been reported to develop strong perpendicular magnetic anisotropy~\citep{s41467-021-22777-x,s41467-022-30738-1}. Furthermore, in the Z-AFM phase, spins are predicted to cant away from the basal plane, residing in the $yz$ plane at an angle of approximately $70^{\circ}$ relative to the $z$-axis~\citep{s41467-021-27834-z}.  These diverse observations are likely driven by the substantial epitaxial strain imposed during molecular beam epitaxy growth, as 2D materials can accommodate a significant strain that fundamentally alters their magnetic landscape. 

Although 2D \ce{CrTe2} possesses a higher magnetic ordering temperature than many counterparts, robust room-temperature magnetism in the true 2D limit remains elusive~\citep{Gibertini2019,acsnano.1c09150}. It is therefore critical to explore experimentally feasible routes to manipulate and enhance magnetism through external perturbations such as strain, chemical doping, optical excitation, and electrostatic gating. Among these, charge carrier modulation induced by electrostatic or ionic gating offers a particularly effective and controllable means to tune magnetic interactions~\citep{s41565-018-0121-3,s41565-018-0135-x,acs.nanolett.8b02806, s41563-018-0040-6,s41586-018-0626-9, acs.nanolett.9b03316,  PhysRevB.103.214411, PhysRevMaterials.6.084407}.  Such electrical control of magnetism is especially appealing for nanoscale magnetic devices. For instance, magnetism in both insulating chromium trihalides and the itinerant ferromagnet \ce{Fe3GeTe2} can be significantly modified through carrier injection. While ferromagnetic materials have historically been the primary focus for such manipulation, analogous control in antiferromagnetic systems remains significantly less explored, despite their considerable potential for high-speed, high-density spintronic applications.

Building upon experimental observations~\citep{s41467-021-22777-x,s41467-022-30738-1,Freitas_2015,s12274-020-3021-4,s12274-020-3021-4,PhysRevMaterials.5.034008,acsami.0c07017,s41467-021-27834-z},  we systematically explore the interplay between magnetism, the underlying triangular lattice of \ce{CrTe2}, and the effects of charge-carrier modulation. Calculations reveal an exceptionally strong spin-lattice coupling, driving the emergence of distinct magnetic phases as a function of the lattice parameter. We further investigate the complexities of carrier doping and present detailed phase diagrams mapping magnetic order against both strain and carrier density. This phase space exhibits a rich tapestry of FM and intricate AFM states, with ordering temperatures that are highly tunable via doping, ultimately enabling both room-temperature ferromagnetism and antiferromagnetism. By analyzing the competition between multiple exchange mechanisms, we provide a microscopic framework for the evolution of magnetism under external perturbations. Furthermore, we quantify phonon renormalization and evaluate the spin-phonon and spin-lattice coupling constants. Finally, we examine Brillouin zone-folded phonon modes, demonstrating how these emergent vibrational signatures below the ordering temperature can uniquely identify the underlying magnetic configuration through Raman spectroscopy.  

\section{Spin Hamiltonian and Computational details}
The effective spin Hamiltonian describing long-range magnetism in the 2D limit can be modeled using an anisotropic bilinear Heisenberg framework~\citep{PhysRevB.103.214411,PhysRevMaterials.6.084407},
\begin{eqnarray}
\altmathcal{H}_{\rm spin} = &-& \frac{1}{2}\sum_{\mathclap{{\langle i, j\rangle}}} J_{1}\bm{S}_i \cdot \bm{S}_j \ - \  \frac{1}{2}\sum_{\mathclap{{\langle \langle i, j\rangle\rangle}}} J_{2}\bm{S}_i \cdot \bm{S}_j \nonumber \\ 
&-& \frac{1}{2}\sum_{\mathclap{{\langle \langle\langle i, j\rangle\rangle\rangle}}} J_{3}\bm{S}_i \cdot \bm{S}_j + \altmathcal{H}_{\rm SIA}, \nonumber
\end{eqnarray}
where $J_1$, $J_2$, and $J_3$ denote isotropic first, second, and third neighbor exchange interactions, respectively. Positive values of $J$ indicate FM coupling between $\bm{S}_i$ and $\bm{S}_j$, while negative values represent AFM interactions. The final term, $\altmathcal{H}_{\rm SIA}$, accounts for the on-site single-ion anisotropy (SIA). For a system with an out-of-plane magnetic easy axis, this contribution reduces to,
\begin{equation}
\altmathcal{H}_{\rm SIA} = - \sum_i A_z (S_i^z)^2, \nonumber
\end{equation}
where $A_z > 0$ stabilizes an out-of-plane (easy-axis) orientation, while $A_z < 0$ favors an in-plane (easy-plane) configuration. In cases where the magnetic easy axis lies within the basal plane, the rotational symmetry within the $xy$-plane is explicitly lifted by an additional in-plane anisotropy term. The total single-ion anisotropy Hamiltonian is then expressed as, 
\begin{equation}
\altmathcal{H}_{\rm SIA} = - \sum_i A_z (S_i^z)^2 - \sum_i A_{xy}\left[ (S_i^x)^2 - (S_i^y)^2 \right], \nonumber
\end{equation}
where $A_{xy}$ parameterizes the in-plane anisotropy that lifts the continuous azimuthal degeneracy, thereby selecting a preferred crystallographic spin orientation within the $xy$-plane.
%

\begin{figure*}[!t]
\centering
\includegraphics[scale=0.75]{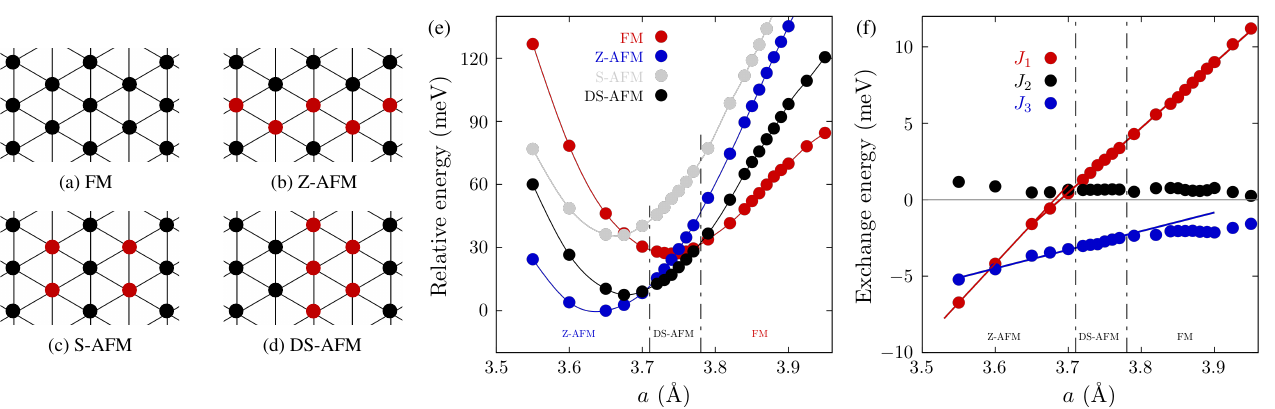}
\caption{
Magnetic exchange interactions are computed through energy mapping of various in-plane spin-ordered phases, such as (a) ferromagnetic (FM), (b) zigzag antiferromagnetic (Z-AFM), (c) stripe antiferromagnetic (S-AFM), and (d) double-stripe antiferromagnetic (DS-AFM) configurations. Black and red dots denote antiparallel collinear spins. Magnetism depends strongly on the in-plane lattice parameter $a$ of monolayer \ce{CrTe2}. 
(e) Magnetic phase stability as a function of the in-plane lattice parameter $a$. The Z-AFM state is favored below a critical value of $a = 3.70$ \AA, while the DS-AFM and FM phases become progressively more stable at larger lattice constants, consistent with experimental reports. 
(f) Calculated exchange interactions $J$s exhibit significant variation with lattice parameter, showing significant strain sensitivity. 
}
\label{fig:figure1}
\end{figure*}

This model has been successfully applied to 2D magnetic insulators, including chromium trihalides and \ce{CrGeTe3}~\citep{PhysRevB.103.214411,PhysRevMaterials.6.084407}. Incorporating exchange interactions beyond the first neighbor is essential for accurately predicting and reproducing experimental results, both qualitatively and quantitatively~\cite{nature22060,nature22391,acs.nanolett.9b00553,s41565-018-0135-x,Kim11131,science.abd5146}.  We compute these isotropic exchange interactions through energy mapping of various spin-ordered phases, including FM, Z-AFM, stripe antiferromagnetic (S-AFM), and double-stripe antiferromagnetic (DS-AFM) configurations [Figure~\ref{fig:figure1}(a)-(d)]. The corresponding energies on a triangular magnetic lattice are expressed as,
\begin{equation}
\begin{split}
E_{\mathrm {FM}} &= E_0 - \frac{fS^2}{2}\left[+6J_1 + 6J_2 + 6J_3 \right], \nonumber \\
E_{\mathrm {S\mbox{-}AFM}} &= E_0 - \frac{fS^2}{2}\left[-2J_1 - 2J_2 + 6J_3 \right], \nonumber \\
E_{\mathrm {{Z\mbox{-}AFM}/{DS\mbox{-}AFM}}} &= E_0 - \frac{fS^2}{2}\left[\mp 2J_1 \pm 2J_2 - 2J_3 \right].
\end{split}
\end{equation}
$E_0$ is the energy of the paramagnetic state, and $f$ is the number of magnetic atoms in the supercell. Spin-orbit coupling breaks the continuous rotational symmetry of interacting spins, leading to magnetocrystalline anisotropy. $A_z$ is evaluated from the energy difference between in-plane $[010]$ and out-of-plane $[001]$ magnetization direction, $A_z = (E_{\mathrm [010]} - E_{\mathrm [001]})/S^2$, while the in-plane component is determined by $A_{xy} = (E_{\mathrm [100]} - E_{\mathrm [010]})/2S^2$. For multilayer or bulk systems, this Hamiltonian is extended by incorporating the interlayer exchange interaction $J_{\perp}$.

In metallic magnets, exchange interactions are often mediated or modified by itinerant carriers, facilitating long-range magnetic coupling. Although the Heisenberg Hamiltonian traditionally describes localized spins in insulators, it effectively captures the magnetization, ordering temperatures, and excitation spectra of metals where local moments persist within a conducting background~\citep{1.2710181,PhysRevB.107.224411,PhysRevB.103.045114}. This dual localized-itinerant character is well documented in similar 2D systems. In the ferromagnets \ce{Fe3GeTe2} and \ce{Fe3GaTe2}, the weak temperature dependence of exchange-split bands provides clear spectroscopic evidence that robust local moments coexist with itinerant electrons~\citep{PhysRevB.101.201104,PhysRevB.109.104410}. Similarly, measurements confirm a localized Heisenberg magnetism in atomically thin \ce{Cr2Te3}~\citep{s41467-023-40997-1}. For \ce{CrTe2}, dynamical mean-field theory confirms this dual character~\citep{PhysRevB.111.035118}, matching the spin-split bands resolved by photoemission in the ordered phase~\citep{s41467-021-22777-x}. Consequently, mapping the magnetic energy landscape of \ce{CrTe2} onto a spin-only effective Hamiltonian provides a valid description of the essential exchange physics. However, because the itinerant background introduces Fermi-surface and multi-spin couplings, capturing the full thermodynamic stability of its ground state requires supplementing the standard bilinear framework with higher-order exchange corrections, such as biquadratic interaction.


The structural, electronic, and magnetic properties were investigated using first-principles density functional theory (DFT)~\citep{PhysRev.136.B864,PhysRev.140.A1133}, as implemented in the Vienna Ab initio Simulation Package (VASP)~\citep{PhysRevB.47.558,PhysRevB.54.11169}. To identify the magnetic ground state and exchange constants, we calculated the total energies of various magnetic configurations [Figure~\ref{fig:figure1}(a)-(d)] and mapped them onto the Heisenberg Hamiltonian $\altmathcal{H}_{\rm spin}$. Furthermore, we incorporated relativistic spin-orbit coupling to evaluate the magnetic anisotropy energies by analyzing the total energies for different crystallographic spin orientations. 

Electronic wavefunctions were expanded in a plane-wave basis set with a kinetic energy cutoff of 600 eV, utilizing the projector-augmented wave formalism~\citep{PhysRevB.50.17953}. Exchange-correlation energies are described within the generalized gradient approximation using the Perdew-Burke-Ernzerhof functional~\citep{PhysRevLett.77.3865}. To account for strong electron correlation in the \ce{Cr}-$3d$ orbitals, a Hubbard-type on-site Coulomb interaction of $U_{\text{eff}} = 2$ eV was included via the rotationally invariant Dudarev approach~\citep{PhysRevB.57.1505}. Long-range van der Waals (vdW) interactions in the bulk phase were incorporated using the DFT-D3 semi-empirical dispersion correction~\citep{10.1063/1.3382344}.  Magnetic configurations were modeled using an in-plane $\sqrt{3} \times 2$ supercell for the monolayer, with the exception of the DS-AFM phase, which necessitated a larger $2\sqrt{3} \times 2$ supercell to accommodate the magnetic periodicity. The Brillouin zone was sampled with a $\Gamma$-centered $8 \times 7 \times 1$ Monkhorst-Pack $k$-mesh for the monolayer~\citep{PhysRevB.13.5188}, while denser meshes were employed to ensure the convergence of phonon frequencies. To prevent spurious interactions between periodic images, a vacuum spacing of at least 20 \AA\ was maintained along the out-of-plane direction. As monolayer properties were investigated as a function of the lattice parameter $a$, internal ionic positions for each magnetic configuration were optimized until the interatomic forces are reduced below a threshold of 0.01 eV/\AA, with a total energy convergence of 10$^{-8}$ eV. For bulk calculations, both lattice parameters and atomic positions are fully relaxed. To elucidate the microscopic exchange mechanisms, the electronic structure is projected onto a tight-binding Hamiltonian by transforming the Bloch orbitals into maximally localized Wannier functions using the Wannier90 code~\citep{PhysRevB.56.12847,Pizzi2020}. This transformation provides orbital-resolved electronic hopping parameters, which are critical for analyzing ligand-mediated and direct exchange pathways.

To investigate magnetic phase transitions and determine ordering temperatures, we performed classical Monte Carlo (MC) simulations on a 2D triangular lattice. A periodic lattice of $10^4$ spins was utilized to minimize finite-size effects. The Metropolis algorithm was employed with a single-site update scheme~\citep{10.1063/1.1699114}, where at each step, a random spin vector $\mathbf{S}_i = (S^x_i, S^y_i, S^z_i)$ is assigned a new direction in three-dimensional space according to the Marsaglia procedure~\citep{marsaglia1972}. For each temperature point, we performed $2 \times 10^8$ MC steps to ensure thermal equilibrium, with physical observables averaging over 192 independent simulations to suppress statistical fluctuations. Bulk simulations were performed using the VAMPIRE code~\citep{Evans_2014}, which also served as a benchmark to validate our monolayer MC results.

\section{Results and Discussion}
We begin by discussing the properties of bulk \ce{CrTe2}, where the close agreement between our calculations and prior experimental observations serves as a benchmark for our theoretical framework. Building on this foundation, we investigate the magnetic phase diagrams of the \ce{CrTe2} monolayer as a function of lattice parameter and charge-carrier density. This approach allows us to elucidate the complex interplay between competing exchange mechanisms and the resulting spin-phonon coupling.

\subsection{Bulk \ce{CrTe2}}
Bulk $1T$-\ce{CrTe2} crystallizes in a layered \ce{CdI2}-type structure with $P\bar{3}m1$ space group, where the magnetic \ce{Cr} atoms are positioned at the centers of edge-sharing octahedra, forming triangular networks sandwiched between \ce{Te} sheets. Optimized lattice parameters are in excellent agreement with experimental values~\cite{Freitas_2015,supple}. Within this octahedral crystal field, the $\ce{Cr^{3+}}$ ions adopt a $t_{2g}^3 e_g^0$ electronic configuration, which theoretically hosts a localized spin magnetic moment of 3 $\mu_{\text{B}}$/\ce{Cr}. The calculated value between 3 and 3.3 $\mu_{\text B}$/\ce{Cr} aligns well with both experimental measurements~\citep{s41467-021-22777-x}, and earlier theoretical work~\citep{acs.jpcc.0c04913}. Furthermore, the half-filled $t_{2g}$ shell leaves the orbital magnetic moment negligible~\cite{s41467-021-22777-x}, consistent with a quenched orbital state.

While experiments characterize bulk $\ce{CrTe2}$ as an itinerant ferromagnet with a $T_{\rm{C}}$ of 310 K~\cite{Freitas_2015,s12274-020-3021-4}, present calculations reveal that the stability of this layered FM ground state depends on two distinct physical mechanisms. The in-plane magnetic stability is tightly intertwined with variations in the in-plane lattice parameter, highlighting a robust spin-lattice coupling. Notably, this in-plane interaction remains qualitatively independent of the on-site Coulomb interaction $U$. In contrast, the out-of-plane interlayer coupling $J_\perp$ is exceptionally sensitive to electronic correlations. As $U$ decreases from 2 eV, $J_\perp$ initially undergoes a transition from AFM to FM behavior, with the FM coupling strengthening continuously as $U$ is lowered further. This correlation-driven modulation of $J_\perp$ shifts the predicted ordering temperature between 398 K and 272 K at optimized lattice parameters (Supplemental Material~\citep{supple}). To validate this trend, we repeated the analysis using the experimental lattice parameters, and the qualitative sensitivity of $J_\perp$ to $U$ persists, with the calculated ordering temperature ranging from 325 K to 292 K, effectively bounding the experimental $T_{\rm{C}}$. Furthermore, we predict a persistent in-plane magnetic easy axis ($A_z < 0$) across all considered values of $U$, consistent with experimental observations~\cite{s41467-021-22777-x,s12274-020-3021-4,s41467-021-21072-z}. This in-plane spin orientation distinguishes \ce{CrTe2} from other stoichiometric chromium tellurides, such as \ce{Cr2Te3}~\citep{s41467-023-40997-1} and \ce{Cr5Te8}~\citep{PhysRevB.100.024434}, which exhibit out-of-plane easy axes. 

\subsection{Monolayer \ce{CrTe2}}
Calculations reveal that the magnetic ground state and ordering temperature of epitaxial \ce{CrTe2} monolayers are highly tunable via the lattice parameter $a$  [Figure~\ref{fig:figure1}(e)-(f)]. Above a critical threshold of $a >$ 3.78 \AA,  the FM $J_1$ interaction dominates the AFM $J_3$ ($|J_3/J_1| < 0.63$), therby stabilizing the FM phase. This transition qualitatively aligns with classical Heisenberg $J_1$-$J_2$-$J_3$ model predictions~\citep{RUBIN20121062,PhysRevB.104.184427}. Furthermore, $T_{\rm C}$ increases monotonically with $a$ as the FM $J_1$ coupling strengthens, rising from 98 K to 194 K as $a$ expands from 3.85 to 3.93 \AA. These results show excellent agreement with experimental reports of FM ordering near 200 K for lattice constants exceeding 3.8 \AA~\citep{s41467-021-22777-x,s41467-022-30738-1}.

For lattice constants below 3.78 \AA, two distinct AFM phases emerge. In the intermediate regime, 3.71 $< a <$ 3.78 \AA, the competition between FM $J_1$ and AFM $J_3$, within a weak FM $J_2$ background, stabilizes a previously unreported double-stripe AFM phase, warranting further experimental investigation. At $a < 3.70$ \AA, both $J_1$ and $J_3$ become AFM, favoring a Z-AFM ground state that is consistent with recent experimental reports~\citep{s41467-021-27834-z}. The corresponding N\'eel temperature increases significantly as these AFM couplings strengthen under compression, rising from 150 K at 3.60 \AA\ to 205 K at 3.55 \AA. While a quantum spin liquid phase is predicted in literature for the AFM $J_1$-$J_2$ model on a triangular lattice~\citep{PhysRevB.92.041105,s41567-023-02259-1}, we exclude this possibility for \ce{CrTe2} as calculations yield a FM $J_2$. Additionally, the on-site anisotropy exhibits a non-monotonic dependence on $a$, transitioning from easy-plane to easy-axis magnetism as $a$ exceeds 3.87 \AA~\cite{supple}.  

\begin{figure*}[!t]
\centering
\includegraphics[scale=0.85]{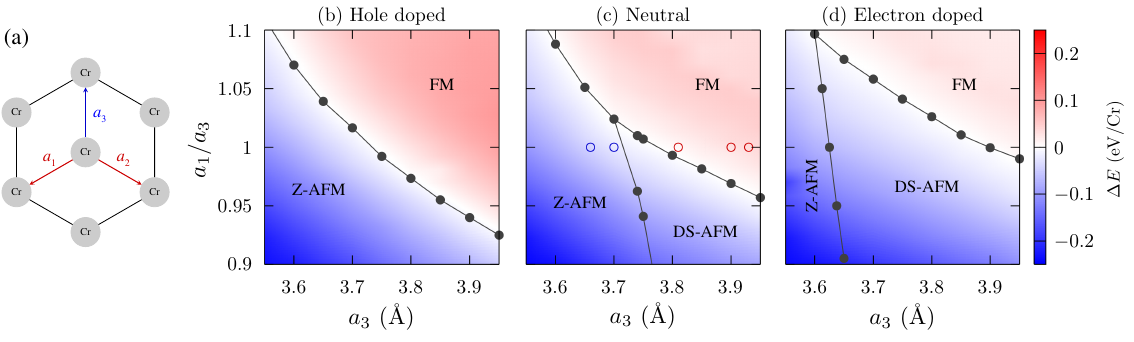}
\caption{The magnetic phase diagrams reveal a fascinating dependence on lattice and carrier doping. 
(a) Schematic of the uniaxially strained triangular lattice with $a_1=a_2\neq a_3$.  
Magnetic phase diagrams are shown for (b) hole-doped (0.1 $h$/f.u.), (c) charge-neutral, and (d) electron-doped (0.1 $e$/f.u.) monolayer. Applied doping levels correspond to a carrier density of approximately $(5-10) \times 10^{13}$ cm$^{-2}$  depending on the specific lattice parameter. The color gradient represents the energy difference between the most stable AFM and the FM solutions, $\Delta E = E(\rm{AFM}) - E(\rm{FM})$. 
Open circles denote experimental phases observed for isotropic lattices, identifying the Z-AFM (blue) and FM (red) phases~\citep{s41467-021-22777-x, s41467-022-30738-1, s41535-025-00772-5,aelm.202400720,s41563-026-02537-2}.
While hole doping stabilizes the FM and Z-AFM phases, electron doping suppresses the FM ground state and promotes a DS-AFM configuration across a significant portion of the phase space.
}
\label{fig:figure2}
\end{figure*}

To assess the impact of geometric frustration on the triangular lattice, we compared the energetic stability of the non-collinear \SI{120}{\degree} AFM phase against the collinear Z-AFM state. Despite the underlying lattice frustration, electronic structure calculations indicate that the collinear Z-AFM configuration remains the true ground state, with the \SI{120}{\degree} AFM phase consistently residing higher in energy (Supplemental Material~\citep{supple}). Notably, this energy penalty for the \SI{120}{\degree} phase grows as the lattice expands, reflecting the increasing dominance of further-neighbor $J_2$ and $J_3$ couplings in stabilizing the zigzag-like magnetic topology. These results underscore the necessity of incorporating a biquadratic exchange term into the effective spin Hamiltonian $\altmathcal{H}_{\rm spin}$, expressed as, 
$$
\altmathcal{H}_{\rm bq} = -\frac{1}{2}\sum_{\mathclap{{\langle i, j\rangle}}} K_1 (\bm{S}_i \cdot \bm{S}_j)^2,
$$
where $K_1$ denotes the nearest-neighbor biquadratic exchange interaction. This higher-order interaction is crucial for capturing the accurate magnetic ground state predicted by DFT. Under this convention, a positive biquadratic coupling ($K_1>0$) drives spin collinearity, stabilizing the Z-AFM ground state by contributing an energy of  $-3KS^4$ per site compared to only $-\frac{3}{4}KS^4$ for the \SI{120}{\degree} phase. Within the Z-AFM regime [Figure~\ref{fig:figure1}(f)], the calculated $K_1$  is positive and $|K_1/J_3|$ ranges between 0.3 and 0.4, corroborating the robust thermodynamic stability of the collinear zigzag order against geometric frustration. Biquadratic exchange interactions of this magnitude are comparable to those observed in related vdW magnets, such as nickel dihalides~\citep{PhysRevLett.127.247204}.

We note that the qualitative nature of the magnetic phases in the monolayer remains robust against the variations in $U$ (Supplemental Material~\citep{supple}). Quantitatively, however, the individual exchange parameters within the $J_1$-$J_2$-$J_3$ triangular lattice model exhibit a complex evolution with $U$, and drive a shift toward higher ordering temperatures. Specifically, as $U$ is lowered, the calculated $T_{\rm{N}}$ for the Z-AFM phase ($a = 3.55$ \AA) rises from 210 K to 256 K, while the $T_{\rm{C}}$ for the FM phase ($a = 3.85$ \AA) increases from 100 K to 200 K.

\subsection{Magnetic phase diagram: Lattice strain and charge doping}
Building upon the analysis of the uniformly strained triangular lattice, we now investigate the magnetic phases that emerge under uniaxial strain, where the lattice symmetry is lowered such that $a_1 = a_2 \neq a_3$ [Figure~\ref{fig:figure2}(a)]. Such anisotropic strain is frequently induced by epitaxial growth on specific substrates, as exemplified by \ce{CrTe_2} monolayers grown on graphene/\ce{SiC(0001)}. These monolayers exhibit a significant uniaxial strain, with reported in-plane lattice constants of $a_1 = a_2 = 3.7$ \AA\ and $a_3=3.4$ \AA~\citep{s41467-021-27834-z}. 

The magnetic phase diagram for the anisotropic lattice reveals a fascinating interplay between lattice parameter and magnetic order [Figure~\ref{fig:figure2}(c)]. Tensile strain, defined by $a_1/a_3 > 1$, progressively stabilizes the FM phase across a wide range of lattice constants. In contrast, the Z-AFM phase is increasingly favored for compressed lattices, $a_1/a_3 < 1$, when the reference lattice parameter $a_3$ is below 3.75 \AA. Interestingly, a significant region of the phase space hosts a DS-AFM ground state when $a_3$ exceeds 3.75 \AA\ under compressive conditions, $a_1/a_3 < 1$.

Manipulating carrier density is a well-established strategy for tuning magnetism in 2D materials~\cite{PhysRevB.90.035403, PhysRevB.95.174419, s41586-018-0626-9, PhysRevB.103.214411, PhysRevMaterials.6.084407, PhysRevLett.125.267205}. For example, significant modulation of the ordering temperature, including room-temperature ferromagnetism, has been demonstrated in ultrathin \ce{Fe3GeTe2}~\citep{s41586-018-0626-9} and monolayer \ce{CrBr3}~\citep{PhysRevMaterials.6.084407} through carrier injection. In epitaxial systems, the substrate can inherently shift the magnetic layer away from the charge-neutrality point via interfacial charge transfer, introducing significant electron or hole doping. Furthermore, the carrier concentration can be dynamically modulated by gate voltage within a device architecture.

\begin{figure*}[!t]
\centering
\includegraphics[scale=0.70]{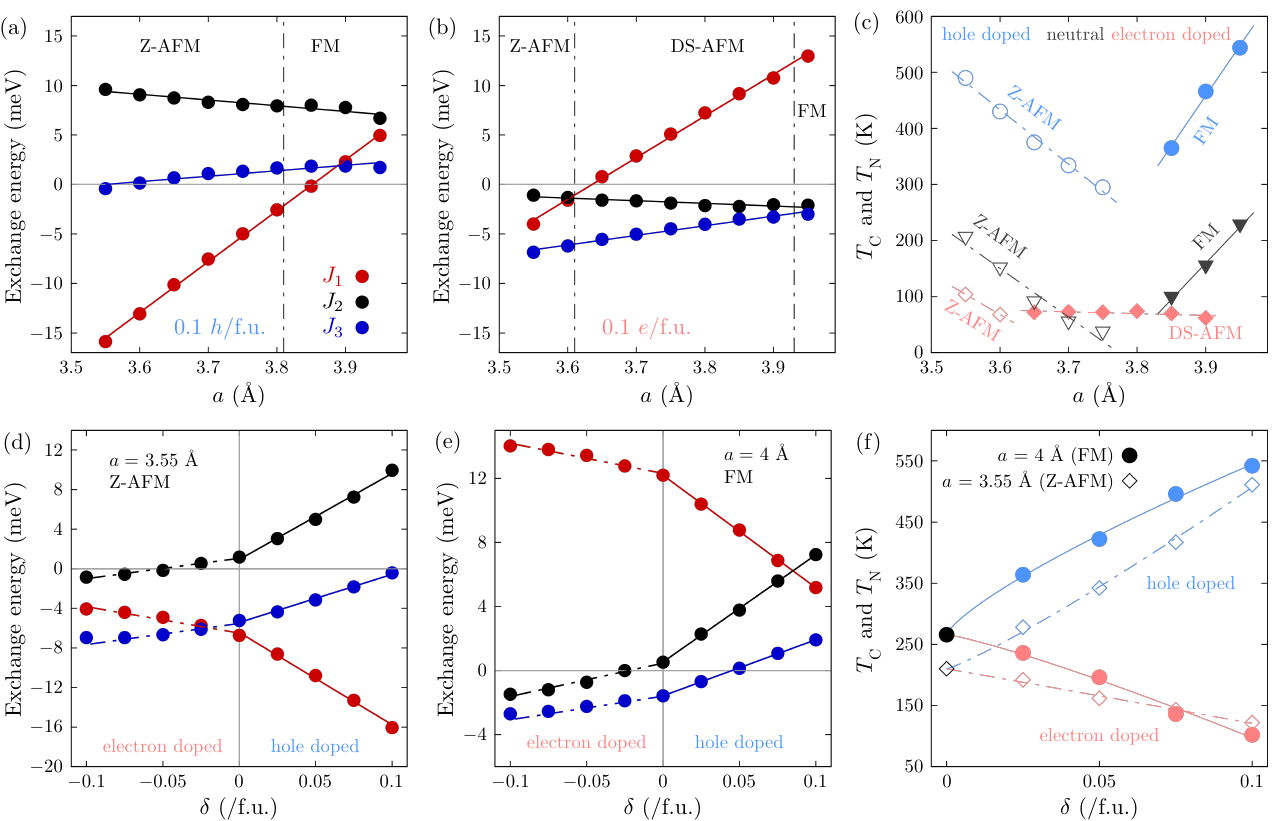}
\caption{
Calculations reveal a strong dependence of exchange interactions and thus, the transition temperatures on the in-plane lattice parameter and carrier density. Exchange interactions for an isotropic lattice ($a_1 = a_3 = a$) in (a) hole-doped (0.1 $h$/f.u.) and (b) electron-doped (0.1 $e$/f.u.) \ce{CrTe2} monolayers. illustrating the profound effect of carrier doping on the magnetic exchange hierarchy. 
(c) In hole-doped monolayers, the calculated Curie and N\'eel temperatures indicate the persistence of high-temperature ferromagnetism and antiferromagnetism, depending on the lattice parameter $a$. The DS-AFM state is stabilized under electron doping, though the ordering temperature remains unaffected over a range of $a$.  
(d)-(e) Evolution of exchange interactions with doped carrier density $\delta$ for two limiting lattice parameters, 3.55 \AA\ (Z-AFM) and 4 \AA\  (FM). 
(f) Dependence of magnetic ordering temperatures on carrier densities for fixed lattice parameters of 3.55 \AA\ (Z-AFM) and 4  \AA\  (FM). Both magnetic phases remain robust across all doping levels, exhibiting a characteristic increase in ordering temperature with hole doping and a decrease with electron doping.   
}
\label{fig:figure3}
\end{figure*}

Carrier doping exerts a dramatic influence on the magnetic phase diagram of monolayer \ce{CrTe2} [Figure~\ref{fig:figure2}]. Upon introducing a modest hole concentration of 0.1 $h$/f.u. [Figure~\ref{fig:figure2}(b)], the FM phase is progressively reinforced, particularly at a larger and isotropic lattice with $a_1/a_3 = 1$. Furthermore, the stability of FM and Z-AFM orders is highly sensitive to anisotropic lattice strain.  While tensile strain ($a_1/a_3 > 1$)  significantly stabilizes the FM phase, the compressed regime ($a_1/a_3 < 1$)  progressively favors Z-AFM ordering. Notably, under hole doping, the DS-AFM phase vanishes entirely, indicating a profound reconfiguration of the magnetic energy landscape.  Electron doping induces a contrasting magnetic evolution [Figure~\ref{fig:figure2}(d)]. While the FM state persists as the ground state across a significant portion of the phase diagram, its stability is notably suppressed relative to the charge-neutral case. Interestingly, electron doping selectively stabilizes the double-stripe configuration, with the DS-AFM phase expanding to occupy a dominant portion of the AFM region.  Consequently, the phase space hosting the Z-AFM order shrinks significantly.  

To elucidate the microscopic mechanisms driving these phases, we calculated the exchange interactions for the doped systems on a uniformly strained lattice [Figure~\ref{fig:figure3}(a)-\ref{fig:figure3}(b)], corresponding to the isotropic limit $a_1/a_3=1$ in Figure~\ref{fig:figure2}. Significant modifications occur in the exchange profile of the charged systems relative to the charge-neutral monolayer. At a constant doping density of $\delta = 0.1$/f.u., $J_1$ exhibits high sensitivity to the in-plane lattice parameter $a$ for both hole- and electron-doped cases, whereas the variations in $J_2$ and $J_3$ are comparatively moderate. In the hole-doped regime [Figure~\ref{fig:figure3}(a)], the exchange landscape undergoes a significant transformation compared to the neutral case [Figure~\ref{fig:figure1}(f)]. Notably, both $J_2$ and $J_3$ shift to FM character, with $J_2 \gg J_3$. While the AFM to FM transition in $J_1$ is shifted toward a larger lattice constant, all exchange interactions converge to an FM state once the lattice surpasses 3.85 \AA. This collectively results in robust ferromagnetism at larger values of $a$.  Conversely, at smaller lattice constants, the persistent AFM nature of $J_1$ stabilizes the Z-AFM phase, despite a strong FM second-neighbor contribution.  

The exchange landscape is markedly different under electron doping [Figure~\ref{fig:figure3}(b)], where $J_1$ remains predominantly FM, except at very small lattice constants. Crucially,  both $J_2$ and $J_3$ remain AFM, with $J_3$ emerging as the dominant interaction. The resulting competition between these exchange channels stabilizes the DS-AFM structure over a wide range of $a$. While FM and Z-AFM orders have been experimentally confirmed in \ce{CrTe2}~\citep{s41467-021-22777-x, s41467-022-30738-1, s41535-025-00772-5,aelm.202400720,s41563-026-02537-2,s41467-021-27834-z,h4h8-j473}, the DS-AFM phase remains elusive. Although such an order is rare, reported primarily in systems such as monoclinic \ce{FeTe}~\citep{science.1251682}. The results indicate that the DS-AFM phase is highly accessible in electron-doped \ce{CrTe2} monolayers, offering a new platform for exploring exotic magnetic order.

Mapping the relative strengths of the exchange couplings reveals the criteria governing the magnetic phase boundaries (Supplemental Material~\citep{supple}). In the AFM $J_1$ regime, the Z-AFM phase is favored by ferromagnetic further-neighbor couplings ($J_2/J_1,  J_3/J_1 < 0$) and persists as these interactions transition to AFM character, up to critical thresholds of $J_3/J_1 \sim 4$ and $J_2/J_1 \sim 0.5$. This transition reflects the competition inherent in frustrated triangular lattices, where a dominant $J_3$ coupling forces the magnetic order to conform to the periodicity of the third-neighbor lattice, effectively producing a Z-AFM state over more frustrated configurations. Conversely, in the FM-$J_1$ regime, the magnetic landscape is governed by competition between the primary FM $J_1$ and further-neighbor frustration. Here, the DS-AFM phase emerges under strong magnetic frustration ($J_3/J_1 < -0.5$ and $J_2/J_1 < 0.5$), whereas the FM order remains stable within the regime defined by $J_3/J_1 > -0.4$ and $J_2/J_1 > -0.2$.

Utilizing the exchange constants [Figure~\ref{fig:figure1}(f) and Figure~\ref{fig:figure3}(a)-(b)],  we calculated the magnetic ordering temperatures for electron- and hole-doped monolayers and compared them with the neutral case [Figure~\ref{fig:figure3}(c)]. In the hole-doped monolayer ($\delta$ = 0.1 $h$/f.u.), we predict that both room-temperature antiferromagnetism and ferromagnetism are achievable, depending on the in-plane lattice parameter [Figure~\ref{fig:figure3}(c)]. In contrast, electron doping generally suppresses magnetic ordering, leading to lower transition temperatures across the lattice range. Notably, while the DS-AFM structure is stabilized over a wide range of $a$ under electron doping, the calculated $T_{\rm N}$ remains remarkably constant despite significant underlying variations in the individual exchange parameters.

Having established the effects of constant carrier density across varying lattice parameters, we now examine the magnetic evolution under varied doping density $\delta$ for the structural limits corresponding to the Z-AFM and FM regimes [Figure~\ref{fig:figure3}(d)-(f)]. Hole doping significantly reinforces magnetic stability for both phases, with ordering temperatures increasing steadily and exceeding 500 K as $\delta$ rises [Figure~\ref{fig:figure3}(e)]. This enhancement is driven by a stronger AFM $J_1$ coupled with synergistic FM $J_2$ contributions in the Z-AFM phase [Figure~\ref{fig:figure3}(d)], while in the FM phase, the stabilizing role of $J_2$ and $J_3$ outweighs the reduction in nearest-neighbor exchange [Figure~\ref{fig:figure3}(e)]. Conversely, electron doping exerts a comparatively weaker influence, progressively destabilizing both magnetic orders and reducing the ordering temperatures. These results are consistent with the experimental studies on \ce{NaCrTe2}~\citep{acs.nanolett.4c01542}, where \ce{Na} layers are intercalated between successive \ce{CeTe2} layers~\citep{PhysRevMaterials.5.L091401}. In this configuration, the \ce{Na} atoms act as electron donors to the magnetic layers, and our calculations correctly corroborate the observed decrease in the ordering temperature. Further, this prediction aligns with the experimentally observed reduction in $T_{\rm C}$ for both bulk and surface-layer samples, the latter of which are electronically equivalent to monolayer \ce{NaCrTe2}~\citep{acs.nanolett.4c01542,PhysRevMaterials.5.L091401}.

\begin{figure*}[!t]
\centering
\includegraphics[scale=0.92]{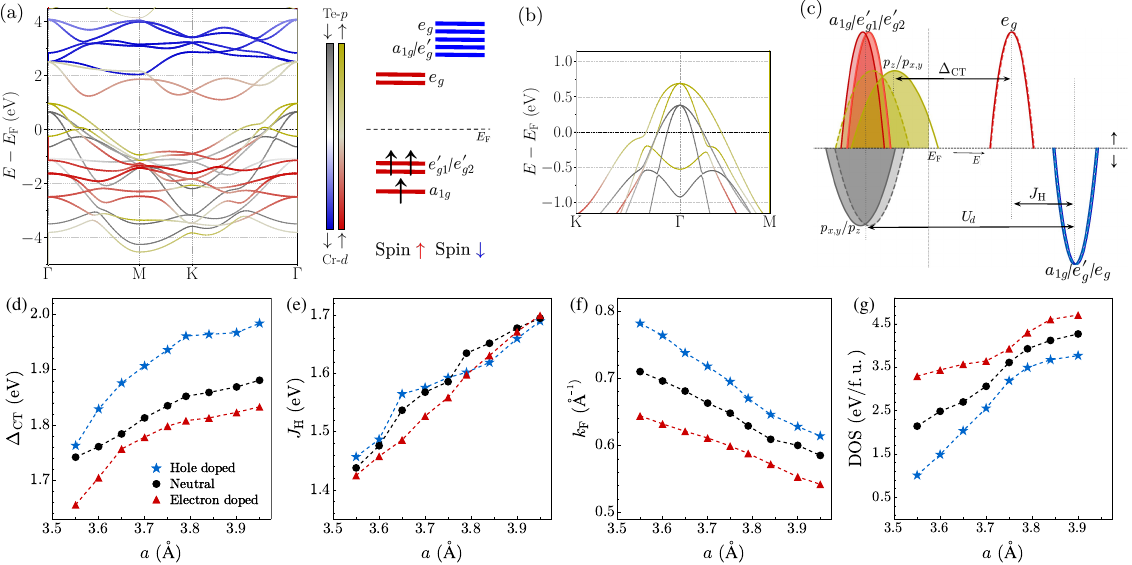}
\caption{
(a) Spin-polarized electronic band structure of monolayer $1T$-\ce{CrTe2} calculated for an in-plane lattice parameter $a = \SI{3.95}{\angstrom}$. The color scale indicates the orbital character, with contributions from \ce{Te}-$p$ and \ce{Cr}-$d$ states as labeled. Strong $p$-$d$ hybridization leads to metallic behavior. The trigonal distortion induced $a_{1g}$, $e_g^{\prime}$, and $e_g$ manifolds are schematically indicated, with spin-majority (red) and spin-minority (blue) components.
(b) An enlarged view of the electronic structure near the Fermi level reveals a spin-polarized ligand hole pocket at the $\Gamma$ point, giving rise to a small magnetic moment on the \ce{Te} sites.
(c) Schematic illustration of the spin-polarized density of states showing the relative positions of the electronic subbands. The key energy scales governing the exchange interactions, including the charge-transfer energy $\Delta_{\mathrm{CT}}$, the $d$-electron correlation $U_d$, and Hund's coupling $J_{\text H}$, are indicated.
The evolution of the key parameters (d) $\Delta_{\mathrm{CT}}$, (e) $J_{\text H}$, together with (f) the Fermi momentum $k_{\text F}$ and (g) the density of states at the Fermi level $D_{\text F}$, shows a strong dependence on the lattice parameter and charge doping, thereby dictating the dominant exchange mechanisms. The systematic variation of these quantities reflects the interplay among lattice structure, electronic degrees of freedom, and magnetism under charge doping. The charge densities considered here are 0.1 $e$/f.u. and 0.1 $h$/f.u.
}
\label{fig:figure4}
\end{figure*}

\subsection{Competing exchange mechanisms}
We investigate the microscopic mechanisms underlying the evolution of magnetism as a function of lattice parameter and carrier density, focusing on monolayers with a uniform lattice for clarity. The exchange landscape is inherently complex, arising from competing interactions, direct exchange (DE), superexchange (SE)~\citep{PhysRev.79.350,GOODENOUGH1958287,KANAMORI195987}, Ruderman-Kittel-Kasuya-Yosida (RKKY) exchange~\citep{PhysRev.96.99,10.1143/PTP.16.45,PhysRev.106.893}, and ligand-hole mediated double exchange (\uline{$L$}-DE)~\citep{PhysRevB.89.024416,PhysRevB.111.L180404}. Such competition is a hallmark of  metallic ferromagnets, where the interplay between itinerant and localized electrons dictates the magnetic ground state~\citep{s41586-018-0626-9,PhysRevB.108.L041401,aelm.202300646}. 

Before discussing these exchange mechanisms, we describe the electronic structure of the \ce{CrTe2} monolayer, which defines the relevant energy scales governing the magnetic interactions (Figure~\ref{fig:figure4}).  The electronic band structure reveals strong $p-d$ hybridization [Figure~\ref{fig:figure4}(a)], alongside a ligand hole pocket at the $\Gamma$-point derived from spin-polarized ligand $p$ states [Figure~\ref{fig:figure4}(b)]. The local octahedral coordination of \ce{Cr} exhibits a trigonal distortion along the crystallographic $[111]$ direction~\cite{supple}, which lowers the local symmetry and induces mixing between the nominal $t_{2g}$ and $e_{g}$ manifolds. Consequently, the \ce{Cr}-$d$ states reorganize into symmetry-adapted linear combinations belonging to the $a_{1g}$, $e_g^{\prime}$, and $e_{g}$ irreducible representations~\citep{khomskii2014transition,PhysRevB.77.125106}. The $a_{1g}$ singlet and the $e_g^{\prime}$ doublets are half-filled [Figure~\ref{fig:figure4}(a)], consistent with the $d^3$ electronic configuration of \ce{Cr^{3+}}. This yields a magnetic moment of about 3.3 $\mu_{\text B}$/\ce{Cr}, in agreement with experimental values. Additionally, a finite moment of approximately 0.2 $\mu_{\text B}$/\ce{Te} is induced due to the presence of ligand holes (Supplemental Material~\cite{supple}). 

The parameters governing the various exchange interactions are extracted from the electronic structure [Figure~\ref{fig:figure4}(c)], with their evolution tracked as a function of the lattice parameter. The charge-transfer energy $\Delta_{\text{CT}}$ plays an important role in both superexchange and $\uline{L}$-DE by controlling the strength of the $pd$ hybridization. We define $\Delta_{\text{CT}} = \epsilon_d - \epsilon_p$, where $\epsilon_d$ and $\epsilon_p$ denote the band centers of the unoccupied $d$ and occupied $p$ states, respectively [Figure~\ref{fig:figure4}(c)]. As lattice parameter increases, $\Delta_{\text{CT}}$ increases slightly, indicating a gradual weakening of the $pd$ hybridization [Figure~\ref{fig:figure4}(d)]. The on-site Coulomb interaction, which governs both DE and SE, is estimated to be about 4.5 eV. This high value highlights the strongly correlated nature of the system and remains essentially unchanged with $a$. Similar to $\Delta_{\text {CT}}$, the Hund's coupling $J_{\text H}$, which renormalizes the FM superexchange, increases with $a$ [Figure~\ref{fig:figure4}(e)]. 

Despite the correlated nature of the \ce{Cr}-$d$ electrons, $1T$-\ce{CrTe2} remains metallic due to significant $pd$ hybridization. Two key parameters,  the Fermi momentum $k_{\text F}$ and the density of states at the Fermi level $D_{\text F}$, are central to the RKKY interaction. The Fermi momentum, estimated from the hole pocket in the spin-polarized band structure, decreases as $a$ increases [Figure~\ref{fig:figure4}(f)]. This trend is consistent with the reduction in electron density $n$ and the two-dimensional scaling relation $k_{\rm F} \propto n^{1/2}$. The calculated values are in good agreement with angle-resolved photoemission spectroscopy (ARPES) measurements, which report $k_{\rm F} \sim 0.5~\text{\AA}^{-1}$~\citep{s41467-021-22777-x,s41535-025-00772-5}. In contrast, the metallicity is enhanced with $a$ through an increase in $D_{\text F}$ [Figure~\ref{fig:figure4}(g)], which counteracts the reduction in $k_{\text F}$.

\begin{figure}[t]
\centering
\includegraphics[scale=0.9]{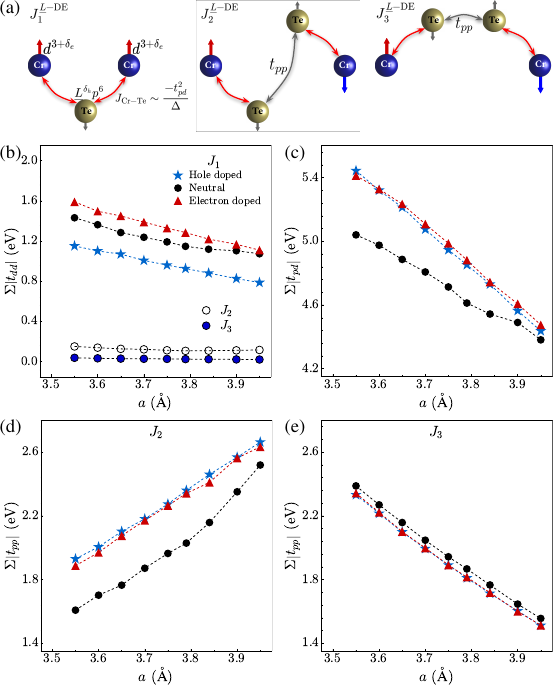}
\caption{
(a) Schematic representation of the ligand-hole--mediated double exchange interaction pathways. Finite \ce{Te} moments are strong AFM coupled to \ce{Cr}-spins, $J_{\ce{Cr-Te}} < 0$, analogous to the Zhang-Rice mechanism. Ligand-ligand hopping $t_{pp}$ enforces AFM coupling between \ce{Te} moments along the long-range \ce{Cr-Te-Te-Cr} paths, yielding weak AFM $J_2^{\uline{L}\rm -DE}$ and $J_3^{\uline{L}\rm -DE}$ interactions.
 Evolution of the key hopping amplitudes with lattice parameter and charge doping, which govern the direct exchange, superexchange, super-superexchange and \uline{$L$}-DE, (b) metal-metal hopping $t_{dd}$, (c) ligand-metal hopping $t_{pd}$, and (d, e) are ligand-ligand hopping $t_{pp}$ along the exchange pathways for the second and third neighbour interactions.   Since $t_{dd}$ is negligibly small for $J_2$ and $J_3$, its variation with charge doping is insignificant and therefore not shown in panel (b).  
While cumulative hopping parameters are presented, the orbital-resolved hoppings are provided in the Supplemental Material. The charge densities considered are 0.1 $e$/f.u. and 0.1 $h$/f.u.
}
\label{fig:figure5}
\end{figure}

Having established the relevant parameters [Figure~\ref{fig:figure4}(d)-(g)], we now evaluate the specific mechanisms that together govern the magnetic exchange interactions. 
Direct exchange is determined by the interplay between Coulomb repulsion and kinetic energy associated with electron delocalization. At small \ce{Cr-Cr} separations, substantial hopping between $d$-orbitals, $t_{dd}$, enhances kinetic exchange, favoring AFM coupling in strongly correlated \ce{CrTe2} ($t_{dd}/U_d \ll $  1). With increasing $a$, the narrowing of $d$-orbital bandwidth enhances the effective electron correlation $U_d^\prime = U_d/W_d$. Simultaneously, the rapid decay of the $t_{dd}$ hopping (Figure~\ref{fig:figure5}) further suppresses the AFM direct exchange contribution, $J_1^{\rm DE} = -t_{dd}^2/U_d^\prime$, facilitating the transiton toward a FM regime in $J_1$  [Figure~\ref{fig:figure1}(f)]. 

Following the Goodenough-Kanamori-Anderson rules~\citep{PhysRev.79.350, GOODENOUGH1958287, KANAMORI195987}, a nearly \SI{90}{\degree} \ce{Cr-Te-Cr} superexchange pathway yields a robust FM contribution to $J_1$ (Supplemental Material~\citep{supple}). This arises from the coupling between half-filled \ce{Cr}-($a_{1g}+e^\prime_{g}$) and empty \ce{Cr}-$e_g^0$ states mediated by \ce{Te}-$p$ orbitals [Figure~\ref{fig:figure4}(a)].  Although the \ce{Cr-Te-Cr} bond angle varies between 80-\SI{90}{\degree} with increasing $a$, the concurrent increase in the \ce{Cr-Te} distance renders the $p-d$ hopping amplitude $t_{pd}$ only weakly dependent on $a$. Dominant contributions originate from hopping between $p_x/p_y$  and $a_{1g}/e^\prime_{g}$ orbitals (Supplemental Material~\citep{supple}). This is consistent with the calculated bandwidth $W_d$, which narrows only slightly with $a$, while $U_d$ remains essentially unchanged.  Upon lattice expansion, the increase in $\Delta_{\rm CT}$ and the concurrent reduction in $t_{pd}$ decrease the effective hopping, $t'_{dd} = t_{pd}^2 / \Delta_{\rm CT}$ [Figure~\ref{fig:figure4}(d) and Figure~\ref{fig:figure5}]. Relative to AFM SE, the FM superexchange is renormalized by the ratio $J_{\text H}/\Delta_{\rm CT}$, with Hund's coupling $J_{\text H}$ ranging from 1.4 to 1.7 eV [Figure~\ref{fig:figure4}(e)]. Notably, while the typical 3$d$ transition metal systems exhibit $J_{\text H}/\Delta_{\rm CT} \sim 0.2$~\citep{khomskii2014transition}, monolayer \ce{CrTe2} possesses a much larger ratio of 0.8. Consequently, the FM SE contribution, $J_1^{\rm SE} \sim (t'^2_{dd}/\Delta_{\rm CT})\cdot(J_{\text H}/\Delta_{\rm CT})$, remains dominant, though it gradually weakens as the lattice expands.  

In the metallic \ce{CrTe2} monolayer, the localized \ce{Cr^3+} spins interact via  RKKY indirect exchange mediated by conduction electrons. This interaction is long-range and oscillatory, and in two dimensions, it is expressed in terms of Bessel functions of the first and second kind, $J_n$ and $Y_n$, respectively~\citep{PhysRevB.11.2025,PhysRevB.36.8835,PhysRevB.58.3584}, 
$$
J_{\rm RKKY} \propto - D_{\rm F} k_{\rm F}^2 \left[ J_0(k_{\rm F}r)Y_0(k_{\rm F}r) + J_1(k_{\rm F}r)Y_1(k_{\rm F}r) \right], 
$$
where $r$ denotes the spin separation. In the asymptotic limit ($2k_{\rm F}r \gg 1$), the RKKY interaction scales as $J_{\rm RKKY} \propto - D_{\rm F} k_{\rm F}^2 \sin(2k_{\rm F}r)/(2k_{\rm F}r)^2$, resulting in an oscillatory behavior with alternating AFM and FM couplings. Although the metallicity improves with lattice expansion [Figure~\ref{fig:figure4}(g)], its influence is largely offset by the concurrent decrease in $k_{\rm F}$ [Figure~\ref{fig:figure4}(f)]. Consequently, the RKKY contribution to $J_1$ remains AFM over the entire range of lattice parameters~\citep{supple}, providing a persistent AFM background that competes with the FM superexchange. 

The presence of ligand holes drives metallic conductivity through the hybridized $p-d$ bands and simultaneously induces magnetic moments on the \ce{Te} sites, thereby exerting a profound influence on the exchange interactions [Figure~\ref{fig:figure5}(a)]~\citep{PhysRevB.89.024416,PhysRevB.111.L180404}. Analogous to the Zhang-Rice singlet in cuprates~\citep{PhysRevB.37.3759}, the itinerant ligand hole couples antiferromagnetically to the $d$-electrons via strong $p-d$ hybridization and delocalizes across neighboring \ce{Cr} sites through the shared \ce{Te} ligand. This delocalization is kinetically favored when the \ce{Cr} spins are parallel, thereby lowering the kinetic energy of the system and producing an effective ferromagnetic \uline{$L$}-DE.  This interaction is comparatively strong, since  $J_{\rm{Cr-Te}} \sim -t_{pd}^2/\Delta_{\rm CT}$ significantly exceeds the strength of \ce{Cr-Cr} superexchange, $J_1^{\rm SE}$. Despite the larger hole-pocket observed in the zigzag AFM phase, the ligand moment $\mu_{\rm Te}$ is substantially smaller than in the FM phase (Supplemental Material~\citep{supple}), diminishing the effective FM contribution of this mechanism in the AFM regime. With lattice expansion, the modest reduction in $t_{pd}$ [Figure~\ref{fig:figure5}(c)] results in a slight weakening of the \uline{$L$}-DE interaction.

Overall, $J_1$ emerges from a delicate balance of multiple competing channels. While direct AFM exchange dominates at short \ce{Cr-Cr} separations, it weakens rapidly with increasing $a$. The AFM RKKY contribution remains largely insensitive to $a$, whereas both  FM superexchange and FM \uline{$L$}-DE exhibit a gradual, weakly decreasing trend. The interplay of these components drives a magnetic crossover near $a \sim $ 3.7 \AA\ [Figure~\ref{fig:figure1}(f)], where the FM  \uline{$L$}-DE and superexchange mechanisms collectively overtake the AFM direct exchange. 

We now extend our analysis to the higher-order exchange interactions, $J_2$ and $J_3$. The significant suppression of $t_{dd}$ at second-neighbor distance renders the AFM direct exchange contribution to $J_2$ very weak [Figure~\ref{fig:figure5}(b)].  Instead, a fourth-order super-superexchange (SSE) pathway becomes the relevant mechanism, mediated by a ligand-ligand bridge connecting \ce{Te} ions from opposite sublayers. With \ce{Cr-Te-Te} and \ce{Te-Te-Cr} bond angles close to \SI{90}{\degree}, the SSE pathway favours FM coupling, and scales as $J_2^{\rm SSE} \sim (t_{pd}^4 t_{pp}^2)/(\Delta_{\rm CT}^4U_p)$. Direct \ce{Te-Te} hopping $t_{pp}$ increases with $a$ due to shorter \ce{Te-Te} separations, and remains substantial [Figure~\ref{fig:figure5}(d)], with the $t_{pp}/t_{pd}$ ratio ranging between 0.3 and 0.6. Consequently, the narrowing of the ligand bandwidth $W_p$ causes the on-site  Coulomb interaction $U_p$ to increase with $a$ (Supplemental Material~\citep{supple}). Overall, the SSE is weaker than the second-order $J_{1}^{\rm SE}$, and exhibits only a weakly decreasing dependence on $a$. The RKKY contribution to $J_2$ is FM but strongly suppressed by the increased spin separation. Furthermore, the \uline{$L$}-DE mechanism introduces a weak AFM contribution to $J_2$ through the longer \ce{M-L-L-M} exchange pathway [Figure~\ref{fig:figure5}(a)]. While \uline{$L$}-DE promotes metal-ligand singlet formation, a finite $t_{pp}$ drives AFM coupling between ligand moments, thereby stabilizing the \ce{Cr_{$\uparrow$}-Te_{$\downarrow$}-Te_{$\uparrow$}-Cr_{$\downarrow$}} configuration over \ce{Cr_{$\uparrow$}-Te_{$\downarrow$}-Te_{$\downarrow$}-Cr_{$\uparrow$}}. This AFM contribution is weak at small $a$ but strengthens with increasing lattice parameter, tracking the upward trend in $t_{pp}$ [Figure~\ref{fig:figure5}(d)]. The interplay of these competing channels renders a weakly ferromagnetic $J_2$ that persists across the entire range of $a$ [Fig.~\ref{fig:figure1}(f)].

As $t_{dd} \rightarrow 0$ at third-neighbour distances [Figure~\ref{fig:figure5}(b)], kinetic exchange makes no contribution to $J_3$.  Instead, $J_3$ is dominated by the SSE pathway, which differs fundamentally from $J_2^{\rm SSE}$. It is mediated by \ce{Te-Te} bridges within the same ligand sublayer, where the \ce{Cr-Te-Te} and \ce{Te-Te-Cr} bond angles of about \SI{130}{\degree} favor AFM coupling~\citep{supple}. Unlike the $J_2$ case, the inter-ligand hopping $t_{pp}$ involved in $J_3^{\text{SSE}}$ decreases with increasing $a$ $= d_{\rm Te-Te})$ [Figure~\ref{fig:figure5}(e)]. Consequently, the AFM $J_3^{\rm SSE}$ is stronger than the FM $J_2^{\rm SSE}$ at smaller lattice constants but becomes weaker as $a$ expands. The RKKY contribution remains AFM but is notably weak, consistent with the rapid spatial decay of interaction strength at the larger spin separation of $2a$. Similar to $J_2$, the \uline{$L$}-DE mechanism contributes an AFM component to $J_3$. However, unlike the $J_2$ contribution, this term steadily diminishes with increasing $a$. Overall, $J_3$ remains antiferromagnetic across the entire investigated range of lattice parameters [Figure~\ref{fig:figure1}(f)], with its magnitude gradually reduced at larger $a$.

\subsection{Electronic control of exchange interactions} 
To bridge the discussion from lattice strain to the influence of charge density, we examine how carrier injection reconfigures the underlying exchange physics. Carrier doping, achieved by electron injection or removal, profoundly modifies the magnetic exchange landscape [Figure~\ref{fig:figure3}(a) and ~\ref{fig:figure3}(b)] by altering the electronic structure, specifically orbital occupancies, hopping amplitudes, and electronic correlations. This provides a powerful route to engineer and control magnetism at the microscopic level. Such tuning is experimentally accessible via (dual) electrostatic or ionic liquid gating, which can induce 2D carrier densities as high as 10$^{14}$ cm$^{-2}$~\citep{s41578-022-00473-6, adma.201607054, annurev-matsci-080619-012219}.

We begin by discussing the consequences of hole doping on $J_1$. The AFM DE component in $J_1$ [Figure~\ref{fig:figure3}(a)] is significantly strengthened under hole-doping. This enhancement originates from a broadening of the $d$-orbital bandwidth $W_d$~\citep{supple}, particularly for the $e'_g$ manifold, which enhances $t_{dd}$ hopping [Figure~\ref{fig:figure5}(b)] while leaving the on-site Coulomb interaction $U_d$ largely unchanged. The resulting reduction in effective correlation $U_d^\prime$ further amplifies the AFM contribution, $J_1^{\rm DE} \sim -t_{dd}^2/U_d^\prime$. The qualitative dependence of $J_1^{\rm DE}$ on $a$ remains unchanged, being sharply suppressed with narrowing $W_d$ and the concomitant decrease in $t_{dd}$ upon lattice expansion. Simultaneously, the SE component of $J_1$ remains FM and is strengthened by improved $t_{pd}$ hopping amplitude [Figure~\ref{fig:figure5}(c)], an effect particularly pronounced at smaller lattice constants. Other key parameters, such as $J_{\text H}$ and $\Delta_{\rm CT}$, are largely unaffected [Figure~\ref{fig:figure4}(d) and (e)], preserving the overall $a$-dependence of the superexchange. The RKKY interaction remains AFM but is suppressed upon hole doping. Although $k_{\rm F}$ increases with hole concentration [Figure~\ref{fig:figure4}(f)], a sharp reduction in $D_{\rm F}$ limits the density of conduction electrons mediating the interaction [Figure~\ref{fig:figure4}(g)], thereby suppressing the AFM RKKY contribution.  Furthermore, the FM \uline{$L$}-DE contribution is strengthened by the increased $t_{pd}$, particularly at smaller $a$ [Figure~\ref{fig:figure5}(c)].  Collectively, these modifications maintain an $a$-dependence similar to the neutral case [Figure~\ref{fig:figure1}(f)], but the amplified AFM direct exchange shifts the total $J_1(a)$ downward, effectively pushing the AFM-FM crossover to larger $a$ [Figure~\ref{fig:figure3}(a)].

Similar to the neutral case, the AFM DE contribution to $J_2$ remains negligible under hole doping. Instead, the $J_2^{\rm SSE}$  channel remains FM and is reinforced by relative enhancement of $t_{pd}$ and $t_{pp}$, which outweighs the reduction in FM RKKY contribution due to reduced $D_{\rm F}$. The second-neighbor \uline{$L$}-DE interaction also remains weakly AFM, with its strength slightly amplified by the enhanced $t_{pd}$ and $t_{pp}$  hoppings along the \ce{M-L-L-M} channel (Figure~\ref{fig:figure5}). Consequently, hole doping yields a marked enhancement of the total FM exchange in $J_2$ [Figure~\ref{fig:figure3}(a)], primarily driven by the enhancement in FM SSE component. The overall $a$-dependence continues to be governed by the gradual reduction of $t_{pd}$ with lattice expansion, maintaining a robust but weakly decreasing throughout the studied range. 

Hole doping drives the AFM $J_3$ into a weakly FM regime [Figure~\ref{fig:figure3}(a)], a transition that can be understood qualitatively by the simultaneous weakening of all the AFM channels. The already small AFM  $J_3^{\rm SSE}$ is further reduced due to the increased $\Delta_{\rm CT}$ [Figure~\ref{fig:figure4}(d)]. Similarly, the AFM contribution from $J_3^{\rm RKKY}$  is suppressed by the significant depletion of $D_\text F$ under hole injection. Furthermore, the reduction in $t_{pp}$ along the $J_3$ exchange pathway [Figure~\ref{fig:figure5}(e)] further weakens the AFM \uline{$L$}-DE contribution.  

Turning our attention to electron doping, we find that $J_1(a)$ retains its characteristic lattice dependence but exhibits a significantly enhanced effective FM interaction. This enhancement shifts the AFM-FM crossover to a lower lattice constant $a$  [Figure~\ref{fig:figure3}(b)], in direct contrast to the hole-doped case. While $U_d$ remains largely unaffected by moderate electron doping, an increase in $t_{dd}$, particularly at smaller $a$ [Figure~\ref{fig:figure5}(b)], enhances the AFM $J_1^{\rm DE}$. However, as the lattice expands, the FM components $J_1^{\rm SE}$ and   $J_1^{\uline{L}\rm -DE}$ are both strengthened. This is driven by an increase in the $t_{pd}$ hopping [Figure~\ref{fig:figure5}(c)] accompanied by a concurrent reduction in $\Delta_{\rm CT}$. Furthermore, the addition of electrons shrinks the hole pocket, thereby reducing $k_{\text F}$ [Figure~\ref{fig:figure4}(f)], but its impact on $J_1^{\rm RKKY}$ is largely compensated by a moderate increase in $D_{\text F}$ [Figure~\ref{fig:figure4}(g)]. Consequently, the AFM RKKY interaction remains essentially unchanged under electron doping. Overall, the enhanced FM SE and \uline{$L$}-DE channels account for the observed upward shift in the total $J_1$ relative to the neutral case. 

The weak FM character of $J_2$ observed in the neutral case evolves into a weakly AFM coupling under electron doping [Figure~\ref{fig:figure3}(b)].  While the AFM DE becomes slightly stronger, the increase in $t_{pd}$ [Figure~\ref{fig:figure5}(c)], coupled with the reduced $\Delta_{{\rm{CT}}}$ and enhanced $t_{pp}$, strengthen the FM $J_2^{{\rm{SSE}}}$ channel while simultaneously amplifies the AFM  \uline{$L$}-DE contribution. In contrast, the relatively weak FM RKKY component remains largely unchanged. Ultimately, the AFM \uline{$L$}-DE interaction dominates over both the SSE and RKKY channels, governing the overall behavior of $J_2(a)$ and stabilizing the observed antiferromagnetic shift.

The AFM $J_3$ interaction is further reinforced under electron doping [Figure~\ref{fig:figure3}(b)] as all contributing exchange mechanisms are AFM, and generally become stronger, except for the RKKY interaction, which remains essentially unchanged. The observed trends across varying $a$ are consistent with the corresponding variations in $t_{pd}$, $t_{pp}$, and $\Delta_{{\rm{CT}}}$ established in earlier analysis. 

Collectively, these findings reveal that the magnetic ground state of monolayer $\text{CrTe}_2$ is dictated by a complex, highly tunable competition among multiple exchange channels. This complexity underscores that magnetism in 2D metallic systems often transcends the conventional conduction-electron picture, involving a sophisticated interplay of orbital-selective hopping and local-moment interactions. The subtle sensitivity to structural strain and charge-carrier density enables precise control over the sign and magnitude of the effective magnetic interactions, thereby not only triggering distinct magnetic phases, but also tuning their ordering temperatures. Such mechanisms are likely a defining feature across a broader class of emerging 2D metallic magnets.

\begin{figure*}[t]
\centering
\includegraphics[scale=0.87]{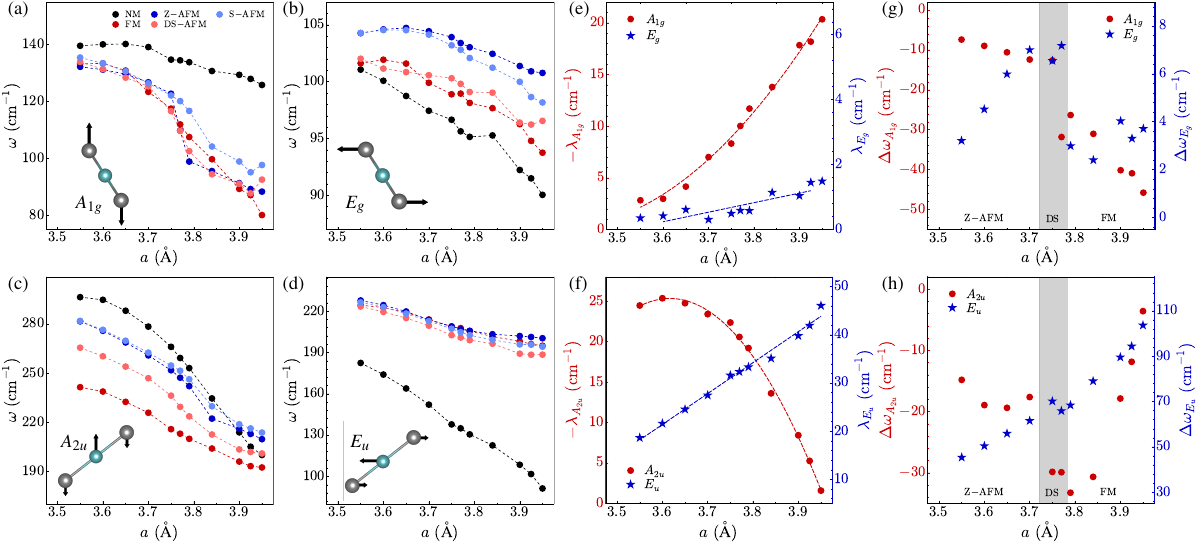}
\caption{
Phonon frequency evolution and spin-phonon coupling in monolayer \ce{CrTe2}. Evolution of the Raman-active $A_{1g}$ and $E_g$ modes (a, b), and the IR-active $A_{2u}$ and $E_u$ modes (c, d) as a function of lattice parameter $a$ for various magnetic phases compared to the non-magnetic (NM) state. Insets illustrate the atomic displacement vectors for the corresponding vibrational modes.%
(e, f) Mode-specific spin-phonon coupling constants $\lambda_{\nu}$ for Raman and IR modes, calculated from the frequency shifts between the NM ($\langle \mathbf{S}_i \cdot \mathbf{S}_j \rangle = 0$) and FM ($\langle \mathbf{S}_i \cdot \mathbf{S}_j \rangle = S^2$) states. The $\lambda_{\nu}$ indicates strong spin-phonon coupling, with a pronounced dependence on $a$. Negative values of $\lambda_{\nu}$ for out-of-plane modes ($A_{1g}$, $A_{2u}$) indicate a redshift (softening) upon ferromagnetic ordering, while positive values for in-plane modes ($E_g$, $E_u$) indicate a blueshift (hardening). %
(g, h) Frequency renormalization $\Delta\omega_{\nu}$ relative to the NM state for the magnetic ground state, highlighting these shifts as robust spectroscopic markers for identifying the underlying magnetic order.
}
\label{fig:figure6}
\end{figure*}

\subsection{Spin-phonon and spin-lattice couplings}
Strain engineering in 2D magnets typically yields only modest changes in magnetic order~\citep{adma.202205714}, transition temperature~\citep{PhysRevLett.127.217203,s41699-022-00315-7}, magnetic anisotropy~\citep{s41565-021-00885-5,s41699-024-00463-y,acsnano.4c16603}, or interlayer exchange~\citep{s41565-021-01052-6}.  In \ce{CrTe2}, however, the strain response is fundamentally more robust, driving transitions between competing magnetic phases (Figure~\ref{fig:figure2} and Figure~\ref{fig:figure3}), consistent with diverse magnetic orders observed experimentally~\citep{s41467-021-22777-x,s41467-022-30738-1,s41467-021-27834-z}. This pronounced magnetoelastic response underscores a profound entanglement between structural and magnetic degrees of freedom. We quantify this coupling through phonon renormalization across magnetic transitions, evaluating both the spin-phonon and spin-lattice coupling parameters.  

The onset of magnetic ordering often lowers the crystallographic symmetry, folding zone-boundary phonons back to the Brillouin zone center and rendering them Raman active~\citep{nanolett.6b03052,tian2016magneto,1.342186,s41565-019-0598-4,s41467-018-07547-6,s41467-018-08284-6, s41467-018-08284-6, PhysRevB.103.235411, PhysRevB.107.075421}. The emergence of these additional modes provides a unique spectroscopic fingerprint of the underlying magnetic configuration, enabling the differentiation of various AFM orders that carry no net moment and are thus indistinguishable by conventional magnetometry. Consequently, Raman spectroscopy serves as a sophisticated indirect probe to investigate magnetism in 2D materials. We investigate these zone-folded modes that provide unique markers for the magnetic orderings across the phase diagram. 
 
\subsubsection{Spin-phonon coupling $\lambda_{\nu}$}
Magnetic ordering induces significant phonon renormalization below the transition temperature, reflecting the underlying spin-phonon interactions. The change in phonon frequency for each eigenmode $\nu$ induced by magnetic ordering can be described as~\citep{seals-113910},
\begin{equation}
\omega_{\nu} = \omega_{\nu}^0 + \lambda_{\nu} \langle \mathbf{S}_i \cdot \mathbf{S}_j \rangle, \nonumber 
\end{equation}
where $\omega_{\nu}$ is the observed phonon frequency in the magnetic phase, typically measured via Raman and Infrared (IR) spectroscopy, $\omega_{\nu}^0$ is the harmonic phonon frequency in the absence of magnetic interaction, $\lambda_{\nu}$ is the mode-dependent spin-phonon coupling constant, and $\langle \mathbf{S}_i \cdot \mathbf{S}_j \rangle$ is the spin-spin correlation function between neighbouring magnetic ions.  To isolate intrinsic spin-phonon coupling from macroscopic magnetostriction, we compute the zone-center optical phonon frequencies for all magnetic configurations using a fixed nonmagnetic crystal structure~\citep{PhysRevLett.82.430,PhysRevLett.96.205505}. Any resulting phonon renormalization therefore arises solely from exchange-driven modifications of the interatomic force constants, with spin-induced lattice distortions explicitly excluded.

Based on the $D_{3d}$ point-group symmetry of $1T$-\ce{CrTe2}, the zone-center optical phonons decompose as, $\Gamma = n \left(A_{1g} + 2E_g + A_{2u} + 2E_u \right)$,  where $n$ is the number of layers. The $A_{1g}$ and $E_g$ modes are inversion-symmetry preserving and Raman active [Figure~\ref{fig:figure6}(a) and ~\ref{fig:figure6}(b)], whereas the $A_{2u}$ and $E_u$ modes break inversion symmetry and are IR active [Figure~\ref{fig:figure6}(c) and ~\ref{fig:figure6}(d)].  Both Raman modes involve only \ce{Te} vibrations, with the $A_{1g}$ mode corresponds to out-of-plane breathing, while the doubly degenerate $E_g$ modes involve in-plane shearing. In contrast, the IR active modes arise from antisymmetric displacements of both \ce{Cr} and \ce{Te} with out-of-plane $A_{2u}$ and in-plane $E_u$ character.    

The calculated frequencies of the $A_{1g}$ and $E_g$ modes, at 140 and 100 cm$^{-1}$, respectively, for $a=3.6$ \AA, are in good agreement with the experimental benchmarks of 134 and 102 cm$^{-1}$ established for high-purity $1T$-\ce{CrTe2} flakes~\citep{s12274-020-3021-4,acsami.0c07017}. The higher frequencies often reported for the $E_g$ manifold, around $124 \text{ cm}^{-1}$,~\citep{s41467-021-21072-z} are attributed to the presence of self-intercalated \ce{Cr_{1+x}Te2} polymorphs~\citep{acsaelm.2c01256,s41467-025-59266-4,qppq-qsx7}, such as \ce{Cr5Te8}, in which excess \ce{Cr} atoms occupying the vdW gap stiffen the lattice vibrations. The IR-active modes appear at higher frequencies due to their lower vibrational reduced mass and the larger force constants $\phi$ associated with direct metal-chalcogen bond stretching [Figure~\ref{fig:figure6}(c) and ~\ref{fig:figure6}(d)].  All phonon modes soften systematically with increasing lattice parameter, consistent with the reduction in bond stiffness expected from  $\omega_{\nu} \propto \sqrt{\phi}$. 

The observed phonon renormalization arises solely from the onset of magnetic ordering (Figure~\ref{fig:figure6}), without corresponding shifts in the equilibrium atomic positions. This indicates that the system is governed by a dominant spin-phonon coupling mechanism rather than conventional magnetostriction. Upon the emergence of magnetic order, the out-of-plane $A_{1g}$ and $A_{2u}$ modes exhibit redshift [Figure~\ref{fig:figure6}(a) and ~\ref{fig:figure6}(c)], while the in-plane $E_g$ and $E_u$ modes undergo a systematic blueshift [Figure~\ref{fig:figure6}(b) and ~\ref{fig:figure6}(d)]. The calculated coupling constants $\lambda_{\nu}$ exhibit intriguing trends [Figure~\ref{fig:figure6}(e) and ~\ref{fig:figure6}(f)]. First, their strong dependence on the lattice parameter indicates a strong sensitivity to the underlying exchange interactions. Second, distinct phonon symmetries modulate these exchange pathways in disparate ways, resulting in the observed mode-dependent renormalization of the phonon frequencies and, consequently, mode-specific coupling strengths $\lambda_{\nu}$. This is governed by how each phonon eigenvector modulates the \ce{Cr-Te} exchange pathways, as we now discuss.

For the $A_{1g}$ mode [Figure~\ref{fig:figure6}(a)], the breathing-like motion of the \ce{Te} sublayers increases the \ce{Cr-Te} bond length, thereby reducing the metal-ligand $t_{pd}$ hopping critical to both FM superexchange and FM ligand-hole-mediated double exchange. The associated weakening of ferromagnetic interactions reduces the magnetic contribution to the restoring force, leading to phonon softening. This effect becomes increasingly pronounced at larger lattice parameters [Figure~\ref{fig:figure6}(a)], where ferromagnetic exchange dominates the magnetic energy landscape. Consequently, $-\lambda_{A_{1g}}$ increases  sharply with lattice parameter [Figure~\ref{fig:figure6}(e)] , reaching magnitudes as high as  $|\lambda_{A_{1g}}| > $ 12 cm$^{-1}$ within the FM region. Such values are exceptionally large compared to other prototypical 2D magnets. For instance,  \ce{CrI3} and \ce{Cr2Ge2Te6} exhibit substantially lower $\lambda_{A_{1g}}$ values, typically below 4 cm$^{-1}$~\citep{tian2016magneto,10.1063/5.0074848}. 

The other Raman-active $E_g$ mode involves an in-plane shearing displacement of the \ce{Te} sublayers that reduces the lateral offset between \ce{Cr} and \ce{Te}   layers, which enhances the overlap between \ce{Te}-$p_z$ and \ce{Cr}-$d_{z^2}$ orbitals. This strengthens the ferromagnetic exchange pathways. The resulting increase in the force constants manifests as a systematic blueshift of the $E_g$ mode upon the magnetic ordering. Since the phonon renormalizaton is relatively weaker for the  $E_g$ mode, the calculated $\lambda_{{E_g}}$ is correspondingly small in magnitude, below 2 cm$^{-1}$  [Figure~\ref{fig:figure6}(e)], comparable to values reported for \ce{CrI3} and \ce{Cr2Ge2Te6}~\citep{tian2016magneto,10.1063/5.0074848}.

A similar mechanism governs the behavior of the infrared active modes. The $A_{2u}$ mode [Figure~\ref{fig:figure6}(c)], characterized by out-of-plane relative displacements of the \ce{Cr} and \ce{Te} layers, primarily modulates the ferromagnetic superexchange interaction by driving the \ce{Cr-Te-Cr} bond angles away from the near-$90^{\circ}$ geometry favorable for ferromagnetic SE coupling. The resulting reduction of FM superexchange manifests as phonon softening, analogous to the $A_{1g}$ mode. Correspondingly, the strong coupling observed at smaller lattice parameters, with a magnitude of $|\lambda_{A_{2u}}| \sim 25$ cm$^{-1}$, gradually decreases with increasing $a$, consistent with the weakening of superexchange at larger $a$  [Figure~\ref{fig:figure6}(f)].  
The $E_u$ mode affects both superexchange and ligand-hole-mediated double exchange in a manner similar to the $E_g$ mode but with a substantially stronger coupling.  The opposite motion of the \ce{Cr} layer relative to the \ce{Te} sublayers leads to a more pronounced enhancement of the $p_z-d_{z^2}$ orbital overlap, resulting in a significant increase in the magnetic contribution to the force constants and a marked hardening of the mode [Figure~\ref{fig:figure6}(d)]. As $a$ increases, the ligand-hole-mediated double exchange becomes increasingly dominant, resulting in a nearly linear increase of $\lambda_{E_u}$, reaching $\sim$ 45 cm$^{-1}$ [Figure~\ref{fig:figure6}(f)]. 

Phonon frequency renormalization $\Delta\omega_{\nu}$ in the magnetic ground state [Figure~\ref{fig:figure6}(g) and ~\ref{fig:figure6}(h)] exhibits distinct signatures that are experimentally accessible through temperature-dependent Raman and IR spectroscopies, providing a useful probe of magnetic order. In the low-$a$ regime characterized by Z-AFM order, $\Delta\omega_{A_{1g}}$ remains modest at approximately  10 cm$^{-1}$, which increases sharply to $|\Delta\omega_{A_{1g}}| > 35$ cm$^{-1}$ upon the transition to FM order at larger lattice parameters [Figure~\ref{fig:figure6}(g)]. This indicates that tracking the $A_{1g}$ Raman mode provides a primary metric for identifying the magnetic state. A similar trend is observed for the IR-active $E_u$  mode [Figure~\ref{fig:figure6}(h)]. Relatively smaller renormalization, $\Delta\omega_{E_u} = 30 - 65$ cm$^{-1}$, signifies the underlying Z-AFM order, whereas higher values of $75 - 110$ cm$^{-1}$ indicate the FM phase.  Such large phonon renormalization substantially exceeds values reported for other 2D magnets, including the strongly coupled systems such as \ce{Cr2Ge2Te6} and \ce{CoPS3}~\citep{tian2016magneto, PhysRevB.103.235411}. Instead, it is comparable to phonon anomalies observed in bulk perovskites such as \ce{NaOsO3} and \ce{Sr2CrReO6}, where large spin-orbit coupling and strong electron correlations cooperatively drive the spin-phonon interaction~\citep{ncomms9916,PhysRevLett.108.177202}.  While the $E_g$ and $A_{2u}$ modes exhibit clear spin-phonon coupling, they lack a definitive trend for differentiating between magnetic configurations. Furthermore, as the DS-AFM phase occupies a narrow region of the phase diagram [Figure~\ref{fig:figure2}(c)], its associated $\Delta\omega$ values are nearly indistinguishable from those of the Z-AFM and FM phases near the phase boundaries, limiting its unambiguous identification based on phonon renormalization alone. 

Carrier doping does not qualitatively alter the lattice parameter dependence $\omega_{\nu}(a)$ or the relative phonon renormalization $\Delta\omega_{\nu}$ across magnetic phases, confirming that the mode-dependent spin-phonon coupling is robust against charge doping (Supplemental Information~\cite{supple}).  Although carrier doping modifies the strain-dependent magnetic phase diagram (Figure~\ref{fig:figure2}), the characteristic spectroscopic markers remain largely unchanged under both electron and hole doping. 

\subsubsection{Spin-lattice coupling}
While the phonon frequency renormalization and the associated spin-phonon coupling constant $\lambda_{\nu}$ provide clear signatures of spin-lattice coupling, they offer only an indirect view of its microscopic origin. Further quantitative understanding can be obtained by directly examining how the magnetic exchange interactions respond to lattice distortions. In particular, the strain derivatives of the exchange couplings govern the evolution of magnetic ground states under lattice deformation. The second derivatives with respect to the atom displacements, projected onto the phonon eigenvectors, control the magnetic contribution to the interatomic force constants, and hence the phonon renormalization $\Delta\omega_{\nu}$. 

\begin{table}[!t]
\centering
\caption{Derivatives of exchange parameters $J$ with respect to the lattice parameter $a$, $J_k^{\prime} \equiv \partial J_k/\partial a$ (\SI[per-mode=symbol]{}{\milli\electronvolt\per\angstrom}), computed for different conditions. The results indicate strong spin-lattice coupling, with $J_1$ playing the dominant role in the strain-induced magnetic phase transitions.}
\label{tab:dJ}
\renewcommand{\arraystretch}{1.2}
\setlength{\tabcolsep}{8pt}
\begin{tabular}{crrr}
\hline\hline
& Neutral & Hole doped & Electron doped \\
\hline
$J_1^{\prime}$ & 43.6 & 51.5 & 41.9 \\
$J_2^{\prime}$ & $-$0.9 & $-$5.5 & $-$2.4 \\
$J_3^{\prime}$ & 11.8 & 5.7 & 9.6 \\
\hline\hline
\end{tabular}
\end{table}

Within the nearest-neighbor Heisenberg spin model, we expand the corresponding magnetic energy $E_{\altmathcal{H}}(u_m, \eta_k, \bm{S}_i)$ to the second order in small atomic displacement $u_m$ and strain $\eta_k$, 
\begin{equation}
\begin{split}
&E_{\altmathcal{H}} = E_{\altmathcal{H}}^0 + \sum_{\mathclap{\substack{\langle i, j \rangle\\ m}}}\frac{\partial J_1}{\partial u_m} \bm{S}_i \cdot \bm{S}_j u_m + \sum_{\mathclap{\substack{\langle i, j \rangle \\ k}}}\frac{\partial J_1}{\partial \eta_k} \bm{S}_i \cdot \bm{S}_j \eta_k \nonumber \\
 &+ \frac{1}{2}\sum_{\mathclap{\substack{\langle i, j \rangle \\ m, n}}}\frac{\partial^2 J_1}{\partial u_m \partial u_n} \bm{S}_i \cdot \bm{S}_j u_m u_n + \frac{1}{2}\sum_{\mathclap{\substack{\langle i, j \rangle \\ k, l}}}\frac{\partial^2 J_1}{\partial \eta_k \partial \eta_l} \bm{S}_i\cdot\bm{S}_j  \eta_k \eta_l \nonumber \\
 &+ \sum_{\mathclap{\substack{\langle i, j \rangle \\ m, k}}} \frac{\partial^2 J_1}{\partial u_m \partial \eta_k} \bm{S}_i \cdot \bm{S}_j  u_m \eta_k + \altmathcal{O}(\geqslant 3).\nonumber
\end{split}
\end{equation}
$E_{\altmathcal{H}}^0$ represents the magnetic energy in the absence of atomic displacements, $u_m = 0$, and strain, $\eta_k=0$. The same formalism can be readily extended to incorporate the second- and third-neighbor exchange interactions, $J_2$ and $J_3$.

Rather than computing the strain derivative $\partial J/\partial \eta$ directly, we evaluate $\partial J/\partial a$ (Table~\ref{tab:dJ}). In the case of uniform biaxial strain, these quantities are strictly proportional and carry equivalent magnetoelastic information. For the neutral monolayer [Figure~\ref{fig:figure1}(f)], the large positive derivative $\partial J_1/\partial a \sim $ 43 meV/\AA\ indicates strong magnetoelastic coupling, implying the FM contribution in $J_1$ increases rapidly with lattice expansion and drives an AFM to FM transition at $a\sim 3.7$ \AA.  Such strong magnetoelastic coupling substantially exceeds the values reported for other vdW magnets, such as \ce{CrI3} and \ce{Cr2Ge2Te6}~\cite{PhysRevB.105.104418, PhysRevB.100.224427}.  In contrast, $J_2$ remains largely insensitive to strain with $\partial J_2/\partial a \sim $ $-$1 meV/\AA.  Although $J_3$ shows a somewhat larger response with  $\partial J_3/\partial a \sim $ 12 meV/\AA\, it is significantly weaker than $J_1$. These trends underscore the dominant role of  $J_1$ in driving the strain-induced magnetic phase transitions in monolayer \ce{CrTe_2}, a qualitative behavior that persists under carrier doping.

 \begin{table}[!t]
\centering
\caption{Calculated second derivatives of the exchange parameters,  $J_k^{\prime\prime} \equiv \partial^2 J_k/\partial Q^2_{\nu}$ (\SI[per-mode=symbol]{}{\milli\electronvolt\per\angstrom^2}) for each phonon mode $\nu$. For the Raman-active $A_{1g}$ mode, $J^{\prime\prime}_k = \partial^2 J_k/\partial u_{\rm Te}\partial u_{\rm Te}$ and  $J^{\prime\prime}_k = \partial^2 J_k/\partial u_{\rm Cr}\partial u_{\rm Te}$ for the IR-active $A_{2u}$ and $E_u$ modes. 
}
\label{tab:d2J}
\renewcommand{\arraystretch}{1.2}
\setlength{\tabcolsep}{6pt}
\begin{tabular}{cccccccc}
\hline\hline

\multirow{2}{*}{Mode} 
& \multicolumn{3}{c}{$a = 3.60$ \AA\ (Z-AFM)} &
& \multicolumn{3}{c}{$a = 3.90$ \AA\ (FM)} \\
\cline{2-4} \cline{6-8}

& $J_1^{\prime\prime}$ & $J_2^{\prime\prime}$ & $J_3^{\prime\prime}$ &
& $J_1^{\prime\prime}$ & $J_2^{\prime\prime}$ & $J_3^{\prime\prime}$ \\
\hline
$A_{1g}$
& 633 & 220   & 284 &
& 858 & --    & 603   \\
$A_{2u}$
& 1637 & 591 & 404 &
& 1276 & 474 & 450  \\
$E_u$
& 386 & --    & 253 &
& 464  & --    & 324   \\
\hline\hline
\end{tabular}
\end{table}

Having established the dominant role of $J_1$ in the strain response, we now turn to its second derivatives with respect to the phonon normal modes $Q_{\nu}$ (Table~\ref{tab:d2J}). For each eigenmode $\nu$, the term $\partial^2 J/\partial Q^2_{\nu}$ governs the magnetic contribution to the phonon force constants and is thus directly proportional to the spin-phonon coupling constant, $\lambda_{\nu}$, since $\Delta\omega_{\nu} \propto \sum_{ij}(\partial^2J_{ij}/\partial Q^2_{\nu}) \langle \bm{S}_i \cdot \bm{S}_j\rangle / \omega_{\nu}$. Consistent with the trends observed in $\lambda_{\nu}$, modes exhibit sizeable $\partial^2 J/\partial Q^2_{\nu}$, aligning with the qualitative mode-dependent modulation of ferromagnetic exchange discussed earlier. In particular, $\partial^2 J_1/\partial u_{\rm Te}\partial u_{\rm Te}$ associated with the $A_{1g}$ mode increases sharply with lattice parameter, reaching values as large as \SI{858}{\milli\electronvolt\per\angstrom^2}. This pronounced enhancement reflects the growing sensitivity of the ferromagnetic exchange to the symmetric breathing distortions of the \ce{Cr-Te} bonds at expanded lattice constants. Furthermore, the first-neighbor exchange couples strongly to the IR-active $A_{2u}$ and $E_u$ modes. The calculated  $\partial^2 J_1/\partial u_{\rm Cr}\partial u_{\rm Te} \sim $ \SI{1637}{\milli\electronvolt\per\angstrom^2} for the $A_{2u}$ mode is largest in the low-$a$ regime and decreases with increasing $a$, whereas the $E_u$ mode exhibits the opposite trend. Overall, the consistent behavior of $\partial^2 J/\partial Q^2_{\nu}$ and $\lambda_{\nu}$ across all modes corroborates the microscopic exchange mechanisms governing the sensitivity of $J_1$ to lattice distortions.

A similar qualitative trend in $\partial^2 J/\partial Q^2_{\nu}$ is observed for $J_2$ and $J_3$ exchanges (Table~\ref{tab:d2J}). However, the magnitudes are substantially smaller across all modes, once again underscoring the dominant role of the nearest-neighbor exchange interaction in mediating the spin-lattice couplings in monolayer \ce{CrTe2}. Notably, the computed $\partial^2 J/\partial Q^2_{\nu}$ values are orders of magnitude larger than those reported for other 2D magnets, such as \ce{Cr2Ge2Te6}~\citep{PhysRevB.100.224427}, as well as prototypical strongly correlated oxides like \ce{LaMnO3}~\citep{PhysRevB.60.11879}. These results place monolayer \ce{CrTe2} among the most strongly spin-phonon-coupled materials characterized to date.

\begin{figure}[t]
\centering
\includegraphics[scale=1]{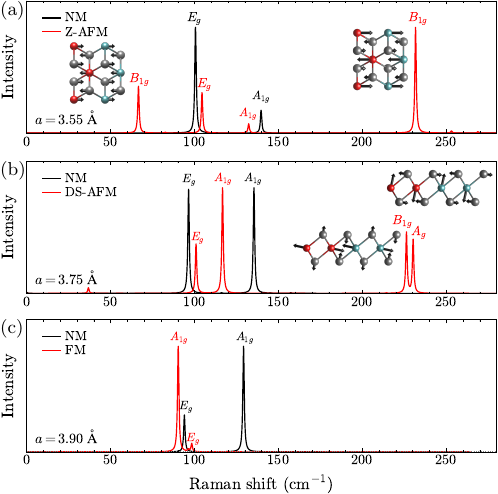}
\caption{
Simulated Raman spectra for monolayer \ce{CrTe2} in the (a) zigzag AFM ($a$ = 3.55 \AA), (b) double-stripe  AFM ($a$ = 3.75 \AA) and (c) FM ($a$ = 3.90 \AA) phases. Spectra for the corresponding non-magnetic structures are provided for comparison. Beyond the characteristic $A_{1g}$ and $E_g$ structural modes, the Z-AFM and DS-AFM phases exhibit unique zone-folded Raman-active modes arising from the enlargement of the magnetic unit cell.  In contrast, the FM phase retains the translational symmetry of the chemical unit cell, resulting in no additional modes.  These results demonstrate that Raman spectroscopy can effectively distinguish between competing AFM phases. 
}
\label{fig:Raman}
\end{figure}

\subsubsection{Zone-folded phonon modes and magnetic ordering}
The emergence of zone-folded Raman-active modes provides a unique, symmetry-sensitive fingerprint of the underlying magnetic order, offering resolving capabilities beyond those of phonon renormalization alone (Figure~\ref{fig:Raman}). A similar approach has been successfully employed to identify interlayer AFM order in \ce{CrI3}~\citep{s41565-019-0598-4,s41467-018-07547-6}  and intralayer Z-AFM order in \ce{FePS3}~\citep{nanolett.6b03052}, \ce{NiPS3}~\citep{s41467-018-08284-6}, and XY-type AFM order in \ce{CoPS3}~\citep{PhysRevB.103.235411}. 

Two zone-folded $B_{1g}$ modes are identified in the Z-AFM phase [Figure~\ref{fig:Raman}(a)], both sharing the same displacement topology with in-plane displacements directed perpendicular to the FM chains.  The intrachain \ce{Cr} displacements $\mathbf{u}_{\ce{Cr}}$ are antiparallel within the FM chains and parallel across the AFM-coupled chains. The magnetic cell doubling splits the single \ce{Te} Wyckoff site in the structural space group $P\bar{3}m1$ into two inequivalent sites, intrachain and interchain bridging \ce{Te}, both participating in the mode eigenvectors. Both intrachain and interchain bridging \ce{Te} displacements $\mathbf{u}_{\ce{Te}}$ are antiparallel. These modes modulate both direct \ce{Cr-Cr} and indirect \ce{Cr-Te-Cr} exchange channels. The high-frequency $B_{1g}$ mode at 230 cm$^{-1}$ is dominated by \ce{Cr} displacements ($\lvert \mathbf{u}_{\ce{Cr}}\rvert > \lvert \mathbf{u}_{\ce{Te}}\rvert $), whereas the low-frequency mode at 70 cm$^{-1}$ exhibits comparable displacement amplitudes for both atomic species, $\lvert \mathbf{u}_{\ce{Cr}}\rvert  \approx \lvert \mathbf{u}_{\ce{Te}}\rvert $. These modes have no net dipole moment and are thus Raman active (Figure~\ref{fig:Raman}), and are exclusively observed for the Z-AFM phase, providing an unambiguous signature absent above $T_{\rm N}$. 

In the DS-AFM phase, two zone-folded modes are identified near 230 cm$^{-1}$ [Figure~\ref{fig:Raman}(b)],  both exhibiting mixed eigenvectors that combine in-plane and out-of-plane atomic displacements.  While such a mixing is strictly forbidden in the structural phase, the off-diagonal force constants, specifically $\partial^2 E/\partial u^z_{\ce{Cr}} \partial u^x_{\ce{Te}}$  and $\partial^2 E/\partial u^x_{\ce{Cr}} \partial u^z_{\ce{Te}}$, become symmetry allowed in the DS-AFM phase, driving the mixed eigenvectors through exchange magnetostriction.  The in-plane displacements in both modes are oriented perpendicular to the FM chains. Furthermore, both modes are \ce{Cr}-dominated, with a displacement ratio of $u_{\ce{Cr}}/u_{\ce{Te}} \sim 5$. The first mode transforms as $B_{1g}$ symmetry and couples in-plane $\mathbf{u}_{\ce{Cr}}$ with out-of-plane $\mathbf{u}_{\ce{Te}}$. The complementary $A_g$ mode combines out-of-plane $\mathbf{u}_{\ce{Cr}}$ with in-plane $\mathbf{u}_{\ce{Te}}$. These vibrations modulate the \ce{Cr-Te} bond lengths, thereby altering $t_{pd}$ hopping, particularly through the $d_{z^2}-p_z$ overlap, and consequently, the magnetic exchanges. The two modes remain nearly degenerate with a splitting $\Delta\omega$ of only 5 cm$^{-1}$, consistent with the flat phonon dispersion along $\Gamma-\rm M$ in the structural phase. This near-degeneracy distinguishes the double-stripe phase from the zigzag phase, even though the high-energy vibrational features in the latter appear at a comparable spectral energy. Collectively, these zone-folded modes, activated purely by the DS-AFM order, constitute a definitive spectroscopic fingerprint of this magnetic phase. Notably, this magnetism-induced hybridization of in-plane and out-of-plane modes mirrors the symmetry-breaking in iron pnictides~\citep{PhysRevLett.106.067002}, where the transition to a stripe-AFM phase similarly activates off-diagonal force constants, coupling previously orthogonal vibrational sectors.

In contrast to the zigzag and DS-AFM phases, the FM phase does not generate additional Raman-active modes and retains only the renormalized $A_{1g}$ and $E_g$ modes of the structural phase [Figure~\ref{fig:Raman}(c)]. The ferromagnetic spin configuration preserves the crystallographic unit cell while maximizing the phonon renormalization $\Delta\omega$ of the zone-center $A_{1g}$ mode, as all nearest-neighbor spin correlations $\langle \mathbf{S}_i \cdot \mathbf{S}_j \rangle = S^2$ contribute constructively at $\Gamma$. These results establish Raman spectroscopy as a comprehensive and noninvasive probe for identifying diverse magnetic phases in \ce{CrTe2} monolayers.

\section{Summary and Conclusions}
In summary, we have systematically mapped the magnetic landscape of monolayer \ce{CrTe2}, uncovering a remarkably rich phase diagram governed by the interplay of lattice strain and carrier density. These dual perturbations, experimentally accessible via epitaxial growth and electrostatic gating, serve as powerful knobs to tune the electronic and magnetic properties. We show that the resulting magnetic evolution is dictated by a delicate competition among several microscopic interaction channels, including direct exchange, (higher-order) superexchange, RKKY, and ligand-mediated double exchange. By disentangling these pathways, we provide a unified framework that explains how structural and electronic perturbations reconfigure magnetic order, while specifically highlighting the active role of ligand states in mediating long-range order. This ligand-centric perspective differentiates the physics of \ce{CrTe2} from simpler metal-only models and underscores the importance of $pd$ hybridization in 2D magnetism. Crucially, the results reconcile the seemingly contradictory experimental observations of ferromagnetic and zigzag AFM phases in epitaxial monolayers. Furthermore, we predict the emergence of a novel double-stripe AFM phase that is further stabilized by electron doping. Reminiscent of the complex magnetic textures found in iron pnictide superconductors, this phase offers a new frontier for exploring exotic magnetism in 2D materials. Ultimately, the long-range Heisenberg spin model not only explains current experimental results but also predicts the realization of room-temperature ferromagnetism and antiferromagnetism at specific lattice-parameter limits, making \ce{CrTe2} a prime candidate for high-temperature spintronic applications. 

The high sensitivity of the magnetic ground state to lattice strain reveals an exceptionally strong magnetoelastic coupling in \ce{CrTe2}, placing it among the most mechanically tunable 2D magnets. We have quantified this coupling and elucidated its microscopic origins by analyzing phonon renormalization across the magnetic transitions. By evaluating mode-dependent spin-phonon and spin-lattice coupling constants, we demonstrate how the lattice dynamics are intrinsically intertwined with the underlying spin correlations. Furthermore, we identify specific zone-folded Raman modes that arise directly from the magnetic symmetry breaking. These modes serve as unique experimental fingerprints that distinguish between the FM, zigzag AFM, and double-stripe AFM phases. These findings establish Raman spectroscopy as a definitive probe for navigating the complex magnetic phase space and provide a clear roadmap for the experimental verification of these predictions.

\section{Acknowledgements}
A.S. acknowledges the University Grants Commission, India, for support through a research fellowship. We sincerely acknowledge the support and resources provided by the PARAM Brahma Facility at the Indian Institute of Science Education and Research, Pune, under the National Supercomputing Mission of the Government of India. Additionally, we acknowledge the funding from the National Mission on Interdisciplinary Cyber-Physical Systems (NM-ICPS) of the Department of Science and Technology, Government of India, through the I-HUB Quantum Technology Foundation, Pune, India. 

\bibliography{CrTe2.bib}

\begin{thebibliography}{113}%
\makeatletter
\providecommand \@ifxundefined [1]{%
 \@ifx{#1\undefined}
}%
\providecommand \@ifnum [1]{%
 \ifnum #1\expandafter \@firstoftwo
 \else \expandafter \@secondoftwo
 \fi
}%
\providecommand \@ifx [1]{%
 \ifx #1\expandafter \@firstoftwo
 \else \expandafter \@secondoftwo
 \fi
}%
\providecommand \natexlab [1]{#1}%
\providecommand \enquote  [1]{``#1''}%
\providecommand \bibnamefont  [1]{#1}%
\providecommand \bibfnamefont [1]{#1}%
\providecommand \citenamefont [1]{#1}%
\providecommand \href@noop [0]{\@secondoftwo}%
\providecommand \href [0]{\begingroup \@sanitize@url \@href}%
\providecommand \@href[1]{\@@startlink{#1}\@@href}%
\providecommand \@@href[1]{\endgroup#1\@@endlink}%
\providecommand \@sanitize@url [0]{\catcode `\\12\catcode `\$12\catcode
  `\&12\catcode `\#12\catcode `\^12\catcode `\_12\catcode `\%12\relax}%
\providecommand \@@startlink[1]{}%
\providecommand \@@endlink[0]{}%
\providecommand \url  [0]{\begingroup\@sanitize@url \@url }%
\providecommand \@url [1]{\endgroup\@href {#1}{\urlprefix }}%
\providecommand \urlprefix  [0]{URL }%
\providecommand \Eprint [0]{\href }%
\providecommand \doibase [0]{https://doi.org/}%
\providecommand \selectlanguage [0]{\@gobble}%
\providecommand \bibinfo  [0]{\@secondoftwo}%
\providecommand \bibfield  [0]{\@secondoftwo}%
\providecommand \translation [1]{[#1]}%
\providecommand \BibitemOpen [0]{}%
\providecommand \bibitemStop [0]{}%
\providecommand \bibitemNoStop [0]{.\EOS\space}%
\providecommand \EOS [0]{\spacefactor3000\relax}%
\providecommand \BibitemShut  [1]{\csname bibitem#1\endcsname}%
\let\auto@bib@innerbib\@empty
\bibitem [{\citenamefont {Hohenberg}(1967)}]{PhysRev.158.383}%
  \BibitemOpen
  \bibfield  {author} {\bibinfo {author} {\bibfnamefont {P.~C.}\ \bibnamefont
  {Hohenberg}},\ }\bibfield  {title} {\bibinfo {title} {Existence of long-range
  order in one and two dimensions},\ }\href
  {https://link.aps.org/doi/10.1103/PhysRev.158.383} {\bibfield  {journal}
  {\bibinfo  {journal} {Phys. Rev.}\ }\textbf {\bibinfo {volume} {158}},\
  \bibinfo {pages} {383} (\bibinfo {year} {1967})}\BibitemShut {NoStop}%
\bibitem [{\citenamefont {Mermin}\ and\ \citenamefont
  {Wagner}(1966)}]{PhysRevLett.17.1133}%
  \BibitemOpen
  \bibfield  {author} {\bibinfo {author} {\bibfnamefont {N.~D.}\ \bibnamefont
  {Mermin}}\ and\ \bibinfo {author} {\bibfnamefont {H.}~\bibnamefont
  {Wagner}},\ }\bibfield  {title} {\bibinfo {title} {Absence of ferromagnetism
  or antiferromagnetism in one- or two-dimensional isotropic {H}eisenberg
  models},\ }\href {https://link.aps.org/doi/10.1103/PhysRevLett.17.1133}
  {\bibfield  {journal} {\bibinfo  {journal} {Phys. Rev. Lett.}\ }\textbf
  {\bibinfo {volume} {17}},\ \bibinfo {pages} {1133} (\bibinfo {year}
  {1966})}\BibitemShut {NoStop}%
\bibitem [{\citenamefont {Lee}\ \emph {et~al.}(2016)\citenamefont {Lee},
  \citenamefont {Lee}, \citenamefont {Ryoo}, \citenamefont {Kang},
  \citenamefont {Kim}, \citenamefont {Kim}, \citenamefont {Park}, \citenamefont
  {Park},\ and\ \citenamefont {Cheong}}]{nanolett.6b03052}%
  \BibitemOpen
  \bibfield  {author} {\bibinfo {author} {\bibfnamefont {J.-U.}\ \bibnamefont
  {Lee}}, \bibinfo {author} {\bibfnamefont {S.}~\bibnamefont {Lee}}, \bibinfo
  {author} {\bibfnamefont {J.~H.}\ \bibnamefont {Ryoo}}, \bibinfo {author}
  {\bibfnamefont {S.}~\bibnamefont {Kang}}, \bibinfo {author} {\bibfnamefont
  {T.~Y.}\ \bibnamefont {Kim}}, \bibinfo {author} {\bibfnamefont
  {P.}~\bibnamefont {Kim}}, \bibinfo {author} {\bibfnamefont {C.-H.}\
  \bibnamefont {Park}}, \bibinfo {author} {\bibfnamefont {J.-G.}\ \bibnamefont
  {Park}},\ and\ \bibinfo {author} {\bibfnamefont {H.}~\bibnamefont {Cheong}},\
  }\bibfield  {title} {\bibinfo {title} {Ising-type magnetic ordering in
  atomically thin \ce{FePS3}},\ }\href
  {https://doi.org/10.1021/acs.nanolett.6b03052} {\bibfield  {journal}
  {\bibinfo  {journal} {Nano Lett.}\ }\textbf {\bibinfo {volume} {16}},\
  \bibinfo {pages} {7433} (\bibinfo {year} {2016})}\BibitemShut {NoStop}%
\bibitem [{\citenamefont {Gong}\ \emph {et~al.}(2017)\citenamefont {Gong},
  \citenamefont {Li}, \citenamefont {Li}, \citenamefont {Ji}, \citenamefont
  {Stern}, \citenamefont {Xia}, \citenamefont {Cao}, \citenamefont {Bao},
  \citenamefont {Wang}, \citenamefont {Wang}, \citenamefont {Qiu},
  \citenamefont {Cava}, \citenamefont {Louie}, \citenamefont {Xia},\ and\
  \citenamefont {Zhang}}]{nature22060}%
  \BibitemOpen
  \bibfield  {author} {\bibinfo {author} {\bibfnamefont {C.}~\bibnamefont
  {Gong}}, \bibinfo {author} {\bibfnamefont {L.}~\bibnamefont {Li}}, \bibinfo
  {author} {\bibfnamefont {Z.}~\bibnamefont {Li}}, \bibinfo {author}
  {\bibfnamefont {H.}~\bibnamefont {Ji}}, \bibinfo {author} {\bibfnamefont
  {A.}~\bibnamefont {Stern}}, \bibinfo {author} {\bibfnamefont
  {Y.}~\bibnamefont {Xia}}, \bibinfo {author} {\bibfnamefont {T.}~\bibnamefont
  {Cao}}, \bibinfo {author} {\bibfnamefont {W.}~\bibnamefont {Bao}}, \bibinfo
  {author} {\bibfnamefont {C.~e.}\ \bibnamefont {Wang}}, \bibinfo {author}
  {\bibfnamefont {Y.}~\bibnamefont {Wang}}, \bibinfo {author} {\bibfnamefont
  {Z.~Q.}\ \bibnamefont {Qiu}}, \bibinfo {author} {\bibfnamefont {R.~J.}\
  \bibnamefont {Cava}}, \bibinfo {author} {\bibfnamefont {S.~G.}\ \bibnamefont
  {Louie}}, \bibinfo {author} {\bibfnamefont {J.}~\bibnamefont {Xia}},\ and\
  \bibinfo {author} {\bibfnamefont {X.}~\bibnamefont {Zhang}},\ }\bibfield
  {title} {\bibinfo {title} {Discovery of intrinsic ferromagnetism in
  two-dimensional van der {W}aals crystals},\ }\href
  {https://doi.org/10.1038/nature22060} {\bibfield  {journal} {\bibinfo
  {journal} {Nature}\ }\textbf {\bibinfo {volume} {546}},\ \bibinfo {pages}
  {265} (\bibinfo {year} {2017})}\BibitemShut {NoStop}%
\bibitem [{\citenamefont {Huang}\ \emph {et~al.}(2017)\citenamefont {Huang},
  \citenamefont {Clark}, \citenamefont {Navarro-Moratalla}, \citenamefont
  {Klein}, \citenamefont {Cheng}, \citenamefont {Seyler}, \citenamefont
  {Zhong}, \citenamefont {Schmidgall}, \citenamefont {McGuire}, \citenamefont
  {Cobden}, \citenamefont {Yao}, \citenamefont {Xiao}, \citenamefont
  {Jarillo-Herrero},\ and\ \citenamefont {Xu}}]{nature22391}%
  \BibitemOpen
  \bibfield  {author} {\bibinfo {author} {\bibfnamefont {B.}~\bibnamefont
  {Huang}}, \bibinfo {author} {\bibfnamefont {G.}~\bibnamefont {Clark}},
  \bibinfo {author} {\bibfnamefont {E.}~\bibnamefont {Navarro-Moratalla}},
  \bibinfo {author} {\bibfnamefont {D.~R.}\ \bibnamefont {Klein}}, \bibinfo
  {author} {\bibfnamefont {R.}~\bibnamefont {Cheng}}, \bibinfo {author}
  {\bibfnamefont {K.~L.}\ \bibnamefont {Seyler}}, \bibinfo {author}
  {\bibfnamefont {D.}~\bibnamefont {Zhong}}, \bibinfo {author} {\bibfnamefont
  {E.}~\bibnamefont {Schmidgall}}, \bibinfo {author} {\bibfnamefont {M.~A.}\
  \bibnamefont {McGuire}}, \bibinfo {author} {\bibfnamefont {D.~H.}\
  \bibnamefont {Cobden}}, \bibinfo {author} {\bibfnamefont {W.}~\bibnamefont
  {Yao}}, \bibinfo {author} {\bibfnamefont {D.}~\bibnamefont {Xiao}}, \bibinfo
  {author} {\bibfnamefont {P.}~\bibnamefont {Jarillo-Herrero}},\ and\ \bibinfo
  {author} {\bibfnamefont {X.}~\bibnamefont {Xu}},\ }\bibfield  {title}
  {\bibinfo {title} {Layer-dependent ferromagnetism in a van der {W}aals
  crystal down to the monolayer limit},\ }\href
  {https://doi.org/10.1038/nature22391} {\bibfield  {journal} {\bibinfo
  {journal} {Nature}\ }\textbf {\bibinfo {volume} {546}},\ \bibinfo {pages}
  {270} (\bibinfo {year} {2017})}\BibitemShut {NoStop}%
\bibitem [{\citenamefont {Gibertini}\ \emph {et~al.}(2019)\citenamefont
  {Gibertini}, \citenamefont {Koperski}, \citenamefont {Morpurgo},\ and\
  \citenamefont {Novoselov}}]{Gibertini2019}%
  \BibitemOpen
  \bibfield  {author} {\bibinfo {author} {\bibfnamefont {M.}~\bibnamefont
  {Gibertini}}, \bibinfo {author} {\bibfnamefont {M.}~\bibnamefont {Koperski}},
  \bibinfo {author} {\bibfnamefont {A.~F.}\ \bibnamefont {Morpurgo}},\ and\
  \bibinfo {author} {\bibfnamefont {K.~S.}\ \bibnamefont {Novoselov}},\
  }\bibfield  {title} {\bibinfo {title} {Magnetic {2D} materials and
  heterostructures},\ }\href {https://doi.org/10.1038/s41565-019-0438-6}
  {\bibfield  {journal} {\bibinfo  {journal} {Nat. Nanotechnol.}\ }\textbf
  {\bibinfo {volume} {14}},\ \bibinfo {pages} {408} (\bibinfo {year}
  {2019})}\BibitemShut {NoStop}%
\bibitem [{\citenamefont {Wang}\ \emph {et~al.}(2022)\citenamefont {Wang},
  \citenamefont {Bedoya-Pinto}, \citenamefont {Blei}, \citenamefont {Dismukes},
  \citenamefont {Hamo}, \citenamefont {Jenkins}, \citenamefont {Koperski},
  \citenamefont {Liu}, \citenamefont {Sun}, \citenamefont {Telford},
  \citenamefont {Kim}, \citenamefont {Augustin}, \citenamefont {Vool},
  \citenamefont {Yin}, \citenamefont {Li}, \citenamefont {Falin}, \citenamefont
  {Dean}, \citenamefont {Casanova}, \citenamefont {Evans}, \citenamefont
  {Chshiev}, \citenamefont {Mishchenko}, \citenamefont {Petrovic},
  \citenamefont {He}, \citenamefont {Zhao}, \citenamefont {Tsen}, \citenamefont
  {Gerardot}, \citenamefont {Brotons-Gisbert}, \citenamefont {Guguchia},
  \citenamefont {Roy}, \citenamefont {Tongay}, \citenamefont {Wang},
  \citenamefont {Hasan}, \citenamefont {Wrachtrup}, \citenamefont {Yacoby},
  \citenamefont {Fert}, \citenamefont {Parkin}, \citenamefont {Novoselov},
  \citenamefont {Dai}, \citenamefont {Balicas},\ and\ \citenamefont
  {Santos}}]{acsnano.1c09150}%
  \BibitemOpen
  \bibfield  {author} {\bibinfo {author} {\bibfnamefont {Q.~H.}\ \bibnamefont
  {Wang}}, \bibinfo {author} {\bibfnamefont {A.}~\bibnamefont {Bedoya-Pinto}},
  \bibinfo {author} {\bibfnamefont {M.}~\bibnamefont {Blei}}, \bibinfo {author}
  {\bibfnamefont {A.~H.}\ \bibnamefont {Dismukes}}, \bibinfo {author}
  {\bibfnamefont {A.}~\bibnamefont {Hamo}}, \bibinfo {author} {\bibfnamefont
  {S.}~\bibnamefont {Jenkins}}, \bibinfo {author} {\bibfnamefont
  {M.}~\bibnamefont {Koperski}}, \bibinfo {author} {\bibfnamefont
  {Y.}~\bibnamefont {Liu}}, \bibinfo {author} {\bibfnamefont {Q.-C.}\
  \bibnamefont {Sun}}, \bibinfo {author} {\bibfnamefont {E.~J.}\ \bibnamefont
  {Telford}}, \bibinfo {author} {\bibfnamefont {H.~H.}\ \bibnamefont {Kim}},
  \bibinfo {author} {\bibfnamefont {M.}~\bibnamefont {Augustin}}, \bibinfo
  {author} {\bibfnamefont {U.}~\bibnamefont {Vool}}, \bibinfo {author}
  {\bibfnamefont {J.-X.}\ \bibnamefont {Yin}}, \bibinfo {author} {\bibfnamefont
  {L.~H.}\ \bibnamefont {Li}}, \bibinfo {author} {\bibfnamefont
  {A.}~\bibnamefont {Falin}}, \bibinfo {author} {\bibfnamefont {C.~R.}\
  \bibnamefont {Dean}}, \bibinfo {author} {\bibfnamefont {F.}~\bibnamefont
  {Casanova}}, \bibinfo {author} {\bibfnamefont {R.~F.~L.}\ \bibnamefont
  {Evans}}, \bibinfo {author} {\bibfnamefont {M.}~\bibnamefont {Chshiev}},
  \bibinfo {author} {\bibfnamefont {A.}~\bibnamefont {Mishchenko}}, \bibinfo
  {author} {\bibfnamefont {C.}~\bibnamefont {Petrovic}}, \bibinfo {author}
  {\bibfnamefont {R.}~\bibnamefont {He}}, \bibinfo {author} {\bibfnamefont
  {L.}~\bibnamefont {Zhao}}, \bibinfo {author} {\bibfnamefont {A.~W.}\
  \bibnamefont {Tsen}}, \bibinfo {author} {\bibfnamefont {B.~D.}\ \bibnamefont
  {Gerardot}}, \bibinfo {author} {\bibfnamefont {M.}~\bibnamefont
  {Brotons-Gisbert}}, \bibinfo {author} {\bibfnamefont {Z.}~\bibnamefont
  {Guguchia}}, \bibinfo {author} {\bibfnamefont {X.}~\bibnamefont {Roy}},
  \bibinfo {author} {\bibfnamefont {S.}~\bibnamefont {Tongay}}, \bibinfo
  {author} {\bibfnamefont {Z.}~\bibnamefont {Wang}}, \bibinfo {author}
  {\bibfnamefont {M.~Z.}\ \bibnamefont {Hasan}}, \bibinfo {author}
  {\bibfnamefont {J.}~\bibnamefont {Wrachtrup}}, \bibinfo {author}
  {\bibfnamefont {A.}~\bibnamefont {Yacoby}}, \bibinfo {author} {\bibfnamefont
  {A.}~\bibnamefont {Fert}}, \bibinfo {author} {\bibfnamefont {S.}~\bibnamefont
  {Parkin}}, \bibinfo {author} {\bibfnamefont {K.~S.}\ \bibnamefont
  {Novoselov}}, \bibinfo {author} {\bibfnamefont {P.}~\bibnamefont {Dai}},
  \bibinfo {author} {\bibfnamefont {L.}~\bibnamefont {Balicas}},\ and\ \bibinfo
  {author} {\bibfnamefont {E.~J.~G.}\ \bibnamefont {Santos}},\ }\bibfield
  {title} {\bibinfo {title} {The magnetic genome of two-dimensional van der
  {W}aals materials},\ }\href {https://doi.org/10.1021/acsnano.1c09150}
  {\bibfield  {journal} {\bibinfo  {journal} {ACS Nano}\ }\textbf {\bibinfo
  {volume} {16}},\ \bibinfo {pages} {6960} (\bibinfo {year}
  {2022})}\BibitemShut {NoStop}%
\bibitem [{\citenamefont {Huang}\ \emph {et~al.}(2018)\citenamefont {Huang},
  \citenamefont {Clark}, \citenamefont {Klein}, \citenamefont {MacNeill},
  \citenamefont {Navarro-Moratalla}, \citenamefont {Seyler}, \citenamefont
  {Wilson}, \citenamefont {McGuire}, \citenamefont {Cobden}, \citenamefont
  {Xiao}, \citenamefont {Yao}, \citenamefont {Jarillo-Herrero},\ and\
  \citenamefont {Xu}}]{s41565-018-0121-3}%
  \BibitemOpen
  \bibfield  {author} {\bibinfo {author} {\bibfnamefont {B.}~\bibnamefont
  {Huang}}, \bibinfo {author} {\bibfnamefont {G.}~\bibnamefont {Clark}},
  \bibinfo {author} {\bibfnamefont {D.~R.}\ \bibnamefont {Klein}}, \bibinfo
  {author} {\bibfnamefont {D.}~\bibnamefont {MacNeill}}, \bibinfo {author}
  {\bibfnamefont {E.}~\bibnamefont {Navarro-Moratalla}}, \bibinfo {author}
  {\bibfnamefont {K.~L.}\ \bibnamefont {Seyler}}, \bibinfo {author}
  {\bibfnamefont {N.}~\bibnamefont {Wilson}}, \bibinfo {author} {\bibfnamefont
  {M.~A.}\ \bibnamefont {McGuire}}, \bibinfo {author} {\bibfnamefont {D.~H.}\
  \bibnamefont {Cobden}}, \bibinfo {author} {\bibfnamefont {D.}~\bibnamefont
  {Xiao}}, \bibinfo {author} {\bibfnamefont {W.}~\bibnamefont {Yao}}, \bibinfo
  {author} {\bibfnamefont {P.}~\bibnamefont {Jarillo-Herrero}},\ and\ \bibinfo
  {author} {\bibfnamefont {X.}~\bibnamefont {Xu}},\ }\bibfield  {title}
  {\bibinfo {title} {Electrical control of {2D} magnetism in bilayer
  \ce{CrI3}},\ }\href {https://doi.org/10.1038/s41565-018-0121-3} {\bibfield
  {journal} {\bibinfo  {journal} {Nat. Nanotechnol.}\ }\textbf {\bibinfo
  {volume} {13}},\ \bibinfo {pages} {544} (\bibinfo {year} {2018})}\BibitemShut
  {NoStop}%
\bibitem [{\citenamefont {Jiang}\ \emph
  {et~al.}(2018{\natexlab{a}})\citenamefont {Jiang}, \citenamefont {Li},
  \citenamefont {Wang}, \citenamefont {Mak},\ and\ \citenamefont
  {Shan}}]{s41565-018-0135-x}%
  \BibitemOpen
  \bibfield  {author} {\bibinfo {author} {\bibfnamefont {S.}~\bibnamefont
  {Jiang}}, \bibinfo {author} {\bibfnamefont {L.}~\bibnamefont {Li}}, \bibinfo
  {author} {\bibfnamefont {Z.}~\bibnamefont {Wang}}, \bibinfo {author}
  {\bibfnamefont {K.~F.}\ \bibnamefont {Mak}},\ and\ \bibinfo {author}
  {\bibfnamefont {J.}~\bibnamefont {Shan}},\ }\bibfield  {title} {\bibinfo
  {title} {Controlling magnetism in 2\ce{D} \ce{CrI3} by electrostatic
  doping},\ }\href {https://doi.org/10.1038/s41565-018-0135-x} {\bibfield
  {journal} {\bibinfo  {journal} {Nat. Nanotechnol.}\ }\textbf {\bibinfo
  {volume} {13}},\ \bibinfo {pages} {549} (\bibinfo {year}
  {2018}{\natexlab{a}})}\BibitemShut {NoStop}%
\bibitem [{\citenamefont {Li}\ \emph {et~al.}(2018)\citenamefont {Li},
  \citenamefont {Yang}, \citenamefont {Gong}, \citenamefont {Chopdekar},
  \citenamefont {N'Diaye}, \citenamefont {Turner}, \citenamefont {Chen},
  \citenamefont {Scholl}, \citenamefont {Shafer}, \citenamefont {Arenholz},
  \citenamefont {Schmid}, \citenamefont {Wang}, \citenamefont {Liu},
  \citenamefont {Gao}, \citenamefont {Admasu}, \citenamefont {Cheong},
  \citenamefont {Hwang}, \citenamefont {Li}, \citenamefont {Wang},
  \citenamefont {Zhang},\ and\ \citenamefont {Qiu}}]{acs.nanolett.8b02806}%
  \BibitemOpen
  \bibfield  {author} {\bibinfo {author} {\bibfnamefont {Q.}~\bibnamefont
  {Li}}, \bibinfo {author} {\bibfnamefont {M.}~\bibnamefont {Yang}}, \bibinfo
  {author} {\bibfnamefont {C.}~\bibnamefont {Gong}}, \bibinfo {author}
  {\bibfnamefont {R.~V.}\ \bibnamefont {Chopdekar}}, \bibinfo {author}
  {\bibfnamefont {A.~T.}\ \bibnamefont {N'Diaye}}, \bibinfo {author}
  {\bibfnamefont {J.}~\bibnamefont {Turner}}, \bibinfo {author} {\bibfnamefont
  {G.}~\bibnamefont {Chen}}, \bibinfo {author} {\bibfnamefont {A.}~\bibnamefont
  {Scholl}}, \bibinfo {author} {\bibfnamefont {P.}~\bibnamefont {Shafer}},
  \bibinfo {author} {\bibfnamefont {E.}~\bibnamefont {Arenholz}}, \bibinfo
  {author} {\bibfnamefont {A.~K.}\ \bibnamefont {Schmid}}, \bibinfo {author}
  {\bibfnamefont {S.}~\bibnamefont {Wang}}, \bibinfo {author} {\bibfnamefont
  {K.}~\bibnamefont {Liu}}, \bibinfo {author} {\bibfnamefont {N.}~\bibnamefont
  {Gao}}, \bibinfo {author} {\bibfnamefont {A.~S.}\ \bibnamefont {Admasu}},
  \bibinfo {author} {\bibfnamefont {S.-W.}\ \bibnamefont {Cheong}}, \bibinfo
  {author} {\bibfnamefont {C.}~\bibnamefont {Hwang}}, \bibinfo {author}
  {\bibfnamefont {J.}~\bibnamefont {Li}}, \bibinfo {author} {\bibfnamefont
  {F.}~\bibnamefont {Wang}}, \bibinfo {author} {\bibfnamefont {X.}~\bibnamefont
  {Zhang}},\ and\ \bibinfo {author} {\bibfnamefont {Z.}~\bibnamefont {Qiu}},\
  }\bibfield  {title} {\bibinfo {title} {Patterning-induced ferromagnetism of
  \ce{Fe3GeTe2} van der {Waals} materials beyond room temperature},\ }\href
  {https://doi.org/10.1021/acs.nanolett.8b02806} {\bibfield  {journal}
  {\bibinfo  {journal} {Nano Lett.}\ }\textbf {\bibinfo {volume} {18}},\
  \bibinfo {pages} {5974} (\bibinfo {year} {2018})}\BibitemShut {NoStop}%
\bibitem [{\citenamefont {Jiang}\ \emph
  {et~al.}(2018{\natexlab{b}})\citenamefont {Jiang}, \citenamefont {Shan},\
  and\ \citenamefont {Mak}}]{s41563-018-0040-6}%
  \BibitemOpen
  \bibfield  {author} {\bibinfo {author} {\bibfnamefont {S.}~\bibnamefont
  {Jiang}}, \bibinfo {author} {\bibfnamefont {J.}~\bibnamefont {Shan}},\ and\
  \bibinfo {author} {\bibfnamefont {K.~F.}\ \bibnamefont {Mak}},\ }\bibfield
  {title} {\bibinfo {title} {Electric-field switching of two-dimensional van
  der {Waals} magnets},\ }\href {https://doi.org/10.1038/s41563-018-0040-6}
  {\bibfield  {journal} {\bibinfo  {journal} {Nat. Mater.}\ }\textbf {\bibinfo
  {volume} {17}},\ \bibinfo {pages} {406} (\bibinfo {year}
  {2018}{\natexlab{b}})}\BibitemShut {NoStop}%
\bibitem [{\citenamefont {Deng}\ \emph {et~al.}(2018)\citenamefont {Deng},
  \citenamefont {Yu}, \citenamefont {Song}, \citenamefont {Zhang},
  \citenamefont {Wang}, \citenamefont {Sun}, \citenamefont {Yi}, \citenamefont
  {Wu}, \citenamefont {Wu}, \citenamefont {Zhu}, \citenamefont {Wang},
  \citenamefont {Chen},\ and\ \citenamefont {Zhang}}]{s41586-018-0626-9}%
  \BibitemOpen
  \bibfield  {author} {\bibinfo {author} {\bibfnamefont {Y.}~\bibnamefont
  {Deng}}, \bibinfo {author} {\bibfnamefont {Y.}~\bibnamefont {Yu}}, \bibinfo
  {author} {\bibfnamefont {Y.}~\bibnamefont {Song}}, \bibinfo {author}
  {\bibfnamefont {J.}~\bibnamefont {Zhang}}, \bibinfo {author} {\bibfnamefont
  {N.~Z.}\ \bibnamefont {Wang}}, \bibinfo {author} {\bibfnamefont
  {Z.}~\bibnamefont {Sun}}, \bibinfo {author} {\bibfnamefont {Y.}~\bibnamefont
  {Yi}}, \bibinfo {author} {\bibfnamefont {Y.~Z.}\ \bibnamefont {Wu}}, \bibinfo
  {author} {\bibfnamefont {S.}~\bibnamefont {Wu}}, \bibinfo {author}
  {\bibfnamefont {J.}~\bibnamefont {Zhu}}, \bibinfo {author} {\bibfnamefont
  {J.}~\bibnamefont {Wang}}, \bibinfo {author} {\bibfnamefont {X.~H.}\
  \bibnamefont {Chen}},\ and\ \bibinfo {author} {\bibfnamefont
  {Y.}~\bibnamefont {Zhang}},\ }\bibfield  {title} {\bibinfo {title}
  {Gate-tunable room-temperature ferromagnetism in two-dimensional
  \ce{Fe3GeTe2}},\ }\href {https://doi.org/10.1038/s41586-018-0626-9}
  {\bibfield  {journal} {\bibinfo  {journal} {Nature}\ }\textbf {\bibinfo
  {volume} {563}},\ \bibinfo {pages} {94} (\bibinfo {year} {2018})}\BibitemShut
  {NoStop}%
\bibitem [{\citenamefont {Park}\ \emph {et~al.}(2020)\citenamefont {Park},
  \citenamefont {Kim}, \citenamefont {Liu}, \citenamefont {Hwang},
  \citenamefont {Kim}, \citenamefont {Kim}, \citenamefont {Kim}, \citenamefont
  {Petrovic}, \citenamefont {Hwang}, \citenamefont {Mo}, \citenamefont {Kim},
  \citenamefont {Min}, \citenamefont {Koo}, \citenamefont {Chang},
  \citenamefont {Jang}, \citenamefont {Choi},\ and\ \citenamefont
  {Ryu}}]{acs.nanolett.9b03316}%
  \BibitemOpen
  \bibfield  {author} {\bibinfo {author} {\bibfnamefont {S.~Y.}\ \bibnamefont
  {Park}}, \bibinfo {author} {\bibfnamefont {D.~S.}\ \bibnamefont {Kim}},
  \bibinfo {author} {\bibfnamefont {Y.}~\bibnamefont {Liu}}, \bibinfo {author}
  {\bibfnamefont {J.}~\bibnamefont {Hwang}}, \bibinfo {author} {\bibfnamefont
  {Y.}~\bibnamefont {Kim}}, \bibinfo {author} {\bibfnamefont {W.}~\bibnamefont
  {Kim}}, \bibinfo {author} {\bibfnamefont {J.-Y.}\ \bibnamefont {Kim}},
  \bibinfo {author} {\bibfnamefont {C.}~\bibnamefont {Petrovic}}, \bibinfo
  {author} {\bibfnamefont {C.}~\bibnamefont {Hwang}}, \bibinfo {author}
  {\bibfnamefont {S.-K.}\ \bibnamefont {Mo}}, \bibinfo {author} {\bibfnamefont
  {H.-j.}\ \bibnamefont {Kim}}, \bibinfo {author} {\bibfnamefont {B.-C.}\
  \bibnamefont {Min}}, \bibinfo {author} {\bibfnamefont {H.~C.}\ \bibnamefont
  {Koo}}, \bibinfo {author} {\bibfnamefont {J.}~\bibnamefont {Chang}}, \bibinfo
  {author} {\bibfnamefont {C.}~\bibnamefont {Jang}}, \bibinfo {author}
  {\bibfnamefont {J.~W.}\ \bibnamefont {Choi}},\ and\ \bibinfo {author}
  {\bibfnamefont {H.}~\bibnamefont {Ryu}},\ }\bibfield  {title} {\bibinfo
  {title} {Controlling the magnetic anisotropy of the van der {Waals}
  ferromagnet \ce{Fe3GeTe2} through hole doping},\ }\href
  {https://doi.org/10.1021/acs.nanolett.9b03316} {\bibfield  {journal}
  {\bibinfo  {journal} {Nano Lett.}\ }\textbf {\bibinfo {volume} {20}},\
  \bibinfo {pages} {95} (\bibinfo {year} {2020})}\BibitemShut {NoStop}%
\bibitem [{\citenamefont {Singh}\ and\ \citenamefont
  {Kabir}(2021)}]{PhysRevB.103.214411}%
  \BibitemOpen
  \bibfield  {author} {\bibinfo {author} {\bibfnamefont {C.~K.}\ \bibnamefont
  {Singh}}\ and\ \bibinfo {author} {\bibfnamefont {M.}~\bibnamefont {Kabir}},\
  }\bibfield  {title} {\bibinfo {title} {Long-range anisotropic {Heisenberg}
  ferromagnets and electrically tunable ordering},\ }\href
  {https://link.aps.org/doi/10.1103/PhysRevB.103.214411} {\bibfield  {journal}
  {\bibinfo  {journal} {Phys. Rev. B}\ }\textbf {\bibinfo {volume} {103}},\
  \bibinfo {pages} {214411} (\bibinfo {year} {2021})}\BibitemShut {NoStop}%
\bibitem [{\citenamefont {Singh}\ and\ \citenamefont
  {Kabir}(2022)}]{PhysRevMaterials.6.084407}%
  \BibitemOpen
  \bibfield  {author} {\bibinfo {author} {\bibfnamefont {C.~K.}\ \bibnamefont
  {Singh}}\ and\ \bibinfo {author} {\bibfnamefont {M.}~\bibnamefont {Kabir}},\
  }\bibfield  {title} {\bibinfo {title} {Room-temperature ferromagnetism in
  two-dimensional \ce{CrBr3}},\ }\href
  {https://link.aps.org/doi/10.1103/PhysRevMaterials.6.084407} {\bibfield
  {journal} {\bibinfo  {journal} {Phys. Rev. Mater.}\ }\textbf {\bibinfo
  {volume} {6}},\ \bibinfo {pages} {084407} (\bibinfo {year}
  {2022})}\BibitemShut {NoStop}%
\bibitem [{\citenamefont {Bedoya-Pinto}\ \emph {et~al.}(2021)\citenamefont
  {Bedoya-Pinto}, \citenamefont {Ji}, \citenamefont {Pandeya}, \citenamefont
  {Gargiani}, \citenamefont {Valvidares}, \citenamefont {Sessi}, \citenamefont
  {Taylor}, \citenamefont {Radu}, \citenamefont {Chang},\ and\ \citenamefont
  {Parkin}}]{science.abd5146}%
  \BibitemOpen
  \bibfield  {author} {\bibinfo {author} {\bibfnamefont {A.}~\bibnamefont
  {Bedoya-Pinto}}, \bibinfo {author} {\bibfnamefont {J.-R.}\ \bibnamefont
  {Ji}}, \bibinfo {author} {\bibfnamefont {A.~K.}\ \bibnamefont {Pandeya}},
  \bibinfo {author} {\bibfnamefont {P.}~\bibnamefont {Gargiani}}, \bibinfo
  {author} {\bibfnamefont {M.}~\bibnamefont {Valvidares}}, \bibinfo {author}
  {\bibfnamefont {P.}~\bibnamefont {Sessi}}, \bibinfo {author} {\bibfnamefont
  {J.~M.}\ \bibnamefont {Taylor}}, \bibinfo {author} {\bibfnamefont
  {F.}~\bibnamefont {Radu}}, \bibinfo {author} {\bibfnamefont {K.}~\bibnamefont
  {Chang}},\ and\ \bibinfo {author} {\bibfnamefont {S.~S.~P.}\ \bibnamefont
  {Parkin}},\ }\bibfield  {title} {\bibinfo {title} {Intrinsic 2\ce{D}-\ce{XY}
  ferromagnetism in a van der {Waals} monolayer},\ }\href
  {https://www.science.org/doi/abs/10.1126/science.abd5146} {\bibfield
  {journal} {\bibinfo  {journal} {Science}\ }\textbf {\bibinfo {volume}
  {374}},\ \bibinfo {pages} {616} (\bibinfo {year} {2021})}\BibitemShut
  {NoStop}%
\bibitem [{\citenamefont {Kim}\ \emph {et~al.}(2019{\natexlab{a}})\citenamefont
  {Kim}, \citenamefont {Yang}, \citenamefont {Li}, \citenamefont {Jiang},
  \citenamefont {Jin}, \citenamefont {Tao}, \citenamefont {Nichols},
  \citenamefont {Sfigakis}, \citenamefont {Zhong}, \citenamefont {Li},
  \citenamefont {Tian}, \citenamefont {Cory}, \citenamefont {Miao},
  \citenamefont {Shan}, \citenamefont {Mak}, \citenamefont {Lei}, \citenamefont
  {Sun}, \citenamefont {Zhao},\ and\ \citenamefont {Tsen}}]{Kim11131}%
  \BibitemOpen
  \bibfield  {author} {\bibinfo {author} {\bibfnamefont {H.~H.}\ \bibnamefont
  {Kim}}, \bibinfo {author} {\bibfnamefont {B.}~\bibnamefont {Yang}}, \bibinfo
  {author} {\bibfnamefont {S.}~\bibnamefont {Li}}, \bibinfo {author}
  {\bibfnamefont {S.}~\bibnamefont {Jiang}}, \bibinfo {author} {\bibfnamefont
  {C.}~\bibnamefont {Jin}}, \bibinfo {author} {\bibfnamefont {Z.}~\bibnamefont
  {Tao}}, \bibinfo {author} {\bibfnamefont {G.}~\bibnamefont {Nichols}},
  \bibinfo {author} {\bibfnamefont {F.}~\bibnamefont {Sfigakis}}, \bibinfo
  {author} {\bibfnamefont {S.}~\bibnamefont {Zhong}}, \bibinfo {author}
  {\bibfnamefont {C.}~\bibnamefont {Li}}, \bibinfo {author} {\bibfnamefont
  {S.}~\bibnamefont {Tian}}, \bibinfo {author} {\bibfnamefont {D.~G.}\
  \bibnamefont {Cory}}, \bibinfo {author} {\bibfnamefont {G.-X.}\ \bibnamefont
  {Miao}}, \bibinfo {author} {\bibfnamefont {J.}~\bibnamefont {Shan}}, \bibinfo
  {author} {\bibfnamefont {K.~F.}\ \bibnamefont {Mak}}, \bibinfo {author}
  {\bibfnamefont {H.}~\bibnamefont {Lei}}, \bibinfo {author} {\bibfnamefont
  {K.}~\bibnamefont {Sun}}, \bibinfo {author} {\bibfnamefont {L.}~\bibnamefont
  {Zhao}},\ and\ \bibinfo {author} {\bibfnamefont {A.~W.}\ \bibnamefont
  {Tsen}},\ }\bibfield  {title} {\bibinfo {title} {Evolution of interlayer and
  intralayer magnetism in three atomically thin chromium trihalides},\ }\href
  {https://www.pnas.org/content/116/23/11131} {\bibfield  {journal} {\bibinfo
  {journal} {Proc. Natl. Acad. Sci. USA}\ }\textbf {\bibinfo {volume} {116}},\
  \bibinfo {pages} {11131} (\bibinfo {year} {2019}{\natexlab{a}})}\BibitemShut
  {NoStop}%
\bibitem [{\citenamefont {Zhang}\ \emph
  {et~al.}(2019{\natexlab{a}})\citenamefont {Zhang}, \citenamefont {Shang},
  \citenamefont {Jiang}, \citenamefont {Rasmita}, \citenamefont {Gao},\ and\
  \citenamefont {Yu}}]{acs.nanolett.9b00553}%
  \BibitemOpen
  \bibfield  {author} {\bibinfo {author} {\bibfnamefont {Z.}~\bibnamefont
  {Zhang}}, \bibinfo {author} {\bibfnamefont {J.}~\bibnamefont {Shang}},
  \bibinfo {author} {\bibfnamefont {C.}~\bibnamefont {Jiang}}, \bibinfo
  {author} {\bibfnamefont {A.}~\bibnamefont {Rasmita}}, \bibinfo {author}
  {\bibfnamefont {W.}~\bibnamefont {Gao}},\ and\ \bibinfo {author}
  {\bibfnamefont {T.}~\bibnamefont {Yu}},\ }\bibfield  {title} {\bibinfo
  {title} {Direct photoluminescence probing of ferromagnetism in monolayer
  two-dimensional \ce{CrBr3}},\ }\href
  {https://doi.org/10.1021/acs.nanolett.9b00553} {\bibfield  {journal}
  {\bibinfo  {journal} {Nano Lett.}\ }\textbf {\bibinfo {volume} {19}},\
  \bibinfo {pages} {3138} (\bibinfo {year} {2019}{\natexlab{a}})}\BibitemShut
  {NoStop}%
\bibitem [{\citenamefont {Wang}\ \emph
  {et~al.}(2019{\natexlab{a}})\citenamefont {Wang}, \citenamefont {Gibertini},
  \citenamefont {Dumcenco}, \citenamefont {Taniguchi}, \citenamefont
  {Watanabe}, \citenamefont {Giannini},\ and\ \citenamefont
  {Morpurgo}}]{s41565-019-0565-0}%
  \BibitemOpen
  \bibfield  {author} {\bibinfo {author} {\bibfnamefont {Z.}~\bibnamefont
  {Wang}}, \bibinfo {author} {\bibfnamefont {M.}~\bibnamefont {Gibertini}},
  \bibinfo {author} {\bibfnamefont {D.}~\bibnamefont {Dumcenco}}, \bibinfo
  {author} {\bibfnamefont {T.}~\bibnamefont {Taniguchi}}, \bibinfo {author}
  {\bibfnamefont {K.}~\bibnamefont {Watanabe}}, \bibinfo {author}
  {\bibfnamefont {E.}~\bibnamefont {Giannini}},\ and\ \bibinfo {author}
  {\bibfnamefont {A.~F.}\ \bibnamefont {Morpurgo}},\ }\bibfield  {title}
  {\bibinfo {title} {Determining the phase diagram of atomically thin layered
  antiferromagnet \ce{CrCl3}},\ }\href
  {https://doi.org/10.1038/s41565-019-0565-0} {\bibfield  {journal} {\bibinfo
  {journal} {Nat. Nanotechnol.}\ }\textbf {\bibinfo {volume} {14}},\ \bibinfo
  {pages} {1116} (\bibinfo {year} {2019}{\natexlab{a}})}\BibitemShut {NoStop}%
\bibitem [{\citenamefont {Klein}\ \emph {et~al.}(2019)\citenamefont {Klein},
  \citenamefont {MacNeill}, \citenamefont {Song}, \citenamefont {Larson},
  \citenamefont {Fang}, \citenamefont {Xu}, \citenamefont {Ribeiro},
  \citenamefont {Canfield}, \citenamefont {Kaxiras}, \citenamefont {Comin},\
  and\ \citenamefont {Jarillo-Herrero}}]{s41567-019-0651-0}%
  \BibitemOpen
  \bibfield  {author} {\bibinfo {author} {\bibfnamefont {D.~R.}\ \bibnamefont
  {Klein}}, \bibinfo {author} {\bibfnamefont {D.}~\bibnamefont {MacNeill}},
  \bibinfo {author} {\bibfnamefont {Q.}~\bibnamefont {Song}}, \bibinfo {author}
  {\bibfnamefont {D.~T.}\ \bibnamefont {Larson}}, \bibinfo {author}
  {\bibfnamefont {S.}~\bibnamefont {Fang}}, \bibinfo {author} {\bibfnamefont
  {M.}~\bibnamefont {Xu}}, \bibinfo {author} {\bibfnamefont {R.~A.}\
  \bibnamefont {Ribeiro}}, \bibinfo {author} {\bibfnamefont {P.~C.}\
  \bibnamefont {Canfield}}, \bibinfo {author} {\bibfnamefont {E.}~\bibnamefont
  {Kaxiras}}, \bibinfo {author} {\bibfnamefont {R.}~\bibnamefont {Comin}},\
  and\ \bibinfo {author} {\bibfnamefont {P.}~\bibnamefont {Jarillo-Herrero}},\
  }\bibfield  {title} {\bibinfo {title} {Enhancement of interlayer exchange in
  an ultrathin two-dimensional magnet},\ }\href
  {https://doi.org/10.1038/s41567-019-0651-0} {\bibfield  {journal} {\bibinfo
  {journal} {Nat. Phys.}\ }\textbf {\bibinfo {volume} {15}},\ \bibinfo {pages}
  {1255} (\bibinfo {year} {2019})}\BibitemShut {NoStop}%
\bibitem [{\citenamefont {Zhang}\ \emph {et~al.}(2021)\citenamefont {Zhang},
  \citenamefont {Lu}, \citenamefont {Liu}, \citenamefont {Niu}, \citenamefont
  {Sun}, \citenamefont {Cook}, \citenamefont {Vaninger}, \citenamefont
  {Miceli}, \citenamefont {Singh}, \citenamefont {Lian}, \citenamefont {Chang},
  \citenamefont {He}, \citenamefont {Du}, \citenamefont {He}, \citenamefont
  {Zhang}, \citenamefont {Bian},\ and\ \citenamefont
  {Xu}}]{s41467-021-22777-x}%
  \BibitemOpen
  \bibfield  {author} {\bibinfo {author} {\bibfnamefont {X.}~\bibnamefont
  {Zhang}}, \bibinfo {author} {\bibfnamefont {Q.}~\bibnamefont {Lu}}, \bibinfo
  {author} {\bibfnamefont {W.}~\bibnamefont {Liu}}, \bibinfo {author}
  {\bibfnamefont {W.}~\bibnamefont {Niu}}, \bibinfo {author} {\bibfnamefont
  {J.}~\bibnamefont {Sun}}, \bibinfo {author} {\bibfnamefont {J.}~\bibnamefont
  {Cook}}, \bibinfo {author} {\bibfnamefont {M.}~\bibnamefont {Vaninger}},
  \bibinfo {author} {\bibfnamefont {P.~F.}\ \bibnamefont {Miceli}}, \bibinfo
  {author} {\bibfnamefont {D.~J.}\ \bibnamefont {Singh}}, \bibinfo {author}
  {\bibfnamefont {S.-W.}\ \bibnamefont {Lian}}, \bibinfo {author}
  {\bibfnamefont {T.-R.}\ \bibnamefont {Chang}}, \bibinfo {author}
  {\bibfnamefont {X.}~\bibnamefont {He}}, \bibinfo {author} {\bibfnamefont
  {J.}~\bibnamefont {Du}}, \bibinfo {author} {\bibfnamefont {L.}~\bibnamefont
  {He}}, \bibinfo {author} {\bibfnamefont {R.}~\bibnamefont {Zhang}}, \bibinfo
  {author} {\bibfnamefont {G.}~\bibnamefont {Bian}},\ and\ \bibinfo {author}
  {\bibfnamefont {Y.}~\bibnamefont {Xu}},\ }\bibfield  {title} {\bibinfo
  {title} {Room-temperature intrinsic ferromagnetism in epitaxial \ce{CrTe2}
  ultrathin films},\ }\href {https://doi.org/10.1038/s41467-021-22777-x}
  {\bibfield  {journal} {\bibinfo  {journal} {Nat. Commun.}\ }\textbf {\bibinfo
  {volume} {12}},\ \bibinfo {pages} {2492} (\bibinfo {year}
  {2021})}\BibitemShut {NoStop}%
\bibitem [{\citenamefont {Ou}\ \emph {et~al.}(2022)\citenamefont {Ou},
  \citenamefont {Yanez}, \citenamefont {Xiao}, \citenamefont {Stanley},
  \citenamefont {Ghosh}, \citenamefont {Zheng}, \citenamefont {Jiang},
  \citenamefont {Huang}, \citenamefont {Pillsbury}, \citenamefont
  {Richardella}, \citenamefont {Liu}, \citenamefont {Low}, \citenamefont
  {Crespi}, \citenamefont {Mkhoyan},\ and\ \citenamefont
  {Samarth}}]{s41467-022-30738-1}%
  \BibitemOpen
  \bibfield  {author} {\bibinfo {author} {\bibfnamefont {Y.}~\bibnamefont
  {Ou}}, \bibinfo {author} {\bibfnamefont {W.}~\bibnamefont {Yanez}}, \bibinfo
  {author} {\bibfnamefont {R.}~\bibnamefont {Xiao}}, \bibinfo {author}
  {\bibfnamefont {M.}~\bibnamefont {Stanley}}, \bibinfo {author} {\bibfnamefont
  {S.}~\bibnamefont {Ghosh}}, \bibinfo {author} {\bibfnamefont
  {B.}~\bibnamefont {Zheng}}, \bibinfo {author} {\bibfnamefont
  {W.}~\bibnamefont {Jiang}}, \bibinfo {author} {\bibfnamefont {Y.-S.}\
  \bibnamefont {Huang}}, \bibinfo {author} {\bibfnamefont {T.}~\bibnamefont
  {Pillsbury}}, \bibinfo {author} {\bibfnamefont {A.}~\bibnamefont
  {Richardella}}, \bibinfo {author} {\bibfnamefont {C.}~\bibnamefont {Liu}},
  \bibinfo {author} {\bibfnamefont {T.}~\bibnamefont {Low}}, \bibinfo {author}
  {\bibfnamefont {V.~H.}\ \bibnamefont {Crespi}}, \bibinfo {author}
  {\bibfnamefont {K.~A.}\ \bibnamefont {Mkhoyan}},\ and\ \bibinfo {author}
  {\bibfnamefont {N.}~\bibnamefont {Samarth}},\ }\bibfield  {title} {\bibinfo
  {title} {\ce{ZrTe2}/\ce{CrTe2}: an epitaxial van der {Waals} platform for
  spintronics},\ }\href {https://doi.org/10.1038/s41467-022-30738-1} {\bibfield
   {journal} {\bibinfo  {journal} {Nat. Commun.}\ }\textbf {\bibinfo {volume}
  {13}},\ \bibinfo {pages} {2972} (\bibinfo {year} {2022})}\BibitemShut
  {NoStop}%
\bibitem [{\citenamefont {Freitas}\ \emph {et~al.}(2015)\citenamefont
  {Freitas}, \citenamefont {Weht}, \citenamefont {Sulpice}, \citenamefont
  {Remenyi}, \citenamefont {Strobel}, \citenamefont {Gay}, \citenamefont
  {Marcus},\ and\ \citenamefont {Nunez-Regueiro}}]{Freitas_2015}%
  \BibitemOpen
  \bibfield  {author} {\bibinfo {author} {\bibfnamefont {D.~C.}\ \bibnamefont
  {Freitas}}, \bibinfo {author} {\bibfnamefont {R.}~\bibnamefont {Weht}},
  \bibinfo {author} {\bibfnamefont {A.}~\bibnamefont {Sulpice}}, \bibinfo
  {author} {\bibfnamefont {G.}~\bibnamefont {Remenyi}}, \bibinfo {author}
  {\bibfnamefont {P.}~\bibnamefont {Strobel}}, \bibinfo {author} {\bibfnamefont
  {F.}~\bibnamefont {Gay}}, \bibinfo {author} {\bibfnamefont {J.}~\bibnamefont
  {Marcus}},\ and\ \bibinfo {author} {\bibfnamefont {M.}~\bibnamefont
  {Nunez-Regueiro}},\ }\bibfield  {title} {\bibinfo {title} {Ferromagnetism in
  layered metastable ${1T}$-\ce{CrTe2}},\ }\href
  {https://dx.doi.org/10.1088/0953-8984/27/17/176002} {\bibfield  {journal}
  {\bibinfo  {journal} {J. Phys.: Condens. Matter}\ }\textbf {\bibinfo {volume}
  {27}},\ \bibinfo {pages} {176002} (\bibinfo {year} {2015})}\BibitemShut
  {NoStop}%
\bibitem [{\citenamefont {Sun}\ \emph {et~al.}(2020)\citenamefont {Sun},
  \citenamefont {Li}, \citenamefont {Wang}, \citenamefont {Sui}, \citenamefont
  {Zhang}, \citenamefont {Wang}, \citenamefont {Liu}, \citenamefont {Li},
  \citenamefont {Feng}, \citenamefont {Zhong}, \citenamefont {Wang},
  \citenamefont {Bouchiat}, \citenamefont {Nunez~Regueiro}, \citenamefont
  {Rougemaille}, \citenamefont {Coraux}, \citenamefont {Purbawati},
  \citenamefont {Hadj-Azzem}, \citenamefont {Wang}, \citenamefont {Dong},
  \citenamefont {Wu}, \citenamefont {Yang}, \citenamefont {Yu}, \citenamefont
  {Wang}, \citenamefont {Han}, \citenamefont {Han},\ and\ \citenamefont
  {Zhang}}]{s12274-020-3021-4}%
  \BibitemOpen
  \bibfield  {author} {\bibinfo {author} {\bibfnamefont {X.}~\bibnamefont
  {Sun}}, \bibinfo {author} {\bibfnamefont {W.}~\bibnamefont {Li}}, \bibinfo
  {author} {\bibfnamefont {X.}~\bibnamefont {Wang}}, \bibinfo {author}
  {\bibfnamefont {Q.}~\bibnamefont {Sui}}, \bibinfo {author} {\bibfnamefont
  {T.}~\bibnamefont {Zhang}}, \bibinfo {author} {\bibfnamefont
  {Z.}~\bibnamefont {Wang}}, \bibinfo {author} {\bibfnamefont {L.}~\bibnamefont
  {Liu}}, \bibinfo {author} {\bibfnamefont {D.}~\bibnamefont {Li}}, \bibinfo
  {author} {\bibfnamefont {S.}~\bibnamefont {Feng}}, \bibinfo {author}
  {\bibfnamefont {S.}~\bibnamefont {Zhong}}, \bibinfo {author} {\bibfnamefont
  {H.}~\bibnamefont {Wang}}, \bibinfo {author} {\bibfnamefont {V.}~\bibnamefont
  {Bouchiat}}, \bibinfo {author} {\bibfnamefont {M.}~\bibnamefont
  {Nunez~Regueiro}}, \bibinfo {author} {\bibfnamefont {N.}~\bibnamefont
  {Rougemaille}}, \bibinfo {author} {\bibfnamefont {J.}~\bibnamefont {Coraux}},
  \bibinfo {author} {\bibfnamefont {A.}~\bibnamefont {Purbawati}}, \bibinfo
  {author} {\bibfnamefont {A.}~\bibnamefont {Hadj-Azzem}}, \bibinfo {author}
  {\bibfnamefont {Z.}~\bibnamefont {Wang}}, \bibinfo {author} {\bibfnamefont
  {B.}~\bibnamefont {Dong}}, \bibinfo {author} {\bibfnamefont {X.}~\bibnamefont
  {Wu}}, \bibinfo {author} {\bibfnamefont {T.}~\bibnamefont {Yang}}, \bibinfo
  {author} {\bibfnamefont {G.}~\bibnamefont {Yu}}, \bibinfo {author}
  {\bibfnamefont {B.}~\bibnamefont {Wang}}, \bibinfo {author} {\bibfnamefont
  {Z.}~\bibnamefont {Han}}, \bibinfo {author} {\bibfnamefont {X.}~\bibnamefont
  {Han}},\ and\ \bibinfo {author} {\bibfnamefont {Z.}~\bibnamefont {Zhang}},\
  }\bibfield  {title} {\bibinfo {title} {Room temperature ferromagnetism in
  ultra-thin van der {Waals} crystals of ${1T}$-\ce{CrTe2}},\ }\href
  {https://doi.org/10.1007/s12274-020-3021-4} {\bibfield  {journal} {\bibinfo
  {journal} {Nano Res.}\ }\textbf {\bibinfo {volume} {13}},\ \bibinfo {pages}
  {3358} (\bibinfo {year} {2020})}\BibitemShut {NoStop}%
\bibitem [{\citenamefont {Fabre}\ \emph {et~al.}(2021)\citenamefont {Fabre},
  \citenamefont {Finco}, \citenamefont {Purbawati}, \citenamefont {Hadj-Azzem},
  \citenamefont {Rougemaille}, \citenamefont {Coraux}, \citenamefont {Philip},\
  and\ \citenamefont {Jacques}}]{PhysRevMaterials.5.034008}%
  \BibitemOpen
  \bibfield  {author} {\bibinfo {author} {\bibfnamefont {F.}~\bibnamefont
  {Fabre}}, \bibinfo {author} {\bibfnamefont {A.}~\bibnamefont {Finco}},
  \bibinfo {author} {\bibfnamefont {A.}~\bibnamefont {Purbawati}}, \bibinfo
  {author} {\bibfnamefont {A.}~\bibnamefont {Hadj-Azzem}}, \bibinfo {author}
  {\bibfnamefont {N.}~\bibnamefont {Rougemaille}}, \bibinfo {author}
  {\bibfnamefont {J.}~\bibnamefont {Coraux}}, \bibinfo {author} {\bibfnamefont
  {I.}~\bibnamefont {Philip}},\ and\ \bibinfo {author} {\bibfnamefont
  {V.}~\bibnamefont {Jacques}},\ }\bibfield  {title} {\bibinfo {title}
  {Characterization of room-temperature in-plane magnetization in thin flakes
  of \ce{CrTe2} with a single-spin magnetometer},\ }\href
  {https://link.aps.org/doi/10.1103/PhysRevMaterials.5.034008} {\bibfield
  {journal} {\bibinfo  {journal} {Phys. Rev. Mater.}\ }\textbf {\bibinfo
  {volume} {5}},\ \bibinfo {pages} {034008} (\bibinfo {year}
  {2021})}\BibitemShut {NoStop}%
\bibitem [{\citenamefont {Kushwaha}\ \emph {et~al.}(2025)\citenamefont
  {Kushwaha}, \citenamefont {Armitage}, \citenamefont {Edwards}, \citenamefont
  {Trzaska}, \citenamefont {Rigden}, \citenamefont {Bencok}, \citenamefont
  {Biswas}, \citenamefont {Lee}, \citenamefont {Sanders}, \citenamefont
  {van~der Laan}, \citenamefont {Wahl}, \citenamefont {King},\ and\
  \citenamefont {Rajan}}]{s41535-025-00772-5}%
  \BibitemOpen
  \bibfield  {author} {\bibinfo {author} {\bibfnamefont {N.}~\bibnamefont
  {Kushwaha}}, \bibinfo {author} {\bibfnamefont {O.}~\bibnamefont {Armitage}},
  \bibinfo {author} {\bibfnamefont {B.}~\bibnamefont {Edwards}}, \bibinfo
  {author} {\bibfnamefont {L.}~\bibnamefont {Trzaska}}, \bibinfo {author}
  {\bibfnamefont {J.}~\bibnamefont {Rigden}}, \bibinfo {author} {\bibfnamefont
  {P.}~\bibnamefont {Bencok}}, \bibinfo {author} {\bibfnamefont
  {D.}~\bibnamefont {Biswas}}, \bibinfo {author} {\bibfnamefont {T.-L.}\
  \bibnamefont {Lee}}, \bibinfo {author} {\bibfnamefont {C.}~\bibnamefont
  {Sanders}}, \bibinfo {author} {\bibfnamefont {G.}~\bibnamefont {van~der
  Laan}}, \bibinfo {author} {\bibfnamefont {P.}~\bibnamefont {Wahl}}, \bibinfo
  {author} {\bibfnamefont {P.~D.~C.}\ \bibnamefont {King}},\ and\ \bibinfo
  {author} {\bibfnamefont {A.}~\bibnamefont {Rajan}},\ }\bibfield  {title}
  {\bibinfo {title} {From ferromagnetic semiconductor to antiferromagnetic
  metal in epitaxial \ce{Cr_xTe_y} monolayers},\ }\href
  {https://doi.org/10.1038/s41535-025-00772-5} {\bibfield  {journal} {\bibinfo
  {journal} {npj Quantum Mater.}\ }\textbf {\bibinfo {volume} {10}},\ \bibinfo
  {pages} {50} (\bibinfo {year} {2025})}\BibitemShut {NoStop}%
\bibitem [{\citenamefont {Miao}\ \emph {et~al.}(2025)\citenamefont {Miao},
  \citenamefont {Gu}, \citenamefont {Sun}, \citenamefont {Chen}, \citenamefont
  {Li}, \citenamefont {Xue}, \citenamefont {Su}, \citenamefont {Su},
  \citenamefont {Zhong}, \citenamefont {Zhang}, \citenamefont {Zhu},
  \citenamefont {Zhang}, \citenamefont {Yao}, \citenamefont {Jiang},
  \citenamefont {Meng}, \citenamefont {Wang},\ and\ \citenamefont
  {Guo}}]{aelm.202400720}%
  \BibitemOpen
  \bibfield  {author} {\bibinfo {author} {\bibfnamefont {G.}~\bibnamefont
  {Miao}}, \bibinfo {author} {\bibfnamefont {M.}~\bibnamefont {Gu}}, \bibinfo
  {author} {\bibfnamefont {H.}~\bibnamefont {Sun}}, \bibinfo {author}
  {\bibfnamefont {P.}~\bibnamefont {Chen}}, \bibinfo {author} {\bibfnamefont
  {J.}~\bibnamefont {Li}}, \bibinfo {author} {\bibfnamefont {S.}~\bibnamefont
  {Xue}}, \bibinfo {author} {\bibfnamefont {N.}~\bibnamefont {Su}}, \bibinfo
  {author} {\bibfnamefont {Z.}~\bibnamefont {Su}}, \bibinfo {author}
  {\bibfnamefont {W.}~\bibnamefont {Zhong}}, \bibinfo {author} {\bibfnamefont
  {Z.}~\bibnamefont {Zhang}}, \bibinfo {author} {\bibfnamefont
  {X.}~\bibnamefont {Zhu}}, \bibinfo {author} {\bibfnamefont {J.}~\bibnamefont
  {Zhang}}, \bibinfo {author} {\bibfnamefont {Y.}~\bibnamefont {Yao}}, \bibinfo
  {author} {\bibfnamefont {W.}~\bibnamefont {Jiang}}, \bibinfo {author}
  {\bibfnamefont {M.}~\bibnamefont {Meng}}, \bibinfo {author} {\bibfnamefont
  {W.}~\bibnamefont {Wang}},\ and\ \bibinfo {author} {\bibfnamefont
  {J.}~\bibnamefont {Guo}},\ }\bibfield  {title} {\bibinfo {title} {Tuning the
  magnetism in ultrathin \ce{Cr_xTe_y} films by lattice dimensionality},\
  }\href
  {https://advanced.onlinelibrary.wiley.com/doi/abs/10.1002/aelm.202400720}
  {\bibfield  {journal} {\bibinfo  {journal} {Adv. Electron. Mater.}\ }\textbf
  {\bibinfo {volume} {11}},\ \bibinfo {pages} {2400720} (\bibinfo {year}
  {2025})}\BibitemShut {NoStop}%
\bibitem [{\citenamefont {Tian}\ \emph {et~al.}(2026)\citenamefont {Tian},
  \citenamefont {Zhong}, \citenamefont {Dong}, \citenamefont {Zhou},
  \citenamefont {Liu}, \citenamefont {Chen}, \citenamefont {Zhang},
  \citenamefont {Cao}, \citenamefont {He}, \citenamefont {Li}, \citenamefont
  {Guo}, \citenamefont {Du}, \citenamefont {Feng}, \citenamefont {Wang},
  \citenamefont {Cheng}, \citenamefont {Zhang}, \citenamefont {Feng},
  \citenamefont {Wu}, \citenamefont {Wei}, \citenamefont {Du}, \citenamefont
  {Lu},\ and\ \citenamefont {Chen}}]{s41563-026-02537-2}%
  \BibitemOpen
  \bibfield  {author} {\bibinfo {author} {\bibfnamefont {D.}~\bibnamefont
  {Tian}}, \bibinfo {author} {\bibfnamefont {S.}~\bibnamefont {Zhong}},
  \bibinfo {author} {\bibfnamefont {J.}~\bibnamefont {Dong}}, \bibinfo {author}
  {\bibfnamefont {S.}~\bibnamefont {Zhou}}, \bibinfo {author} {\bibfnamefont
  {Z.}~\bibnamefont {Liu}}, \bibinfo {author} {\bibfnamefont {K.}~\bibnamefont
  {Chen}}, \bibinfo {author} {\bibfnamefont {W.}~\bibnamefont {Zhang}},
  \bibinfo {author} {\bibfnamefont {L.}~\bibnamefont {Cao}}, \bibinfo {author}
  {\bibfnamefont {X.}~\bibnamefont {He}}, \bibinfo {author} {\bibfnamefont
  {X.}~\bibnamefont {Li}}, \bibinfo {author} {\bibfnamefont {T.}~\bibnamefont
  {Guo}}, \bibinfo {author} {\bibfnamefont {K.}~\bibnamefont {Du}}, \bibinfo
  {author} {\bibfnamefont {H.}~\bibnamefont {Feng}}, \bibinfo {author}
  {\bibfnamefont {Y.}~\bibnamefont {Wang}}, \bibinfo {author} {\bibfnamefont
  {P.}~\bibnamefont {Cheng}}, \bibinfo {author} {\bibfnamefont
  {Y.}~\bibnamefont {Zhang}}, \bibinfo {author} {\bibfnamefont
  {B.}~\bibnamefont {Feng}}, \bibinfo {author} {\bibfnamefont {K.}~\bibnamefont
  {Wu}}, \bibinfo {author} {\bibfnamefont {S.}~\bibnamefont {Wei}}, \bibinfo
  {author} {\bibfnamefont {Y.}~\bibnamefont {Du}}, \bibinfo {author}
  {\bibfnamefont {Y.}~\bibnamefont {Lu}},\ and\ \bibinfo {author}
  {\bibfnamefont {L.}~\bibnamefont {Chen}},\ }\bibfield  {title} {\bibinfo
  {title} {Room-temperature two-dimensional multiferroic metal with
  voltage-controllable magnetic order},\ }\href
  {https://doi.org/10.1038/s41563-026-02537-2} {\bibfield  {journal} {\bibinfo
  {journal} {Nat. Mater.}\ } (\bibinfo {year} {2026})}\BibitemShut {NoStop}%
\bibitem [{\citenamefont {Xian}\ \emph {et~al.}(2022)\citenamefont {Xian},
  \citenamefont {Wang}, \citenamefont {Nie}, \citenamefont {Li}, \citenamefont
  {Han}, \citenamefont {Lin}, \citenamefont {Zhang}, \citenamefont {Liu},
  \citenamefont {Zhang}, \citenamefont {Miao}, \citenamefont {Yi},
  \citenamefont {Wu}, \citenamefont {Chen}, \citenamefont {Han}, \citenamefont
  {Xia}, \citenamefont {Ji},\ and\ \citenamefont {Fu}}]{s41467-021-27834-z}%
  \BibitemOpen
  \bibfield  {author} {\bibinfo {author} {\bibfnamefont {J.-J.}\ \bibnamefont
  {Xian}}, \bibinfo {author} {\bibfnamefont {C.}~\bibnamefont {Wang}}, \bibinfo
  {author} {\bibfnamefont {J.-H.}\ \bibnamefont {Nie}}, \bibinfo {author}
  {\bibfnamefont {R.}~\bibnamefont {Li}}, \bibinfo {author} {\bibfnamefont
  {M.}~\bibnamefont {Han}}, \bibinfo {author} {\bibfnamefont {J.}~\bibnamefont
  {Lin}}, \bibinfo {author} {\bibfnamefont {W.-H.}\ \bibnamefont {Zhang}},
  \bibinfo {author} {\bibfnamefont {Z.-Y.}\ \bibnamefont {Liu}}, \bibinfo
  {author} {\bibfnamefont {Z.-M.}\ \bibnamefont {Zhang}}, \bibinfo {author}
  {\bibfnamefont {M.-P.}\ \bibnamefont {Miao}}, \bibinfo {author}
  {\bibfnamefont {Y.}~\bibnamefont {Yi}}, \bibinfo {author} {\bibfnamefont
  {S.}~\bibnamefont {Wu}}, \bibinfo {author} {\bibfnamefont {X.}~\bibnamefont
  {Chen}}, \bibinfo {author} {\bibfnamefont {J.}~\bibnamefont {Han}}, \bibinfo
  {author} {\bibfnamefont {Z.}~\bibnamefont {Xia}}, \bibinfo {author}
  {\bibfnamefont {W.}~\bibnamefont {Ji}},\ and\ \bibinfo {author}
  {\bibfnamefont {Y.-S.}\ \bibnamefont {Fu}},\ }\bibfield  {title} {\bibinfo
  {title} {Spin mapping of intralayer antiferromagnetism and field-induced spin
  reorientation in monolayer \ce{CrTe2}},\ }\href
  {https://doi.org/10.1038/s41467-021-27834-z} {\bibfield  {journal} {\bibinfo
  {journal} {Nat. Commun.}\ }\textbf {\bibinfo {volume} {13}},\ \bibinfo
  {pages} {257} (\bibinfo {year} {2022})}\BibitemShut {NoStop}%
\bibitem [{\citenamefont {Armitage}\ \emph {et~al.}(2025)\citenamefont
  {Armitage}, \citenamefont {Kushwaha}, \citenamefont {Rajan}, \citenamefont
  {Rhodes}, \citenamefont {Buchberger}, \citenamefont {Saika}, \citenamefont
  {Mo}, \citenamefont {Watson}, \citenamefont {King},\ and\ \citenamefont
  {Wahl}}]{h4h8-j473}%
  \BibitemOpen
  \bibfield  {author} {\bibinfo {author} {\bibfnamefont {O.}~\bibnamefont
  {Armitage}}, \bibinfo {author} {\bibfnamefont {N.}~\bibnamefont {Kushwaha}},
  \bibinfo {author} {\bibfnamefont {A.}~\bibnamefont {Rajan}}, \bibinfo
  {author} {\bibfnamefont {L.~C.}\ \bibnamefont {Rhodes}}, \bibinfo {author}
  {\bibfnamefont {S.}~\bibnamefont {Buchberger}}, \bibinfo {author}
  {\bibfnamefont {B.~K.}\ \bibnamefont {Saika}}, \bibinfo {author}
  {\bibfnamefont {S.}~\bibnamefont {Mo}}, \bibinfo {author} {\bibfnamefont
  {M.~D.}\ \bibnamefont {Watson}}, \bibinfo {author} {\bibfnamefont {P.~D.~C.}\
  \bibnamefont {King}},\ and\ \bibinfo {author} {\bibfnamefont
  {P.}~\bibnamefont {Wahl}},\ }\bibfield  {title} {\bibinfo {title} {Electronic
  structure of monolayer \ce{CrTe2}: An antiferromagnetic two-dimensional van
  der {Waals} material},\ }\href {https://link.aps.org/doi/10.1103/h4h8-j473}
  {\bibfield  {journal} {\bibinfo  {journal} {Phys. Rev. B}\ }\textbf {\bibinfo
  {volume} {112}},\ \bibinfo {pages} {245416} (\bibinfo {year}
  {2025})}\BibitemShut {NoStop}%
\bibitem [{\citenamefont {Purbawati}\ \emph {et~al.}(2020)\citenamefont
  {Purbawati}, \citenamefont {Coraux}, \citenamefont {Vogel}, \citenamefont
  {Hadj-Azzem}, \citenamefont {Wu}, \citenamefont {Bendiab}, \citenamefont
  {Jegouso}, \citenamefont {Renard}, \citenamefont {Marty}, \citenamefont
  {Bouchiat}, \citenamefont {Sulpice}, \citenamefont {Aballe}, \citenamefont
  {Foerster}, \citenamefont {Genuzio}, \citenamefont {Locatelli}, \citenamefont
  {Mente{\c s}}, \citenamefont {Han}, \citenamefont {Sun}, \citenamefont
  {N{\'u}{\~n}ez-Regueiro},\ and\ \citenamefont
  {Rougemaille}}]{acsami.0c07017}%
  \BibitemOpen
  \bibfield  {author} {\bibinfo {author} {\bibfnamefont {A.}~\bibnamefont
  {Purbawati}}, \bibinfo {author} {\bibfnamefont {J.}~\bibnamefont {Coraux}},
  \bibinfo {author} {\bibfnamefont {J.}~\bibnamefont {Vogel}}, \bibinfo
  {author} {\bibfnamefont {A.}~\bibnamefont {Hadj-Azzem}}, \bibinfo {author}
  {\bibfnamefont {N.}~\bibnamefont {Wu}}, \bibinfo {author} {\bibfnamefont
  {N.}~\bibnamefont {Bendiab}}, \bibinfo {author} {\bibfnamefont
  {D.}~\bibnamefont {Jegouso}}, \bibinfo {author} {\bibfnamefont
  {J.}~\bibnamefont {Renard}}, \bibinfo {author} {\bibfnamefont
  {L.}~\bibnamefont {Marty}}, \bibinfo {author} {\bibfnamefont
  {V.}~\bibnamefont {Bouchiat}}, \bibinfo {author} {\bibfnamefont
  {A.}~\bibnamefont {Sulpice}}, \bibinfo {author} {\bibfnamefont
  {L.}~\bibnamefont {Aballe}}, \bibinfo {author} {\bibfnamefont
  {M.}~\bibnamefont {Foerster}}, \bibinfo {author} {\bibfnamefont
  {F.}~\bibnamefont {Genuzio}}, \bibinfo {author} {\bibfnamefont
  {A.}~\bibnamefont {Locatelli}}, \bibinfo {author} {\bibfnamefont {T.~O.}\
  \bibnamefont {Mente{\c s}}}, \bibinfo {author} {\bibfnamefont {Z.~V.}\
  \bibnamefont {Han}}, \bibinfo {author} {\bibfnamefont {X.}~\bibnamefont
  {Sun}}, \bibinfo {author} {\bibfnamefont {M.}~\bibnamefont
  {N{\'u}{\~n}ez-Regueiro}},\ and\ \bibinfo {author} {\bibfnamefont
  {N.}~\bibnamefont {Rougemaille}},\ }\bibfield  {title} {\bibinfo {title}
  {In-plane magnetic domains and {N}{\'e}el-like domain walls in thin flakes of
  the room temperature \ce{CrTe2} van der {Waals} ferromagnet},\ }\href
  {https://doi.org/10.1021/acsami.0c07017} {\bibfield  {journal} {\bibinfo
  {journal} {ACS Appl. Mater. Interfaces}\ }\textbf {\bibinfo {volume} {12}},\
  \bibinfo {pages} {30702} (\bibinfo {year} {2020})}\BibitemShut {NoStop}%
\bibitem [{\citenamefont {Ležaić}\ \emph {et~al.}(2007)\citenamefont
  {Ležaić}, \citenamefont {Mavropoulos},\ and\ \citenamefont
  {Blügel}}]{1.2710181}%
  \BibitemOpen
  \bibfield  {author} {\bibinfo {author} {\bibfnamefont {M.}~\bibnamefont
  {Ležaić}}, \bibinfo {author} {\bibfnamefont {P.}~\bibnamefont
  {Mavropoulos}},\ and\ \bibinfo {author} {\bibfnamefont {S.}~\bibnamefont
  {Blügel}},\ }\bibfield  {title} {\bibinfo {title} {{First-principles
  prediction of high Curie temperature for ferromagnetic bcc-Co and bcc-FeCo
  alloys and its relevance to tunneling magnetoresistance}},\ }\href
  {https://doi.org/10.1063/1.2710181} {\bibfield  {journal} {\bibinfo
  {journal} {Appl. Phys. Lett.}\ }\textbf {\bibinfo {volume} {90}},\ \bibinfo
  {pages} {082504} (\bibinfo {year} {2007})}\BibitemShut {NoStop}%
\bibitem [{\citenamefont {Li}\ \emph {et~al.}(2023)\citenamefont {Li},
  \citenamefont {Zhang}, \citenamefont {You}, \citenamefont {Gu},\ and\
  \citenamefont {Su}}]{PhysRevB.107.224411}%
  \BibitemOpen
  \bibfield  {author} {\bibinfo {author} {\bibfnamefont {J.-W.}\ \bibnamefont
  {Li}}, \bibinfo {author} {\bibfnamefont {Z.}~\bibnamefont {Zhang}}, \bibinfo
  {author} {\bibfnamefont {J.-Y.}\ \bibnamefont {You}}, \bibinfo {author}
  {\bibfnamefont {B.}~\bibnamefont {Gu}},\ and\ \bibinfo {author}
  {\bibfnamefont {G.}~\bibnamefont {Su}},\ }\bibfield  {title} {\bibinfo
  {title} {Two-dimensional {Heisenberg} model with material-dependent
  superexchange interactions},\ }\href
  {https://link.aps.org/doi/10.1103/PhysRevB.107.224411} {\bibfield  {journal}
  {\bibinfo  {journal} {Phys. Rev. B}\ }\textbf {\bibinfo {volume} {107}},\
  \bibinfo {pages} {224411} (\bibinfo {year} {2023})}\BibitemShut {NoStop}%
\bibitem [{\citenamefont {Li}\ \emph {et~al.}(2021)\citenamefont {Li},
  \citenamefont {Wang}, \citenamefont {Tai}, \citenamefont {Wu}, \citenamefont
  {Xiang}, \citenamefont {Sheng},\ and\ \citenamefont
  {Yang}}]{PhysRevB.103.045114}%
  \BibitemOpen
  \bibfield  {author} {\bibinfo {author} {\bibfnamefont {S.}~\bibnamefont
  {Li}}, \bibinfo {author} {\bibfnamefont {S.-S.}\ \bibnamefont {Wang}},
  \bibinfo {author} {\bibfnamefont {B.}~\bibnamefont {Tai}}, \bibinfo {author}
  {\bibfnamefont {W.}~\bibnamefont {Wu}}, \bibinfo {author} {\bibfnamefont
  {B.}~\bibnamefont {Xiang}}, \bibinfo {author} {\bibfnamefont {X.-L.}\
  \bibnamefont {Sheng}},\ and\ \bibinfo {author} {\bibfnamefont {S.~A.}\
  \bibnamefont {Yang}},\ }\bibfield  {title} {\bibinfo {title} {Tunable
  anomalous {Hall} transport in bulk and two-dimensional ${1T}$-\ce{CrTe2}: A
  first-principles study},\ }\href
  {https://link.aps.org/doi/10.1103/PhysRevB.103.045114} {\bibfield  {journal}
  {\bibinfo  {journal} {Phys. Rev. B}\ }\textbf {\bibinfo {volume} {103}},\
  \bibinfo {pages} {045114} (\bibinfo {year} {2021})}\BibitemShut {NoStop}%
\bibitem [{\citenamefont {Xu}\ \emph {et~al.}(2020)\citenamefont {Xu},
  \citenamefont {Li}, \citenamefont {Duan}, \citenamefont {Zhang},
  \citenamefont {Chen}, \citenamefont {Kang}, \citenamefont {Liang},
  \citenamefont {Chen}, \citenamefont {Xia}, \citenamefont {Xu}, \citenamefont
  {Malinowski}, \citenamefont {Xu}, \citenamefont {Chu}, \citenamefont {Li},
  \citenamefont {Guo}, \citenamefont {Liu}, \citenamefont {Yang},\ and\
  \citenamefont {Chen}}]{PhysRevB.101.201104}%
  \BibitemOpen
  \bibfield  {author} {\bibinfo {author} {\bibfnamefont {X.}~\bibnamefont
  {Xu}}, \bibinfo {author} {\bibfnamefont {Y.~W.}\ \bibnamefont {Li}}, \bibinfo
  {author} {\bibfnamefont {S.~R.}\ \bibnamefont {Duan}}, \bibinfo {author}
  {\bibfnamefont {S.~L.}\ \bibnamefont {Zhang}}, \bibinfo {author}
  {\bibfnamefont {Y.~J.}\ \bibnamefont {Chen}}, \bibinfo {author}
  {\bibfnamefont {L.}~\bibnamefont {Kang}}, \bibinfo {author} {\bibfnamefont
  {A.~J.}\ \bibnamefont {Liang}}, \bibinfo {author} {\bibfnamefont
  {C.}~\bibnamefont {Chen}}, \bibinfo {author} {\bibfnamefont {W.}~\bibnamefont
  {Xia}}, \bibinfo {author} {\bibfnamefont {Y.}~\bibnamefont {Xu}}, \bibinfo
  {author} {\bibfnamefont {P.}~\bibnamefont {Malinowski}}, \bibinfo {author}
  {\bibfnamefont {X.~D.}\ \bibnamefont {Xu}}, \bibinfo {author} {\bibfnamefont
  {J.-H.}\ \bibnamefont {Chu}}, \bibinfo {author} {\bibfnamefont
  {G.}~\bibnamefont {Li}}, \bibinfo {author} {\bibfnamefont {Y.~F.}\
  \bibnamefont {Guo}}, \bibinfo {author} {\bibfnamefont {Z.~K.}\ \bibnamefont
  {Liu}}, \bibinfo {author} {\bibfnamefont {L.~X.}\ \bibnamefont {Yang}},\ and\
  \bibinfo {author} {\bibfnamefont {Y.~L.}\ \bibnamefont {Chen}},\ }\bibfield
  {title} {\bibinfo {title} {Signature for {non-Stoner} ferromagnetism in the
  van der {Waals} ferromagnet
  $\mathrm{F}{\mathrm{e}}_{3}\mathrm{GeT}{\mathrm{e}}_{2}$},\ }\href
  {https://link.aps.org/doi/10.1103/PhysRevB.101.201104} {\bibfield  {journal}
  {\bibinfo  {journal} {Phys. Rev. B}\ }\textbf {\bibinfo {volume} {101}},\
  \bibinfo {pages} {201104} (\bibinfo {year} {2020})}\BibitemShut {NoStop}%
\bibitem [{\citenamefont {Wu}\ \emph {et~al.}(2024)\citenamefont {Wu},
  \citenamefont {Hu}, \citenamefont {Xie}, \citenamefont {Jang}, \citenamefont
  {Huang}, \citenamefont {Guo}, \citenamefont {Wu}, \citenamefont {Hu},
  \citenamefont {Yue}, \citenamefont {Shi}, \citenamefont {Basak},
  \citenamefont {Ren}, \citenamefont {Yilmaz}, \citenamefont {Vescovo},
  \citenamefont {Jozwiak}, \citenamefont {Bostwick}, \citenamefont {Rotenberg},
  \citenamefont {Fedorov}, \citenamefont {Denlinger}, \citenamefont {Klewe},
  \citenamefont {Shafer}, \citenamefont {Lu}, \citenamefont {Hashimoto},
  \citenamefont {Kono}, \citenamefont {Frano}, \citenamefont {Birgeneau},
  \citenamefont {Xu}, \citenamefont {Zhu}, \citenamefont {Dai}, \citenamefont
  {Chu},\ and\ \citenamefont {Yi}}]{PhysRevB.109.104410}%
  \BibitemOpen
  \bibfield  {author} {\bibinfo {author} {\bibfnamefont {H.}~\bibnamefont
  {Wu}}, \bibinfo {author} {\bibfnamefont {C.}~\bibnamefont {Hu}}, \bibinfo
  {author} {\bibfnamefont {Y.}~\bibnamefont {Xie}}, \bibinfo {author}
  {\bibfnamefont {B.~G.}\ \bibnamefont {Jang}}, \bibinfo {author}
  {\bibfnamefont {J.}~\bibnamefont {Huang}}, \bibinfo {author} {\bibfnamefont
  {Y.}~\bibnamefont {Guo}}, \bibinfo {author} {\bibfnamefont {S.}~\bibnamefont
  {Wu}}, \bibinfo {author} {\bibfnamefont {C.}~\bibnamefont {Hu}}, \bibinfo
  {author} {\bibfnamefont {Z.}~\bibnamefont {Yue}}, \bibinfo {author}
  {\bibfnamefont {Y.}~\bibnamefont {Shi}}, \bibinfo {author} {\bibfnamefont
  {R.}~\bibnamefont {Basak}}, \bibinfo {author} {\bibfnamefont
  {Z.}~\bibnamefont {Ren}}, \bibinfo {author} {\bibfnamefont {T.}~\bibnamefont
  {Yilmaz}}, \bibinfo {author} {\bibfnamefont {E.}~\bibnamefont {Vescovo}},
  \bibinfo {author} {\bibfnamefont {C.}~\bibnamefont {Jozwiak}}, \bibinfo
  {author} {\bibfnamefont {A.}~\bibnamefont {Bostwick}}, \bibinfo {author}
  {\bibfnamefont {E.}~\bibnamefont {Rotenberg}}, \bibinfo {author}
  {\bibfnamefont {A.}~\bibnamefont {Fedorov}}, \bibinfo {author} {\bibfnamefont
  {J.~D.}\ \bibnamefont {Denlinger}}, \bibinfo {author} {\bibfnamefont
  {C.}~\bibnamefont {Klewe}}, \bibinfo {author} {\bibfnamefont
  {P.}~\bibnamefont {Shafer}}, \bibinfo {author} {\bibfnamefont
  {D.}~\bibnamefont {Lu}}, \bibinfo {author} {\bibfnamefont {M.}~\bibnamefont
  {Hashimoto}}, \bibinfo {author} {\bibfnamefont {J.}~\bibnamefont {Kono}},
  \bibinfo {author} {\bibfnamefont {A.}~\bibnamefont {Frano}}, \bibinfo
  {author} {\bibfnamefont {R.~J.}\ \bibnamefont {Birgeneau}}, \bibinfo {author}
  {\bibfnamefont {X.}~\bibnamefont {Xu}}, \bibinfo {author} {\bibfnamefont
  {J.-X.}\ \bibnamefont {Zhu}}, \bibinfo {author} {\bibfnamefont
  {P.}~\bibnamefont {Dai}}, \bibinfo {author} {\bibfnamefont {J.-H.}\
  \bibnamefont {Chu}},\ and\ \bibinfo {author} {\bibfnamefont {M.}~\bibnamefont
  {Yi}},\ }\bibfield  {title} {\bibinfo {title} {Spectral evidence for
  local-moment ferromagnetism in the van der {Waals} metals \ce{FeGaTe2} and
  \ce{FeGeTe2}},\ }\href {https://link.aps.org/doi/10.1103/PhysRevB.109.104410}
  {\bibfield  {journal} {\bibinfo  {journal} {Phys. Rev. B}\ }\textbf {\bibinfo
  {volume} {109}},\ \bibinfo {pages} {104410} (\bibinfo {year}
  {2024})}\BibitemShut {NoStop}%
\bibitem [{\citenamefont {Zhong}\ \emph {et~al.}(2023)\citenamefont {Zhong},
  \citenamefont {Peng}, \citenamefont {Huang}, \citenamefont {Guan},
  \citenamefont {Hwang}, \citenamefont {Hsu}, \citenamefont {Hu}, \citenamefont
  {Jia}, \citenamefont {Moritz}, \citenamefont {Lu}, \citenamefont {Lee},
  \citenamefont {Jia}, \citenamefont {Devereaux}, \citenamefont {Mo},\ and\
  \citenamefont {Shen}}]{s41467-023-40997-1}%
  \BibitemOpen
  \bibfield  {author} {\bibinfo {author} {\bibfnamefont {Y.}~\bibnamefont
  {Zhong}}, \bibinfo {author} {\bibfnamefont {C.}~\bibnamefont {Peng}},
  \bibinfo {author} {\bibfnamefont {H.}~\bibnamefont {Huang}}, \bibinfo
  {author} {\bibfnamefont {D.}~\bibnamefont {Guan}}, \bibinfo {author}
  {\bibfnamefont {J.}~\bibnamefont {Hwang}}, \bibinfo {author} {\bibfnamefont
  {K.~H.}\ \bibnamefont {Hsu}}, \bibinfo {author} {\bibfnamefont
  {Y.}~\bibnamefont {Hu}}, \bibinfo {author} {\bibfnamefont {C.}~\bibnamefont
  {Jia}}, \bibinfo {author} {\bibfnamefont {B.}~\bibnamefont {Moritz}},
  \bibinfo {author} {\bibfnamefont {D.}~\bibnamefont {Lu}}, \bibinfo {author}
  {\bibfnamefont {J.-S.}\ \bibnamefont {Lee}}, \bibinfo {author} {\bibfnamefont
  {J.-F.}\ \bibnamefont {Jia}}, \bibinfo {author} {\bibfnamefont {T.~P.}\
  \bibnamefont {Devereaux}}, \bibinfo {author} {\bibfnamefont {S.-K.}\
  \bibnamefont {Mo}},\ and\ \bibinfo {author} {\bibfnamefont {Z.-X.}\
  \bibnamefont {Shen}},\ }\bibfield  {title} {\bibinfo {title} {From {Stoner}
  to local moment magnetism in atomically thin \ce{Cr2Te3}},\ }\href
  {https://doi.org/10.1038/s41467-023-40997-1} {\bibfield  {journal} {\bibinfo
  {journal} {Nat. Commun.}\ }\textbf {\bibinfo {volume} {14}},\ \bibinfo
  {pages} {5340} (\bibinfo {year} {2023})}\BibitemShut {NoStop}%
\bibitem [{\citenamefont {Katanin}\ and\ \citenamefont
  {Agapov}(2025)}]{PhysRevB.111.035118}%
  \BibitemOpen
  \bibfield  {author} {\bibinfo {author} {\bibfnamefont {A.~A.}\ \bibnamefont
  {Katanin}}\ and\ \bibinfo {author} {\bibfnamefont {E.~M.}\ \bibnamefont
  {Agapov}},\ }\bibfield  {title} {\bibinfo {title} {Magnetic properties of
  monolayer, multilayer, and bulk ${\mathrm{crte}}_{2}$},\ }\href
  {https://link.aps.org/doi/10.1103/PhysRevB.111.035118} {\bibfield  {journal}
  {\bibinfo  {journal} {Phys. Rev. B}\ }\textbf {\bibinfo {volume} {111}},\
  \bibinfo {pages} {035118} (\bibinfo {year} {2025})}\BibitemShut {NoStop}%
\bibitem [{\citenamefont {Hohenberg}\ and\ \citenamefont
  {Kohn}(1964)}]{PhysRev.136.B864}%
  \BibitemOpen
  \bibfield  {author} {\bibinfo {author} {\bibfnamefont {P.}~\bibnamefont
  {Hohenberg}}\ and\ \bibinfo {author} {\bibfnamefont {W.}~\bibnamefont
  {Kohn}},\ }\bibfield  {title} {\bibinfo {title} {Inhomogeneous electron
  gas},\ }\href {https://link.aps.org/doi/10.1103/PhysRev.136.B864} {\bibfield
  {journal} {\bibinfo  {journal} {Phys. Rev.}\ }\textbf {\bibinfo {volume}
  {136}},\ \bibinfo {pages} {B864} (\bibinfo {year} {1964})}\BibitemShut
  {NoStop}%
\bibitem [{\citenamefont {Kohn}\ and\ \citenamefont
  {Sham}(1965)}]{PhysRev.140.A1133}%
  \BibitemOpen
  \bibfield  {author} {\bibinfo {author} {\bibfnamefont {W.}~\bibnamefont
  {Kohn}}\ and\ \bibinfo {author} {\bibfnamefont {L.~J.}\ \bibnamefont
  {Sham}},\ }\bibfield  {title} {\bibinfo {title} {Self-consistent equations
  including exchange and correlation effects},\ }\href
  {https://link.aps.org/doi/10.1103/PhysRev.140.A1133} {\bibfield  {journal}
  {\bibinfo  {journal} {Phys. Rev.}\ }\textbf {\bibinfo {volume} {140}},\
  \bibinfo {pages} {A1133} (\bibinfo {year} {1965})}\BibitemShut {NoStop}%
\bibitem [{\citenamefont {Kresse}\ and\ \citenamefont
  {Hafner}(1993)}]{PhysRevB.47.558}%
  \BibitemOpen
  \bibfield  {author} {\bibinfo {author} {\bibfnamefont {G.}~\bibnamefont
  {Kresse}}\ and\ \bibinfo {author} {\bibfnamefont {J.}~\bibnamefont
  {Hafner}},\ }\bibfield  {title} {\bibinfo {title} {Ab initio molecular
  dynamics for liquid metals},\ }\href
  {http://link.aps.org/doi/10.1103/PhysRevB.47.558} {\bibfield  {journal}
  {\bibinfo  {journal} {Phys. Rev. B}\ }\textbf {\bibinfo {volume} {47}},\
  \bibinfo {pages} {558} (\bibinfo {year} {1993})}\BibitemShut {NoStop}%
\bibitem [{\citenamefont {Kresse}\ and\ \citenamefont
  {Furthm\"uller}(1996)}]{PhysRevB.54.11169}%
  \BibitemOpen
  \bibfield  {author} {\bibinfo {author} {\bibfnamefont {G.}~\bibnamefont
  {Kresse}}\ and\ \bibinfo {author} {\bibfnamefont {J.}~\bibnamefont
  {Furthm\"uller}},\ }\bibfield  {title} {\bibinfo {title} {Efficient iterative
  schemes for ab initio total-energy calculations using a plane-wave basis
  set},\ }\href {http://link.aps.org/doi/10.1103/PhysRevB.54.11169} {\bibfield
  {journal} {\bibinfo  {journal} {Phys. Rev. B}\ }\textbf {\bibinfo {volume}
  {54}},\ \bibinfo {pages} {11169} (\bibinfo {year} {1996})}\BibitemShut
  {NoStop}%
\bibitem [{\citenamefont {Bl\"ochl}(1994)}]{PhysRevB.50.17953}%
  \BibitemOpen
  \bibfield  {author} {\bibinfo {author} {\bibfnamefont {P.~E.}\ \bibnamefont
  {Bl\"ochl}},\ }\bibfield  {title} {\bibinfo {title} {Projector augmented-wave
  method},\ }\href {http://link.aps.org/doi/10.1103/PhysRevB.50.17953}
  {\bibfield  {journal} {\bibinfo  {journal} {Phys. Rev. B}\ }\textbf {\bibinfo
  {volume} {50}},\ \bibinfo {pages} {17953} (\bibinfo {year}
  {1994})}\BibitemShut {NoStop}%
\bibitem [{\citenamefont {Perdew}\ \emph {et~al.}(1996)\citenamefont {Perdew},
  \citenamefont {Burke},\ and\ \citenamefont
  {Ernzerhof}}]{PhysRevLett.77.3865}%
  \BibitemOpen
  \bibfield  {author} {\bibinfo {author} {\bibfnamefont {J.~P.}\ \bibnamefont
  {Perdew}}, \bibinfo {author} {\bibfnamefont {K.}~\bibnamefont {Burke}},\ and\
  \bibinfo {author} {\bibfnamefont {M.}~\bibnamefont {Ernzerhof}},\ }\bibfield
  {title} {\bibinfo {title} {Generalized gradient approximation made simple},\
  }\href {https://link.aps.org/doi/10.1103/PhysRevLett.77.3865} {\bibfield
  {journal} {\bibinfo  {journal} {Phys. Rev. Lett.}\ }\textbf {\bibinfo
  {volume} {77}},\ \bibinfo {pages} {3865} (\bibinfo {year}
  {1996})}\BibitemShut {NoStop}%
\bibitem [{\citenamefont {Dudarev}\ \emph {et~al.}(1998)\citenamefont
  {Dudarev}, \citenamefont {Botton}, \citenamefont {Savrasov}, \citenamefont
  {Humphreys},\ and\ \citenamefont {Sutton}}]{PhysRevB.57.1505}%
  \BibitemOpen
  \bibfield  {author} {\bibinfo {author} {\bibfnamefont {S.~L.}\ \bibnamefont
  {Dudarev}}, \bibinfo {author} {\bibfnamefont {G.~A.}\ \bibnamefont {Botton}},
  \bibinfo {author} {\bibfnamefont {S.~Y.}\ \bibnamefont {Savrasov}}, \bibinfo
  {author} {\bibfnamefont {C.~J.}\ \bibnamefont {Humphreys}},\ and\ \bibinfo
  {author} {\bibfnamefont {A.~P.}\ \bibnamefont {Sutton}},\ }\bibfield  {title}
  {\bibinfo {title} {Electron-energy-loss spectra and the structural stability
  of nickel oxide: An {LSDA+U} study},\ }\href
  {http://link.aps.org/doi/10.1103/PhysRevB.57.1505} {\bibfield  {journal}
  {\bibinfo  {journal} {Phys. Rev. B}\ }\textbf {\bibinfo {volume} {57}},\
  \bibinfo {pages} {1505} (\bibinfo {year} {1998})}\BibitemShut {NoStop}%
\bibitem [{\citenamefont {Grimme}\ \emph {et~al.}(2010)\citenamefont {Grimme},
  \citenamefont {Antony}, \citenamefont {Ehrlich},\ and\ \citenamefont
  {Krieg}}]{10.1063/1.3382344}%
  \BibitemOpen
  \bibfield  {author} {\bibinfo {author} {\bibfnamefont {S.}~\bibnamefont
  {Grimme}}, \bibinfo {author} {\bibfnamefont {J.}~\bibnamefont {Antony}},
  \bibinfo {author} {\bibfnamefont {S.}~\bibnamefont {Ehrlich}},\ and\ \bibinfo
  {author} {\bibfnamefont {H.}~\bibnamefont {Krieg}},\ }\bibfield  {title}
  {\bibinfo {title} {A consistent and accurate ab initio parametrization of
  density functional dispersion correction ({DFT-D}) for the 94 elements
  \ce{H-Pu}},\ }\href {https://doi.org/10.1063/1.3382344} {\bibfield  {journal}
  {\bibinfo  {journal} {J. Chem. Phys.}\ }\textbf {\bibinfo {volume} {132}},\
  \bibinfo {pages} {154104} (\bibinfo {year} {2010})}\BibitemShut {NoStop}%
\bibitem [{\citenamefont {Monkhorst}\ and\ \citenamefont
  {Pack}(1976)}]{PhysRevB.13.5188}%
  \BibitemOpen
  \bibfield  {author} {\bibinfo {author} {\bibfnamefont {H.~J.}\ \bibnamefont
  {Monkhorst}}\ and\ \bibinfo {author} {\bibfnamefont {J.~D.}\ \bibnamefont
  {Pack}},\ }\bibfield  {title} {\bibinfo {title} {Special points for
  brillouin-zone integrations},\ }\href
  {https://link.aps.org/doi/10.1103/PhysRevB.13.5188} {\bibfield  {journal}
  {\bibinfo  {journal} {Phys. Rev. B}\ }\textbf {\bibinfo {volume} {13}},\
  \bibinfo {pages} {5188} (\bibinfo {year} {1976})}\BibitemShut {NoStop}%
\bibitem [{\citenamefont {Marzari}\ and\ \citenamefont
  {Vanderbilt}(1997)}]{PhysRevB.56.12847}%
  \BibitemOpen
  \bibfield  {author} {\bibinfo {author} {\bibfnamefont {N.}~\bibnamefont
  {Marzari}}\ and\ \bibinfo {author} {\bibfnamefont {D.}~\bibnamefont
  {Vanderbilt}},\ }\bibfield  {title} {\bibinfo {title} {Maximally localized
  generalized {Wannier} functions for composite energy bands},\ }\href
  {https://link.aps.org/doi/10.1103/PhysRevB.56.12847} {\bibfield  {journal}
  {\bibinfo  {journal} {Phys. Rev. B}\ }\textbf {\bibinfo {volume} {56}},\
  \bibinfo {pages} {12847} (\bibinfo {year} {1997})}\BibitemShut {NoStop}%
\bibitem [{\citenamefont {Pizzi}\ \emph {et~al.}(2020)\citenamefont {Pizzi},
  \citenamefont {Vitale}, \citenamefont {Arita}, \citenamefont {Blügel},
  \citenamefont {Freimuth}, \citenamefont {G{\'{e}}ranton}, \citenamefont
  {Gibertini}, \citenamefont {Gresch}, \citenamefont {Johnson}, \citenamefont
  {Koretsune}, \citenamefont {Iba{\~{n}}ez-Azpiroz}, \citenamefont {Lee},
  \citenamefont {Lihm}, \citenamefont {Marchand}, \citenamefont {Marrazzo},
  \citenamefont {Mokrousov}, \citenamefont {Mustafa}, \citenamefont {Nohara},
  \citenamefont {Nomura}, \citenamefont {Paulatto}, \citenamefont
  {Ponc{\'{e}}}, \citenamefont {Ponweiser}, \citenamefont {Qiao}, \citenamefont
  {Thöle}, \citenamefont {Tsirkin}, \citenamefont {Wierzbowska}, \citenamefont
  {Marzari}, \citenamefont {Vanderbilt}, \citenamefont {Souza}, \citenamefont
  {Mostofi},\ and\ \citenamefont {Yates}}]{Pizzi2020}%
  \BibitemOpen
  \bibfield  {author} {\bibinfo {author} {\bibfnamefont {G.}~\bibnamefont
  {Pizzi}}, \bibinfo {author} {\bibfnamefont {V.}~\bibnamefont {Vitale}},
  \bibinfo {author} {\bibfnamefont {R.}~\bibnamefont {Arita}}, \bibinfo
  {author} {\bibfnamefont {S.}~\bibnamefont {Blügel}}, \bibinfo {author}
  {\bibfnamefont {F.}~\bibnamefont {Freimuth}}, \bibinfo {author}
  {\bibfnamefont {G.}~\bibnamefont {G{\'{e}}ranton}}, \bibinfo {author}
  {\bibfnamefont {M.}~\bibnamefont {Gibertini}}, \bibinfo {author}
  {\bibfnamefont {D.}~\bibnamefont {Gresch}}, \bibinfo {author} {\bibfnamefont
  {C.}~\bibnamefont {Johnson}}, \bibinfo {author} {\bibfnamefont
  {T.}~\bibnamefont {Koretsune}}, \bibinfo {author} {\bibfnamefont
  {J.}~\bibnamefont {Iba{\~{n}}ez-Azpiroz}}, \bibinfo {author} {\bibfnamefont
  {H.}~\bibnamefont {Lee}}, \bibinfo {author} {\bibfnamefont {J.-M.}\
  \bibnamefont {Lihm}}, \bibinfo {author} {\bibfnamefont {D.}~\bibnamefont
  {Marchand}}, \bibinfo {author} {\bibfnamefont {A.}~\bibnamefont {Marrazzo}},
  \bibinfo {author} {\bibfnamefont {Y.}~\bibnamefont {Mokrousov}}, \bibinfo
  {author} {\bibfnamefont {J.~I.}\ \bibnamefont {Mustafa}}, \bibinfo {author}
  {\bibfnamefont {Y.}~\bibnamefont {Nohara}}, \bibinfo {author} {\bibfnamefont
  {Y.}~\bibnamefont {Nomura}}, \bibinfo {author} {\bibfnamefont
  {L.}~\bibnamefont {Paulatto}}, \bibinfo {author} {\bibfnamefont
  {S.}~\bibnamefont {Ponc{\'{e}}}}, \bibinfo {author} {\bibfnamefont
  {T.}~\bibnamefont {Ponweiser}}, \bibinfo {author} {\bibfnamefont
  {J.}~\bibnamefont {Qiao}}, \bibinfo {author} {\bibfnamefont {F.}~\bibnamefont
  {Thöle}}, \bibinfo {author} {\bibfnamefont {S.~S.}\ \bibnamefont {Tsirkin}},
  \bibinfo {author} {\bibfnamefont {M.}~\bibnamefont {Wierzbowska}}, \bibinfo
  {author} {\bibfnamefont {N.}~\bibnamefont {Marzari}}, \bibinfo {author}
  {\bibfnamefont {D.}~\bibnamefont {Vanderbilt}}, \bibinfo {author}
  {\bibfnamefont {I.}~\bibnamefont {Souza}}, \bibinfo {author} {\bibfnamefont
  {A.~A.}\ \bibnamefont {Mostofi}},\ and\ \bibinfo {author} {\bibfnamefont
  {J.~R.}\ \bibnamefont {Yates}},\ }\bibfield  {title} {\bibinfo {title}
  {Wannier90 as a community code: new features and applications},\ }\href
  {https://doi.org/10.1088%2F1361-648x%2Fab51ff} {\bibfield  {journal}
  {\bibinfo  {journal} {J. Phys.: Condens. Matter}\ }\textbf {\bibinfo {volume}
  {32}},\ \bibinfo {pages} {165902} (\bibinfo {year} {2020})}\BibitemShut
  {NoStop}%
\bibitem [{\citenamefont {Metropolis}\ \emph {et~al.}(1953)\citenamefont
  {Metropolis}, \citenamefont {Rosenbluth}, \citenamefont {Rosenbluth},
  \citenamefont {Teller},\ and\ \citenamefont {Teller}}]{10.1063/1.1699114}%
  \BibitemOpen
  \bibfield  {author} {\bibinfo {author} {\bibfnamefont {N.}~\bibnamefont
  {Metropolis}}, \bibinfo {author} {\bibfnamefont {A.~W.}\ \bibnamefont
  {Rosenbluth}}, \bibinfo {author} {\bibfnamefont {M.~N.}\ \bibnamefont
  {Rosenbluth}}, \bibinfo {author} {\bibfnamefont {A.~H.}\ \bibnamefont
  {Teller}},\ and\ \bibinfo {author} {\bibfnamefont {E.}~\bibnamefont
  {Teller}},\ }\bibfield  {title} {\bibinfo {title} {Equation of state
  calculations by fast computing machines},\ }\href
  {https://doi.org/10.1063/1.1699114} {\bibfield  {journal} {\bibinfo
  {journal} {J. Chem. Phys.}\ }\textbf {\bibinfo {volume} {21}},\ \bibinfo
  {pages} {1087} (\bibinfo {year} {1953})}\BibitemShut {NoStop}%
\bibitem [{\citenamefont {Marsaglia}(1972)}]{marsaglia1972}%
  \BibitemOpen
  \bibfield  {author} {\bibinfo {author} {\bibfnamefont {G.}~\bibnamefont
  {Marsaglia}},\ }\bibfield  {title} {\bibinfo {title} {Choosing a point from
  the surface of a sphere},\ }\href {https://doi.org/10.1214/aoms/1177692644}
  {\bibfield  {journal} {\bibinfo  {journal} {Ann. Math. Statist.}\ }\textbf
  {\bibinfo {volume} {43}},\ \bibinfo {pages} {645} (\bibinfo {year}
  {1972})}\BibitemShut {NoStop}%
\bibitem [{\citenamefont {Evans}\ \emph {et~al.}(2014)\citenamefont {Evans},
  \citenamefont {Fan}, \citenamefont {Chureemart}, \citenamefont {Ostler},
  \citenamefont {Ellis},\ and\ \citenamefont {Chantrell}}]{Evans_2014}%
  \BibitemOpen
  \bibfield  {author} {\bibinfo {author} {\bibfnamefont {R.~F.~L.}\
  \bibnamefont {Evans}}, \bibinfo {author} {\bibfnamefont {W.~J.}\ \bibnamefont
  {Fan}}, \bibinfo {author} {\bibfnamefont {P.}~\bibnamefont {Chureemart}},
  \bibinfo {author} {\bibfnamefont {T.~A.}\ \bibnamefont {Ostler}}, \bibinfo
  {author} {\bibfnamefont {M.~O.~A.}\ \bibnamefont {Ellis}},\ and\ \bibinfo
  {author} {\bibfnamefont {R.~W.}\ \bibnamefont {Chantrell}},\ }\bibfield
  {title} {\bibinfo {title} {Atomistic spin model simulations of magnetic
  nanomaterials},\ }\href {https://doi.org/10.1088/0953-8984/26/10/103202}
  {\bibfield  {journal} {\bibinfo  {journal} {J. Phys.: Condens. Matter}\
  }\textbf {\bibinfo {volume} {26}},\ \bibinfo {pages} {103202} (\bibinfo
  {year} {2014})}\BibitemShut {NoStop}%
\bibitem [{sup()}]{supple}%
  \BibitemOpen
  \href@noop {} {}\bibinfo {note} {Supplemental Material}\BibitemShut {NoStop}%
\bibitem [{\citenamefont {Otero~Fumega}\ \emph {et~al.}(2020)\citenamefont
  {Otero~Fumega}, \citenamefont {Phillips},\ and\ \citenamefont
  {Pardo}}]{acs.jpcc.0c04913}%
  \BibitemOpen
  \bibfield  {author} {\bibinfo {author} {\bibfnamefont {A.}~\bibnamefont
  {Otero~Fumega}}, \bibinfo {author} {\bibfnamefont {J.}~\bibnamefont
  {Phillips}},\ and\ \bibinfo {author} {\bibfnamefont {V.}~\bibnamefont
  {Pardo}},\ }\bibfield  {title} {\bibinfo {title} {Controlled two-dimensional
  ferromagnetism in ${1T}$-\ce{CrTe2}: The role of charge density wave and
  strain},\ }\href {https://doi.org/10.1021/acs.jpcc.0c04913} {\bibfield
  {journal} {\bibinfo  {journal} {J. Phys. Chem. C}\ }\textbf {\bibinfo
  {volume} {124}},\ \bibinfo {pages} {21047} (\bibinfo {year}
  {2020})}\BibitemShut {NoStop}%
\bibitem [{\citenamefont {Meng}\ \emph {et~al.}(2021)\citenamefont {Meng},
  \citenamefont {Zhou}, \citenamefont {Xu}, \citenamefont {Yang}, \citenamefont
  {Si}, \citenamefont {Liu}, \citenamefont {Wang}, \citenamefont {Jiang},
  \citenamefont {Li}, \citenamefont {Qin}, \citenamefont {Zhang}, \citenamefont
  {Wang}, \citenamefont {Liu}, \citenamefont {Tang}, \citenamefont {Ye},
  \citenamefont {Zhou}, \citenamefont {Bao}, \citenamefont {Gao},\ and\
  \citenamefont {Gong}}]{s41467-021-21072-z}%
  \BibitemOpen
  \bibfield  {author} {\bibinfo {author} {\bibfnamefont {L.}~\bibnamefont
  {Meng}}, \bibinfo {author} {\bibfnamefont {Z.}~\bibnamefont {Zhou}}, \bibinfo
  {author} {\bibfnamefont {M.}~\bibnamefont {Xu}}, \bibinfo {author}
  {\bibfnamefont {S.}~\bibnamefont {Yang}}, \bibinfo {author} {\bibfnamefont
  {K.}~\bibnamefont {Si}}, \bibinfo {author} {\bibfnamefont {L.}~\bibnamefont
  {Liu}}, \bibinfo {author} {\bibfnamefont {X.}~\bibnamefont {Wang}}, \bibinfo
  {author} {\bibfnamefont {H.}~\bibnamefont {Jiang}}, \bibinfo {author}
  {\bibfnamefont {B.}~\bibnamefont {Li}}, \bibinfo {author} {\bibfnamefont
  {P.}~\bibnamefont {Qin}}, \bibinfo {author} {\bibfnamefont {P.}~\bibnamefont
  {Zhang}}, \bibinfo {author} {\bibfnamefont {J.}~\bibnamefont {Wang}},
  \bibinfo {author} {\bibfnamefont {Z.}~\bibnamefont {Liu}}, \bibinfo {author}
  {\bibfnamefont {P.}~\bibnamefont {Tang}}, \bibinfo {author} {\bibfnamefont
  {Y.}~\bibnamefont {Ye}}, \bibinfo {author} {\bibfnamefont {W.}~\bibnamefont
  {Zhou}}, \bibinfo {author} {\bibfnamefont {L.}~\bibnamefont {Bao}}, \bibinfo
  {author} {\bibfnamefont {H.-J.}\ \bibnamefont {Gao}},\ and\ \bibinfo {author}
  {\bibfnamefont {Y.}~\bibnamefont {Gong}},\ }\bibfield  {title} {\bibinfo
  {title} {Anomalous thickness dependence of {Curie} temperature in air-stable
  two-dimensional ferromagnetic ${1T}$-\ce{CrTe2} grown by chemical vapor
  deposition},\ }\href {https://doi.org/10.1038/s41467-021-21072-z} {\bibfield
  {journal} {\bibinfo  {journal} {Nat. Commun.}\ }\textbf {\bibinfo {volume}
  {12}},\ \bibinfo {pages} {809} (\bibinfo {year} {2021})}\BibitemShut
  {NoStop}%
\bibitem [{\citenamefont {Wang}\ \emph
  {et~al.}(2019{\natexlab{b}})\citenamefont {Wang}, \citenamefont {Yan},
  \citenamefont {Li}, \citenamefont {Wang}, \citenamefont {Song}, \citenamefont
  {Song}, \citenamefont {Li}, \citenamefont {Chen}, \citenamefont {Qin},
  \citenamefont {Ling}, \citenamefont {Du}, \citenamefont {Cao}, \citenamefont
  {Luo}, \citenamefont {Xiong},\ and\ \citenamefont
  {Sun}}]{PhysRevB.100.024434}%
  \BibitemOpen
  \bibfield  {author} {\bibinfo {author} {\bibfnamefont {Y.}~\bibnamefont
  {Wang}}, \bibinfo {author} {\bibfnamefont {J.}~\bibnamefont {Yan}}, \bibinfo
  {author} {\bibfnamefont {J.}~\bibnamefont {Li}}, \bibinfo {author}
  {\bibfnamefont {S.}~\bibnamefont {Wang}}, \bibinfo {author} {\bibfnamefont
  {M.}~\bibnamefont {Song}}, \bibinfo {author} {\bibfnamefont {J.}~\bibnamefont
  {Song}}, \bibinfo {author} {\bibfnamefont {Z.}~\bibnamefont {Li}}, \bibinfo
  {author} {\bibfnamefont {K.}~\bibnamefont {Chen}}, \bibinfo {author}
  {\bibfnamefont {Y.}~\bibnamefont {Qin}}, \bibinfo {author} {\bibfnamefont
  {L.}~\bibnamefont {Ling}}, \bibinfo {author} {\bibfnamefont {H.}~\bibnamefont
  {Du}}, \bibinfo {author} {\bibfnamefont {L.}~\bibnamefont {Cao}}, \bibinfo
  {author} {\bibfnamefont {X.}~\bibnamefont {Luo}}, \bibinfo {author}
  {\bibfnamefont {Y.}~\bibnamefont {Xiong}},\ and\ \bibinfo {author}
  {\bibfnamefont {Y.}~\bibnamefont {Sun}},\ }\bibfield  {title} {\bibinfo
  {title} {Magnetic anisotropy and topological {Hall} effect in the trigonal
  chromium tellurides \ce{Cr5Te8}},\ }\href
  {https://link.aps.org/doi/10.1103/PhysRevB.100.024434} {\bibfield  {journal}
  {\bibinfo  {journal} {Phys. Rev. B}\ }\textbf {\bibinfo {volume} {100}},\
  \bibinfo {pages} {024434} (\bibinfo {year} {2019}{\natexlab{b}})}\BibitemShut
  {NoStop}%
\bibitem [{\citenamefont {Rubin}\ \emph {et~al.}(2012)\citenamefont {Rubin},
  \citenamefont {Sherman},\ and\ \citenamefont {Schreiber}}]{RUBIN20121062}%
  \BibitemOpen
  \bibfield  {author} {\bibinfo {author} {\bibfnamefont {P.}~\bibnamefont
  {Rubin}}, \bibinfo {author} {\bibfnamefont {A.}~\bibnamefont {Sherman}},\
  and\ \bibinfo {author} {\bibfnamefont {M.}~\bibnamefont {Schreiber}},\
  }\bibfield  {title} {\bibinfo {title} {Magnetic phase diagram of the spin-1
  two-dimensional ${J}_{1}\text{\ensuremath{-}}{J}_{2}$ {Heisenberg} model on a
  triangular lattice},\ }\href
  {https://www.sciencedirect.com/science/article/pii/S0375960112000916}
  {\bibfield  {journal} {\bibinfo  {journal} {Phys. Lett. A}\ }\textbf
  {\bibinfo {volume} {376}},\ \bibinfo {pages} {1062} (\bibinfo {year}
  {2012})}\BibitemShut {NoStop}%
\bibitem [{\citenamefont {Glittum}\ and\ \citenamefont
  {Sylju\aa{}sen}(2021)}]{PhysRevB.104.184427}%
  \BibitemOpen
  \bibfield  {author} {\bibinfo {author} {\bibfnamefont {C.}~\bibnamefont
  {Glittum}}\ and\ \bibinfo {author} {\bibfnamefont {O.~F.}\ \bibnamefont
  {Sylju\aa{}sen}},\ }\bibfield  {title} {\bibinfo {title} {Arc-shaped
  structure factor in the
  ${J}_{1}\text{\ensuremath{-}}{J}_{2}\text{\ensuremath{-}}{J}_{3}$ classical
  {Heisenberg} model on the triangular lattice},\ }\href
  {https://link.aps.org/doi/10.1103/PhysRevB.104.184427} {\bibfield  {journal}
  {\bibinfo  {journal} {Phys. Rev. B}\ }\textbf {\bibinfo {volume} {104}},\
  \bibinfo {pages} {184427} (\bibinfo {year} {2021})}\BibitemShut {NoStop}%
\bibitem [{\citenamefont {Zhu}\ and\ \citenamefont
  {White}(2015)}]{PhysRevB.92.041105}%
  \BibitemOpen
  \bibfield  {author} {\bibinfo {author} {\bibfnamefont {Z.}~\bibnamefont
  {Zhu}}\ and\ \bibinfo {author} {\bibfnamefont {S.~R.}\ \bibnamefont
  {White}},\ }\bibfield  {title} {\bibinfo {title} {Spin liquid phase of the
  ${S}=\frac{1}{2}\phantom{\rule{4.pt}{0ex}}{J}_{1}\ensuremath{-}{J}_{2}$
  {Heisenberg} model on the triangular lattice},\ }\href
  {https://link.aps.org/doi/10.1103/PhysRevB.92.041105} {\bibfield  {journal}
  {\bibinfo  {journal} {Phys. Rev. B}\ }\textbf {\bibinfo {volume} {92}},\
  \bibinfo {pages} {041105} (\bibinfo {year} {2015})}\BibitemShut {NoStop}%
\bibitem [{\citenamefont {Scheie}\ \emph {et~al.}(2024)\citenamefont {Scheie},
  \citenamefont {Ghioldi}, \citenamefont {Xing}, \citenamefont {Paddison},
  \citenamefont {Sherman}, \citenamefont {Dupont}, \citenamefont {Sanjeewa},
  \citenamefont {Lee}, \citenamefont {Woods}, \citenamefont {Abernathy},
  \citenamefont {Pajerowski}, \citenamefont {Williams}, \citenamefont {Zhang},
  \citenamefont {Manuel}, \citenamefont {Trumper}, \citenamefont {Pemmaraju},
  \citenamefont {Sefat}, \citenamefont {Parker}, \citenamefont {Devereaux},
  \citenamefont {Movshovich}, \citenamefont {Moore}, \citenamefont {Batista},\
  and\ \citenamefont {Tennant}}]{s41567-023-02259-1}%
  \BibitemOpen
  \bibfield  {author} {\bibinfo {author} {\bibfnamefont {A.~O.}\ \bibnamefont
  {Scheie}}, \bibinfo {author} {\bibfnamefont {E.~A.}\ \bibnamefont {Ghioldi}},
  \bibinfo {author} {\bibfnamefont {J.}~\bibnamefont {Xing}}, \bibinfo {author}
  {\bibfnamefont {J.~A.~M.}\ \bibnamefont {Paddison}}, \bibinfo {author}
  {\bibfnamefont {N.~E.}\ \bibnamefont {Sherman}}, \bibinfo {author}
  {\bibfnamefont {M.}~\bibnamefont {Dupont}}, \bibinfo {author} {\bibfnamefont
  {L.~D.}\ \bibnamefont {Sanjeewa}}, \bibinfo {author} {\bibfnamefont
  {S.}~\bibnamefont {Lee}}, \bibinfo {author} {\bibfnamefont {A.~J.}\
  \bibnamefont {Woods}}, \bibinfo {author} {\bibfnamefont {D.}~\bibnamefont
  {Abernathy}}, \bibinfo {author} {\bibfnamefont {D.~M.}\ \bibnamefont
  {Pajerowski}}, \bibinfo {author} {\bibfnamefont {T.~J.}\ \bibnamefont
  {Williams}}, \bibinfo {author} {\bibfnamefont {S.-S.}\ \bibnamefont {Zhang}},
  \bibinfo {author} {\bibfnamefont {L.~O.}\ \bibnamefont {Manuel}}, \bibinfo
  {author} {\bibfnamefont {A.~E.}\ \bibnamefont {Trumper}}, \bibinfo {author}
  {\bibfnamefont {C.~D.}\ \bibnamefont {Pemmaraju}}, \bibinfo {author}
  {\bibfnamefont {A.~S.}\ \bibnamefont {Sefat}}, \bibinfo {author}
  {\bibfnamefont {D.~S.}\ \bibnamefont {Parker}}, \bibinfo {author}
  {\bibfnamefont {T.~P.}\ \bibnamefont {Devereaux}}, \bibinfo {author}
  {\bibfnamefont {R.}~\bibnamefont {Movshovich}}, \bibinfo {author}
  {\bibfnamefont {J.~E.}\ \bibnamefont {Moore}}, \bibinfo {author}
  {\bibfnamefont {C.~D.}\ \bibnamefont {Batista}},\ and\ \bibinfo {author}
  {\bibfnamefont {D.~A.}\ \bibnamefont {Tennant}},\ }\bibfield  {title}
  {\bibinfo {title} {Proximate spin liquid and fractionalization in the
  triangular antiferromagnet \ce{KYbSe2}},\ }\href
  {https://doi.org/10.1038/s41567-023-02259-1} {\bibfield  {journal} {\bibinfo
  {journal} {Nat. Phys.}\ }\textbf {\bibinfo {volume} {20}},\ \bibinfo {pages}
  {74} (\bibinfo {year} {2024})}\BibitemShut {NoStop}%
\bibitem [{\citenamefont {Ni}\ \emph {et~al.}(2021{\natexlab{a}})\citenamefont
  {Ni}, \citenamefont {Li}, \citenamefont {Amoroso}, \citenamefont {He},
  \citenamefont {Feng}, \citenamefont {Kan}, \citenamefont {Picozzi},\ and\
  \citenamefont {Xiang}}]{PhysRevLett.127.247204}%
  \BibitemOpen
  \bibfield  {author} {\bibinfo {author} {\bibfnamefont {J.~Y.}\ \bibnamefont
  {Ni}}, \bibinfo {author} {\bibfnamefont {X.~Y.}\ \bibnamefont {Li}}, \bibinfo
  {author} {\bibfnamefont {D.}~\bibnamefont {Amoroso}}, \bibinfo {author}
  {\bibfnamefont {X.}~\bibnamefont {He}}, \bibinfo {author} {\bibfnamefont
  {J.~S.}\ \bibnamefont {Feng}}, \bibinfo {author} {\bibfnamefont {E.~J.}\
  \bibnamefont {Kan}}, \bibinfo {author} {\bibfnamefont {S.}~\bibnamefont
  {Picozzi}},\ and\ \bibinfo {author} {\bibfnamefont {H.~J.}\ \bibnamefont
  {Xiang}},\ }\bibfield  {title} {\bibinfo {title} {Giant biquadratic exchange
  in {2D} magnets and its role in stabilizing ferromagnetism of \ce{NiCl2}
  monolayers},\ }\href
  {https://link.aps.org/doi/10.1103/PhysRevLett.127.247204} {\bibfield
  {journal} {\bibinfo  {journal} {Phys. Rev. Lett.}\ }\textbf {\bibinfo
  {volume} {127}},\ \bibinfo {pages} {247204} (\bibinfo {year}
  {2021}{\natexlab{a}})}\BibitemShut {NoStop}%
\bibitem [{\citenamefont {Kabir}\ and\ \citenamefont
  {Saha-Dasgupta}(2014)}]{PhysRevB.90.035403}%
  \BibitemOpen
  \bibfield  {author} {\bibinfo {author} {\bibfnamefont {M.}~\bibnamefont
  {Kabir}}\ and\ \bibinfo {author} {\bibfnamefont {T.}~\bibnamefont
  {Saha-Dasgupta}},\ }\bibfield  {title} {\bibinfo {title} {Manipulation of
  edge magnetism in hexagonal graphene nanoflakes},\ }\href
  {https://link.aps.org/doi/10.1103/PhysRevB.90.035403} {\bibfield  {journal}
  {\bibinfo  {journal} {Phys. Rev. B}\ }\textbf {\bibinfo {volume} {90}},\
  \bibinfo {pages} {035403} (\bibinfo {year} {2014})}\BibitemShut {NoStop}%
\bibitem [{\citenamefont {Ganguly}\ \emph {et~al.}(2017)\citenamefont
  {Ganguly}, \citenamefont {Kabir},\ and\ \citenamefont
  {Saha-Dasgupta}}]{PhysRevB.95.174419}%
  \BibitemOpen
  \bibfield  {author} {\bibinfo {author} {\bibfnamefont {S.}~\bibnamefont
  {Ganguly}}, \bibinfo {author} {\bibfnamefont {M.}~\bibnamefont {Kabir}},\
  and\ \bibinfo {author} {\bibfnamefont {T.}~\bibnamefont {Saha-Dasgupta}},\
  }\bibfield  {title} {\bibinfo {title} {Magnetic and electronic crossovers in
  graphene nanoflakes},\ }\href
  {https://link.aps.org/doi/10.1103/PhysRevB.95.174419} {\bibfield  {journal}
  {\bibinfo  {journal} {Phys. Rev. B}\ }\textbf {\bibinfo {volume} {95}},\
  \bibinfo {pages} {174419} (\bibinfo {year} {2017})}\BibitemShut {NoStop}%
\bibitem [{\citenamefont {Liu}\ \emph {et~al.}(2020)\citenamefont {Liu},
  \citenamefont {Liu}, \citenamefont {Yang}, \citenamefont {Chen},
  \citenamefont {Zhang}, \citenamefont {Li}, \citenamefont {Wu}, \citenamefont
  {Ruan}, \citenamefont {Xiu}, \citenamefont {Liu}, \citenamefont {He},
  \citenamefont {Zhang},\ and\ \citenamefont {Xu}}]{PhysRevLett.125.267205}%
  \BibitemOpen
  \bibfield  {author} {\bibinfo {author} {\bibfnamefont {B.}~\bibnamefont
  {Liu}}, \bibinfo {author} {\bibfnamefont {S.}~\bibnamefont {Liu}}, \bibinfo
  {author} {\bibfnamefont {L.}~\bibnamefont {Yang}}, \bibinfo {author}
  {\bibfnamefont {Z.}~\bibnamefont {Chen}}, \bibinfo {author} {\bibfnamefont
  {E.}~\bibnamefont {Zhang}}, \bibinfo {author} {\bibfnamefont
  {Z.}~\bibnamefont {Li}}, \bibinfo {author} {\bibfnamefont {J.}~\bibnamefont
  {Wu}}, \bibinfo {author} {\bibfnamefont {X.}~\bibnamefont {Ruan}}, \bibinfo
  {author} {\bibfnamefont {F.}~\bibnamefont {Xiu}}, \bibinfo {author}
  {\bibfnamefont {W.}~\bibnamefont {Liu}}, \bibinfo {author} {\bibfnamefont
  {L.}~\bibnamefont {He}}, \bibinfo {author} {\bibfnamefont {R.}~\bibnamefont
  {Zhang}},\ and\ \bibinfo {author} {\bibfnamefont {Y.}~\bibnamefont {Xu}},\
  }\bibfield  {title} {\bibinfo {title} {Light-tunable ferromagnetism in
  atomically thin \ce{Fe3GeTe2} driven by femtosecond laser pulse},\ }\href
  {https://link.aps.org/doi/10.1103/PhysRevLett.125.267205} {\bibfield
  {journal} {\bibinfo  {journal} {Phys. Rev. Lett.}\ }\textbf {\bibinfo
  {volume} {125}},\ \bibinfo {pages} {267205} (\bibinfo {year}
  {2020})}\BibitemShut {NoStop}%
\bibitem [{\citenamefont {Enayat}\ \emph {et~al.}(2014)\citenamefont {Enayat},
  \citenamefont {Sun}, \citenamefont {Singh}, \citenamefont {Aluru},
  \citenamefont {Schmaus}, \citenamefont {Yaresko}, \citenamefont {Liu},
  \citenamefont {Lin}, \citenamefont {Tsurkan}, \citenamefont {Loidl},
  \citenamefont {Deisenhofer},\ and\ \citenamefont {Wahl}}]{science.1251682}%
  \BibitemOpen
  \bibfield  {author} {\bibinfo {author} {\bibfnamefont {M.}~\bibnamefont
  {Enayat}}, \bibinfo {author} {\bibfnamefont {Z.}~\bibnamefont {Sun}},
  \bibinfo {author} {\bibfnamefont {U.~R.}\ \bibnamefont {Singh}}, \bibinfo
  {author} {\bibfnamefont {R.}~\bibnamefont {Aluru}}, \bibinfo {author}
  {\bibfnamefont {S.}~\bibnamefont {Schmaus}}, \bibinfo {author} {\bibfnamefont
  {A.}~\bibnamefont {Yaresko}}, \bibinfo {author} {\bibfnamefont
  {Y.}~\bibnamefont {Liu}}, \bibinfo {author} {\bibfnamefont {C.}~\bibnamefont
  {Lin}}, \bibinfo {author} {\bibfnamefont {V.}~\bibnamefont {Tsurkan}},
  \bibinfo {author} {\bibfnamefont {A.}~\bibnamefont {Loidl}}, \bibinfo
  {author} {\bibfnamefont {J.}~\bibnamefont {Deisenhofer}},\ and\ \bibinfo
  {author} {\bibfnamefont {P.}~\bibnamefont {Wahl}},\ }\bibfield  {title}
  {\bibinfo {title} {Real-space imaging of the atomic-scale magnetic structure
  of \ce{FeTe}},\ }\href
  {https://www.science.org/doi/abs/10.1126/science.1251682} {\bibfield
  {journal} {\bibinfo  {journal} {Science}\ }\textbf {\bibinfo {volume}
  {345}},\ \bibinfo {pages} {653} (\bibinfo {year} {2014})}\BibitemShut
  {NoStop}%
\bibitem [{\citenamefont {Li}\ \emph {et~al.}(2024)\citenamefont {Li},
  \citenamefont {Du}, \citenamefont {Wang}, \citenamefont {Xu}, \citenamefont
  {Zhao}, \citenamefont {Zhai}, \citenamefont {Liu}, \citenamefont {Chen},
  \citenamefont {Yang}, \citenamefont {Plumb}, \citenamefont {Ju},
  \citenamefont {Shi}, \citenamefont {Liu}, \citenamefont {Guo}, \citenamefont
  {Chen}, \citenamefont {Chen},\ and\ \citenamefont
  {Yang}}]{acs.nanolett.4c01542}%
  \BibitemOpen
  \bibfield  {author} {\bibinfo {author} {\bibfnamefont {Y.}~\bibnamefont
  {Li}}, \bibinfo {author} {\bibfnamefont {X.}~\bibnamefont {Du}}, \bibinfo
  {author} {\bibfnamefont {J.}~\bibnamefont {Wang}}, \bibinfo {author}
  {\bibfnamefont {R.}~\bibnamefont {Xu}}, \bibinfo {author} {\bibfnamefont
  {W.}~\bibnamefont {Zhao}}, \bibinfo {author} {\bibfnamefont {K.}~\bibnamefont
  {Zhai}}, \bibinfo {author} {\bibfnamefont {J.}~\bibnamefont {Liu}}, \bibinfo
  {author} {\bibfnamefont {H.}~\bibnamefont {Chen}}, \bibinfo {author}
  {\bibfnamefont {Y.}~\bibnamefont {Yang}}, \bibinfo {author} {\bibfnamefont
  {N.~C.}\ \bibnamefont {Plumb}}, \bibinfo {author} {\bibfnamefont
  {S.}~\bibnamefont {Ju}}, \bibinfo {author} {\bibfnamefont {M.}~\bibnamefont
  {Shi}}, \bibinfo {author} {\bibfnamefont {Z.}~\bibnamefont {Liu}}, \bibinfo
  {author} {\bibfnamefont {J.-g.}\ \bibnamefont {Guo}}, \bibinfo {author}
  {\bibfnamefont {X.}~\bibnamefont {Chen}}, \bibinfo {author} {\bibfnamefont
  {Y.}~\bibnamefont {Chen}},\ and\ \bibinfo {author} {\bibfnamefont
  {L.}~\bibnamefont {Yang}},\ }\bibfield  {title} {\bibinfo {title}
  {Quantum-confined tunable ferromagnetism on the surface of a van der {Waals}
  antiferromagnet \ce{NaCrTe2}},\ }\href
  {https://doi.org/10.1021/acs.nanolett.4c01542} {\bibfield  {journal}
  {\bibinfo  {journal} {Nano Lett.}\ }\textbf {\bibinfo {volume} {24}},\
  \bibinfo {pages} {9832} (\bibinfo {year} {2024})}\BibitemShut {NoStop}%
\bibitem [{\citenamefont {Wang}\ \emph {et~al.}(2021)\citenamefont {Wang},
  \citenamefont {Deng}, \citenamefont {Liang}, \citenamefont {Gao},
  \citenamefont {Ying}, \citenamefont {Tian}, \citenamefont {Lei},
  \citenamefont {Song}, \citenamefont {Chen}, \citenamefont {Guo},\ and\
  \citenamefont {Chen}}]{PhysRevMaterials.5.L091401}%
  \BibitemOpen
  \bibfield  {author} {\bibinfo {author} {\bibfnamefont {J.}~\bibnamefont
  {Wang}}, \bibinfo {author} {\bibfnamefont {J.}~\bibnamefont {Deng}}, \bibinfo
  {author} {\bibfnamefont {X.}~\bibnamefont {Liang}}, \bibinfo {author}
  {\bibfnamefont {G.}~\bibnamefont {Gao}}, \bibinfo {author} {\bibfnamefont
  {T.}~\bibnamefont {Ying}}, \bibinfo {author} {\bibfnamefont {S.}~\bibnamefont
  {Tian}}, \bibinfo {author} {\bibfnamefont {H.}~\bibnamefont {Lei}}, \bibinfo
  {author} {\bibfnamefont {Y.}~\bibnamefont {Song}}, \bibinfo {author}
  {\bibfnamefont {X.}~\bibnamefont {Chen}}, \bibinfo {author} {\bibfnamefont
  {J.-g.}\ \bibnamefont {Guo}},\ and\ \bibinfo {author} {\bibfnamefont
  {X.}~\bibnamefont {Chen}},\ }\bibfield  {title} {\bibinfo {title}
  {Spin-flip-driven giant magnetotransport in {A}-type antiferromagnet
  \ce{NaCrTe2}},\ }\href
  {https://link.aps.org/doi/10.1103/PhysRevMaterials.5.L091401} {\bibfield
  {journal} {\bibinfo  {journal} {Phys. Rev. Mater.}\ }\textbf {\bibinfo
  {volume} {5}},\ \bibinfo {pages} {L091401} (\bibinfo {year}
  {2021})}\BibitemShut {NoStop}%
\bibitem [{\citenamefont {Anderson}(1950)}]{PhysRev.79.350}%
  \BibitemOpen
  \bibfield  {author} {\bibinfo {author} {\bibfnamefont {P.~W.}\ \bibnamefont
  {Anderson}},\ }\bibfield  {title} {\bibinfo {title} {Antiferromagnetism.
  {Theory} of superexchange interaction},\ }\href
  {https://link.aps.org/doi/10.1103/PhysRev.79.350} {\bibfield  {journal}
  {\bibinfo  {journal} {Phys. Rev.}\ }\textbf {\bibinfo {volume} {79}},\
  \bibinfo {pages} {350} (\bibinfo {year} {1950})}\BibitemShut {NoStop}%
\bibitem [{\citenamefont {Goodenough}(1958)}]{GOODENOUGH1958287}%
  \BibitemOpen
  \bibfield  {author} {\bibinfo {author} {\bibfnamefont {J.~B.}\ \bibnamefont
  {Goodenough}},\ }\bibfield  {title} {\bibinfo {title} {An interpretation of
  the magnetic properties of the perovskite-type mixed crystals
  \ce{La_{(1-x)}Sr_xCoO_{3-$\lambda$}}},\ }\href
  {http://www.sciencedirect.com/science/article/pii/0022369758901070}
  {\bibfield  {journal} {\bibinfo  {journal} {J. Phys. Chem. Solids}\ }\textbf
  {\bibinfo {volume} {6}},\ \bibinfo {pages} {287 } (\bibinfo {year}
  {1958})}\BibitemShut {NoStop}%
\bibitem [{\citenamefont {Kanamori}(1959)}]{KANAMORI195987}%
  \BibitemOpen
  \bibfield  {author} {\bibinfo {author} {\bibfnamefont {J.}~\bibnamefont
  {Kanamori}},\ }\bibfield  {title} {\bibinfo {title} {Superexchange
  interaction and symmetry properties of electron orbitals},\ }\href
  {http://www.sciencedirect.com/science/article/pii/0022369759900617}
  {\bibfield  {journal} {\bibinfo  {journal} {J. Phys. Chem. Solids}\ }\textbf
  {\bibinfo {volume} {10}},\ \bibinfo {pages} {87 } (\bibinfo {year}
  {1959})}\BibitemShut {NoStop}%
\bibitem [{\citenamefont {Ruderman}\ and\ \citenamefont
  {Kittel}(1954)}]{PhysRev.96.99}%
  \BibitemOpen
  \bibfield  {author} {\bibinfo {author} {\bibfnamefont {M.~A.}\ \bibnamefont
  {Ruderman}}\ and\ \bibinfo {author} {\bibfnamefont {C.}~\bibnamefont
  {Kittel}},\ }\bibfield  {title} {\bibinfo {title} {Indirect exchange coupling
  of nuclear magnetic moments by conduction electrons},\ }\href
  {https://link.aps.org/doi/10.1103/PhysRev.96.99} {\bibfield  {journal}
  {\bibinfo  {journal} {Phys. Rev.}\ }\textbf {\bibinfo {volume} {96}},\
  \bibinfo {pages} {99} (\bibinfo {year} {1954})}\BibitemShut {NoStop}%
\bibitem [{\citenamefont {Kasuya}(1956)}]{10.1143/PTP.16.45}%
  \BibitemOpen
  \bibfield  {author} {\bibinfo {author} {\bibfnamefont {T.}~\bibnamefont
  {Kasuya}},\ }\bibfield  {title} {\bibinfo {title} {A theory of metallic
  ferro- and antiferromagnetism on {Z}ener's model},\ }\href
  {https://doi.org/10.1143/PTP.16.45} {\bibfield  {journal} {\bibinfo
  {journal} {Prog. Theor. Phys.}\ }\textbf {\bibinfo {volume} {16}},\ \bibinfo
  {pages} {45} (\bibinfo {year} {1956})}\BibitemShut {NoStop}%
\bibitem [{\citenamefont {Yosida}(1957)}]{PhysRev.106.893}%
  \BibitemOpen
  \bibfield  {author} {\bibinfo {author} {\bibfnamefont {K.}~\bibnamefont
  {Yosida}},\ }\bibfield  {title} {\bibinfo {title} {Magnetic properties of
  \ce{Cu-Mn} alloys},\ }\href
  {https://link.aps.org/doi/10.1103/PhysRev.106.893} {\bibfield  {journal}
  {\bibinfo  {journal} {Phys. Rev.}\ }\textbf {\bibinfo {volume} {106}},\
  \bibinfo {pages} {893} (\bibinfo {year} {1957})}\BibitemShut {NoStop}%
\bibitem [{\citenamefont {Sakurai}(2014)}]{PhysRevB.89.024416}%
  \BibitemOpen
  \bibfield  {author} {\bibinfo {author} {\bibfnamefont {H.}~\bibnamefont
  {Sakurai}},\ }\bibfield  {title} {\bibinfo {title} {Magnetic and electronic
  properties of \ce{Ca_{1-x}Na_xCr2O4}: Double-exchange interactions and ligand
  holes},\ }\href {https://link.aps.org/doi/10.1103/PhysRevB.89.024416}
  {\bibfield  {journal} {\bibinfo  {journal} {Phys. Rev. B}\ }\textbf {\bibinfo
  {volume} {89}},\ \bibinfo {pages} {024416} (\bibinfo {year}
  {2014})}\BibitemShut {NoStop}%
\bibitem [{\citenamefont {Pradhan}\ \emph {et~al.}(2025)\citenamefont
  {Pradhan}, \citenamefont {Sen}, \citenamefont {Sanyal},\ and\ \citenamefont
  {Saha-Dasgupta}}]{PhysRevB.111.L180404}%
  \BibitemOpen
  \bibfield  {author} {\bibinfo {author} {\bibfnamefont {K.}~\bibnamefont
  {Pradhan}}, \bibinfo {author} {\bibfnamefont {D.}~\bibnamefont {Sen}},
  \bibinfo {author} {\bibfnamefont {P.}~\bibnamefont {Sanyal}},\ and\ \bibinfo
  {author} {\bibfnamefont {T.}~\bibnamefont {Saha-Dasgupta}},\ }\bibfield
  {title} {\bibinfo {title} {Two-sublattice double exchange driven magnetism in
  {Cr}-based two-dimensional magnets},\ }\href
  {https://link.aps.org/doi/10.1103/PhysRevB.111.L180404} {\bibfield  {journal}
  {\bibinfo  {journal} {Phys. Rev. B}\ }\textbf {\bibinfo {volume} {111}},\
  \bibinfo {pages} {L180404} (\bibinfo {year} {2025})}\BibitemShut {NoStop}%
\bibitem [{\citenamefont {Zhu}\ \emph {et~al.}(2023)\citenamefont {Zhu},
  \citenamefont {Pan}, \citenamefont {Ge}, \citenamefont {Fan}, \citenamefont
  {Shi}, \citenamefont {Ma}, \citenamefont {Hu},\ and\ \citenamefont
  {Wu}}]{PhysRevB.108.L041401}%
  \BibitemOpen
  \bibfield  {author} {\bibinfo {author} {\bibfnamefont {Y.}~\bibnamefont
  {Zhu}}, \bibinfo {author} {\bibfnamefont {Y.~F.}\ \bibnamefont {Pan}},
  \bibinfo {author} {\bibfnamefont {L.}~\bibnamefont {Ge}}, \bibinfo {author}
  {\bibfnamefont {J.~Y.}\ \bibnamefont {Fan}}, \bibinfo {author} {\bibfnamefont
  {D.~N.}\ \bibnamefont {Shi}}, \bibinfo {author} {\bibfnamefont {C.~L.}\
  \bibnamefont {Ma}}, \bibinfo {author} {\bibfnamefont {J.}~\bibnamefont
  {Hu}},\ and\ \bibinfo {author} {\bibfnamefont {R.~Q.}\ \bibnamefont {Wu}},\
  }\bibfield  {title} {\bibinfo {title} {Separating {RKKY} interaction from
  other exchange mechanisms in two-dimensional magnetic materials},\ }\href
  {https://link.aps.org/doi/10.1103/PhysRevB.108.L041401} {\bibfield  {journal}
  {\bibinfo  {journal} {Phys. Rev. B}\ }\textbf {\bibinfo {volume} {108}},\
  \bibinfo {pages} {L041401} (\bibinfo {year} {2023})}\BibitemShut {NoStop}%
\bibitem [{\citenamefont {Wang}\ \emph {et~al.}(2024)\citenamefont {Wang},
  \citenamefont {Xu}, \citenamefont {Wang}, \citenamefont {Dai}, \citenamefont
  {Yan}, \citenamefont {Zhou}, \citenamefont {Wang}, \citenamefont {Xu},\ and\
  \citenamefont {He}}]{aelm.202300646}%
  \BibitemOpen
  \bibfield  {author} {\bibinfo {author} {\bibfnamefont {J.}~\bibnamefont
  {Wang}}, \bibinfo {author} {\bibfnamefont {Y.}~\bibnamefont {Xu}}, \bibinfo
  {author} {\bibfnamefont {S.}~\bibnamefont {Wang}}, \bibinfo {author}
  {\bibfnamefont {X.}~\bibnamefont {Dai}}, \bibinfo {author} {\bibfnamefont
  {P.}~\bibnamefont {Yan}}, \bibinfo {author} {\bibfnamefont {J.}~\bibnamefont
  {Zhou}}, \bibinfo {author} {\bibfnamefont {R.}~\bibnamefont {Wang}}, \bibinfo
  {author} {\bibfnamefont {Y.}~\bibnamefont {Xu}},\ and\ \bibinfo {author}
  {\bibfnamefont {L.}~\bibnamefont {He}},\ }\bibfield  {title} {\bibinfo
  {title} {Hole-mediated {RKKY} interaction in {2D} ferromagnetic \ce{CrTe2}
  ultra-thin films},\ }\href
  {https://advanced.onlinelibrary.wiley.com/doi/abs/10.1002/aelm.202300646}
  {\bibfield  {journal} {\bibinfo  {journal} {Adv. Electron. Mater.}\ }\textbf
  {\bibinfo {volume} {10}},\ \bibinfo {pages} {2300646} (\bibinfo {year}
  {2024})}\BibitemShut {NoStop}%
\bibitem [{\citenamefont {Khomskii}(2014)}]{khomskii2014transition}%
  \BibitemOpen
  \bibfield  {author} {\bibinfo {author} {\bibfnamefont {D.~I.}\ \bibnamefont
  {Khomskii}},\ }\href@noop {} {\emph {\bibinfo {title} {Transition metal
  compounds}}}\ (\bibinfo  {publisher} {Cambridge University Press},\ \bibinfo
  {year} {2014})\BibitemShut {NoStop}%
\bibitem [{\citenamefont {Landron}\ and\ \citenamefont
  {Lepetit}(2008)}]{PhysRevB.77.125106}%
  \BibitemOpen
  \bibfield  {author} {\bibinfo {author} {\bibfnamefont {S.}~\bibnamefont
  {Landron}}\ and\ \bibinfo {author} {\bibfnamefont {M.-B.}\ \bibnamefont
  {Lepetit}},\ }\bibfield  {title} {\bibinfo {title} {Importance of
  ${t}_{2g}\text{\ensuremath{-}}{e}_{g}$ hybridization in transition metal
  oxides},\ }\href {https://doi.org/10.1103/PhysRevB.77.125106} {\bibfield
  {journal} {\bibinfo  {journal} {Phys. Rev. B}\ }\textbf {\bibinfo {volume}
  {77}},\ \bibinfo {pages} {125106} (\bibinfo {year} {2008})}\BibitemShut
  {NoStop}%
\bibitem [{\citenamefont {Fischer}\ and\ \citenamefont
  {Klein}(1975)}]{PhysRevB.11.2025}%
  \BibitemOpen
  \bibfield  {author} {\bibinfo {author} {\bibfnamefont {B.}~\bibnamefont
  {Fischer}}\ and\ \bibinfo {author} {\bibfnamefont {M.~W.}\ \bibnamefont
  {Klein}},\ }\bibfield  {title} {\bibinfo {title} {Magnetic and nonmagnetic
  impurities in two-dimensional metals},\ }\href
  {https://link.aps.org/doi/10.1103/PhysRevB.11.2025} {\bibfield  {journal}
  {\bibinfo  {journal} {Phys. Rev. B}\ }\textbf {\bibinfo {volume} {11}},\
  \bibinfo {pages} {2025} (\bibinfo {year} {1975})}\BibitemShut {NoStop}%
\bibitem [{\citenamefont {B\'eal-Monod}(1987)}]{PhysRevB.36.8835}%
  \BibitemOpen
  \bibfield  {author} {\bibinfo {author} {\bibfnamefont {M.~T.}\ \bibnamefont
  {B\'eal-Monod}},\ }\bibfield  {title} {\bibinfo {title}
  {{R}uderman-{K}ittel-{K}asuya-{Y}osida indirect interaction in two
  dimensions},\ }\href {https://link.aps.org/doi/10.1103/PhysRevB.36.8835}
  {\bibfield  {journal} {\bibinfo  {journal} {Phys. Rev. B}\ }\textbf {\bibinfo
  {volume} {36}},\ \bibinfo {pages} {8835} (\bibinfo {year}
  {1987})}\BibitemShut {NoStop}%
\bibitem [{\citenamefont {Litvinov}\ and\ \citenamefont
  {Dugaev}(1998)}]{PhysRevB.58.3584}%
  \BibitemOpen
  \bibfield  {author} {\bibinfo {author} {\bibfnamefont {V.~I.}\ \bibnamefont
  {Litvinov}}\ and\ \bibinfo {author} {\bibfnamefont {V.~K.}\ \bibnamefont
  {Dugaev}},\ }\bibfield  {title} {\bibinfo {title} {{RKKY} interaction in one-
  and two-dimensional electron gases},\ }\href
  {https://link.aps.org/doi/10.1103/PhysRevB.58.3584} {\bibfield  {journal}
  {\bibinfo  {journal} {Phys. Rev. B}\ }\textbf {\bibinfo {volume} {58}},\
  \bibinfo {pages} {3584} (\bibinfo {year} {1998})}\BibitemShut {NoStop}%
\bibitem [{\citenamefont {Zhang}\ and\ \citenamefont
  {Rice}(1988)}]{PhysRevB.37.3759}%
  \BibitemOpen
  \bibfield  {author} {\bibinfo {author} {\bibfnamefont {F.~C.}\ \bibnamefont
  {Zhang}}\ and\ \bibinfo {author} {\bibfnamefont {T.~M.}\ \bibnamefont
  {Rice}},\ }\bibfield  {title} {\bibinfo {title} {Effective {Hamiltonian} for
  the superconducting {Cu} oxides},\ }\href
  {https://link.aps.org/doi/10.1103/PhysRevB.37.3759} {\bibfield  {journal}
  {\bibinfo  {journal} {Phys. Rev. B}\ }\textbf {\bibinfo {volume} {37}},\
  \bibinfo {pages} {3759} (\bibinfo {year} {1988})}\BibitemShut {NoStop}%
\bibitem [{\citenamefont {Wu}\ \emph {et~al.}(2023)\citenamefont {Wu},
  \citenamefont {Li}, \citenamefont {Wu}, \citenamefont {Hwang},\ and\
  \citenamefont {Cui}}]{s41578-022-00473-6}%
  \BibitemOpen
  \bibfield  {author} {\bibinfo {author} {\bibfnamefont {Y.}~\bibnamefont
  {Wu}}, \bibinfo {author} {\bibfnamefont {D.}~\bibnamefont {Li}}, \bibinfo
  {author} {\bibfnamefont {C.-L.}\ \bibnamefont {Wu}}, \bibinfo {author}
  {\bibfnamefont {H.~Y.}\ \bibnamefont {Hwang}},\ and\ \bibinfo {author}
  {\bibfnamefont {Y.}~\bibnamefont {Cui}},\ }\bibfield  {title} {\bibinfo
  {title} {Electrostatic gating and intercalation in {2D} materials},\ }\href
  {https://doi.org/10.1038/s41578-022-00473-6} {\bibfield  {journal} {\bibinfo
  {journal} {Nat. Rev. Mater.}\ }\textbf {\bibinfo {volume} {8}},\ \bibinfo
  {pages} {41} (\bibinfo {year} {2023})}\BibitemShut {NoStop}%
\bibitem [{\citenamefont {Bisri}\ \emph {et~al.}(2017)\citenamefont {Bisri},
  \citenamefont {Shimizu}, \citenamefont {Nakano},\ and\ \citenamefont
  {Iwasa}}]{adma.201607054}%
  \BibitemOpen
  \bibfield  {author} {\bibinfo {author} {\bibfnamefont {S.~Z.}\ \bibnamefont
  {Bisri}}, \bibinfo {author} {\bibfnamefont {S.}~\bibnamefont {Shimizu}},
  \bibinfo {author} {\bibfnamefont {M.}~\bibnamefont {Nakano}},\ and\ \bibinfo
  {author} {\bibfnamefont {Y.}~\bibnamefont {Iwasa}},\ }\bibfield  {title}
  {\bibinfo {title} {Endeavor of iontronics: From fundamentals to applications
  of ion-controlled electronics},\ }\href
  {https://advanced.onlinelibrary.wiley.com/doi/abs/10.1002/adma.201607054}
  {\bibfield  {journal} {\bibinfo  {journal} {Adv. Mater.}\ }\textbf {\bibinfo
  {volume} {29}},\ \bibinfo {pages} {1607054} (\bibinfo {year}
  {2017})}\BibitemShut {NoStop}%
\bibitem [{\citenamefont {Guan}\ \emph {et~al.}(2023)\citenamefont {Guan},
  \citenamefont {Han}, \citenamefont {Li}, \citenamefont {Li},\ and\
  \citenamefont {Parkin}}]{annurev-matsci-080619-012219}%
  \BibitemOpen
  \bibfield  {author} {\bibinfo {author} {\bibfnamefont {Y.}~\bibnamefont
  {Guan}}, \bibinfo {author} {\bibfnamefont {H.}~\bibnamefont {Han}}, \bibinfo
  {author} {\bibfnamefont {F.}~\bibnamefont {Li}}, \bibinfo {author}
  {\bibfnamefont {G.}~\bibnamefont {Li}},\ and\ \bibinfo {author}
  {\bibfnamefont {S.~S.}\ \bibnamefont {Parkin}},\ }\bibfield  {title}
  {\bibinfo {title} {Ionic gating for tuning electronic and magnetic
  properties},\ }\href
  {https://www.annualreviews.org/content/journals/10.1146/annurev-matsci-080619-012219}
  {\bibfield  {journal} {\bibinfo  {journal} {Annu. Rev. Mater. Res.}\ }\textbf
  {\bibinfo {volume} {53}},\ \bibinfo {pages} {25} (\bibinfo {year}
  {2023})}\BibitemShut {NoStop}%
\bibitem [{\citenamefont {Qi}\ \emph {et~al.}(2023)\citenamefont {Qi},
  \citenamefont {Sadi}, \citenamefont {Hu}, \citenamefont {Zheng},
  \citenamefont {Wu}, \citenamefont {Jiang},\ and\ \citenamefont
  {Chen}}]{adma.202205714}%
  \BibitemOpen
  \bibfield  {author} {\bibinfo {author} {\bibfnamefont {Y.}~\bibnamefont
  {Qi}}, \bibinfo {author} {\bibfnamefont {M.~A.}\ \bibnamefont {Sadi}},
  \bibinfo {author} {\bibfnamefont {D.}~\bibnamefont {Hu}}, \bibinfo {author}
  {\bibfnamefont {M.}~\bibnamefont {Zheng}}, \bibinfo {author} {\bibfnamefont
  {Z.}~\bibnamefont {Wu}}, \bibinfo {author} {\bibfnamefont {Y.}~\bibnamefont
  {Jiang}},\ and\ \bibinfo {author} {\bibfnamefont {Y.~P.}\ \bibnamefont
  {Chen}},\ }\bibfield  {title} {\bibinfo {title} {Recent progress in strain
  engineering on van der {Waals} {2D} materials: Tunable electrical,
  electrochemical, magnetic, and optical properties},\ }\href
  {https://doi.org/https://doi.org/10.1002/adma.202205714} {\bibfield
  {journal} {\bibinfo  {journal} {Adv. Mater.}\ }\textbf {\bibinfo {volume}
  {35}},\ \bibinfo {pages} {2205714} (\bibinfo {year} {2023})}\BibitemShut
  {NoStop}%
\bibitem [{\citenamefont {Bhoi}\ \emph {et~al.}(2021)\citenamefont {Bhoi},
  \citenamefont {Gouchi}, \citenamefont {Hiraoka}, \citenamefont {Zhang},
  \citenamefont {Ogita}, \citenamefont {Hasegawa}, \citenamefont {Kitagawa},
  \citenamefont {Takagi}, \citenamefont {Kim},\ and\ \citenamefont
  {Uwatoko}}]{PhysRevLett.127.217203}%
  \BibitemOpen
  \bibfield  {author} {\bibinfo {author} {\bibfnamefont {D.}~\bibnamefont
  {Bhoi}}, \bibinfo {author} {\bibfnamefont {J.}~\bibnamefont {Gouchi}},
  \bibinfo {author} {\bibfnamefont {N.}~\bibnamefont {Hiraoka}}, \bibinfo
  {author} {\bibfnamefont {Y.}~\bibnamefont {Zhang}}, \bibinfo {author}
  {\bibfnamefont {N.}~\bibnamefont {Ogita}}, \bibinfo {author} {\bibfnamefont
  {T.}~\bibnamefont {Hasegawa}}, \bibinfo {author} {\bibfnamefont
  {K.}~\bibnamefont {Kitagawa}}, \bibinfo {author} {\bibfnamefont
  {H.}~\bibnamefont {Takagi}}, \bibinfo {author} {\bibfnamefont {K.~H.}\
  \bibnamefont {Kim}},\ and\ \bibinfo {author} {\bibfnamefont {Y.}~\bibnamefont
  {Uwatoko}},\ }\bibfield  {title} {\bibinfo {title} {Nearly room-temperature
  ferromagnetism in a pressure-induced correlated metallic state of the van der
  {Waals} insulator \ce{CrGeTe3}},\ }\href
  {https://link.aps.org/doi/10.1103/PhysRevLett.127.217203} {\bibfield
  {journal} {\bibinfo  {journal} {Phys. Rev. Lett.}\ }\textbf {\bibinfo
  {volume} {127}},\ \bibinfo {pages} {217203} (\bibinfo {year}
  {2021})}\BibitemShut {NoStop}%
\bibitem [{\citenamefont {{\v S}i{\v s}kins}\ \emph {et~al.}(2022)\citenamefont
  {{\v S}i{\v s}kins}, \citenamefont {Kurdi}, \citenamefont {Lee},
  \citenamefont {Slotboom}, \citenamefont {Xing}, \citenamefont
  {Ma{\~n}as-Valero}, \citenamefont {Coronado}, \citenamefont {Jia},
  \citenamefont {Han}, \citenamefont {van~der Sar}, \citenamefont {van~der
  Zant},\ and\ \citenamefont {Steeneken}}]{s41699-022-00315-7}%
  \BibitemOpen
  \bibfield  {author} {\bibinfo {author} {\bibfnamefont {M.}~\bibnamefont {{\v
  S}i{\v s}kins}}, \bibinfo {author} {\bibfnamefont {S.}~\bibnamefont {Kurdi}},
  \bibinfo {author} {\bibfnamefont {M.}~\bibnamefont {Lee}}, \bibinfo {author}
  {\bibfnamefont {B.~J.~M.}\ \bibnamefont {Slotboom}}, \bibinfo {author}
  {\bibfnamefont {W.}~\bibnamefont {Xing}}, \bibinfo {author} {\bibfnamefont
  {S.}~\bibnamefont {Ma{\~n}as-Valero}}, \bibinfo {author} {\bibfnamefont
  {E.}~\bibnamefont {Coronado}}, \bibinfo {author} {\bibfnamefont
  {S.}~\bibnamefont {Jia}}, \bibinfo {author} {\bibfnamefont {W.}~\bibnamefont
  {Han}}, \bibinfo {author} {\bibfnamefont {T.}~\bibnamefont {van~der Sar}},
  \bibinfo {author} {\bibfnamefont {H.~S.~J.}\ \bibnamefont {van~der Zant}},\
  and\ \bibinfo {author} {\bibfnamefont {P.~G.}\ \bibnamefont {Steeneken}},\
  }\bibfield  {title} {\bibinfo {title} {Nanomechanical probing and strain
  tuning of the {Curie} temperature in suspended \ce{Cr2Ge2Te6}-based
  heterostructures},\ }\href {https://doi.org/10.1038/s41699-022-00315-7}
  {\bibfield  {journal} {\bibinfo  {journal} {npj 2D Mater. Appl.}\ }\textbf
  {\bibinfo {volume} {6}},\ \bibinfo {pages} {41} (\bibinfo {year}
  {2022})}\BibitemShut {NoStop}%
\bibitem [{\citenamefont {Ni}\ \emph {et~al.}(2021{\natexlab{b}})\citenamefont
  {Ni}, \citenamefont {Haglund}, \citenamefont {Wang}, \citenamefont {Xu},
  \citenamefont {Bernhard}, \citenamefont {Mandrus}, \citenamefont {Qian},
  \citenamefont {Mele}, \citenamefont {Kane},\ and\ \citenamefont
  {Wu}}]{s41565-021-00885-5}%
  \BibitemOpen
  \bibfield  {author} {\bibinfo {author} {\bibfnamefont {Z.}~\bibnamefont
  {Ni}}, \bibinfo {author} {\bibfnamefont {A.~V.}\ \bibnamefont {Haglund}},
  \bibinfo {author} {\bibfnamefont {H.}~\bibnamefont {Wang}}, \bibinfo {author}
  {\bibfnamefont {B.}~\bibnamefont {Xu}}, \bibinfo {author} {\bibfnamefont
  {C.}~\bibnamefont {Bernhard}}, \bibinfo {author} {\bibfnamefont {D.~G.}\
  \bibnamefont {Mandrus}}, \bibinfo {author} {\bibfnamefont {X.}~\bibnamefont
  {Qian}}, \bibinfo {author} {\bibfnamefont {E.~J.}\ \bibnamefont {Mele}},
  \bibinfo {author} {\bibfnamefont {C.~L.}\ \bibnamefont {Kane}},\ and\
  \bibinfo {author} {\bibfnamefont {L.}~\bibnamefont {Wu}},\ }\bibfield
  {title} {\bibinfo {title} {Imaging the {N{\'e}el} vector switching in the
  monolayer antiferromagnet \ce{MnPSe3} with strain-controlled {Ising} order},\
  }\href {https://doi.org/10.1038/s41565-021-00885-5} {\bibfield  {journal}
  {\bibinfo  {journal} {Nat. Nanotechnol.}\ }\textbf {\bibinfo {volume} {16}},\
  \bibinfo {pages} {782} (\bibinfo {year} {2021}{\natexlab{b}})}\BibitemShut
  {NoStop}%
\bibitem [{\citenamefont {Pal}\ \emph {et~al.}(2024)\citenamefont {Pal},
  \citenamefont {Pal}, \citenamefont {Mondal}, \citenamefont {Sharma},
  \citenamefont {Das}, \citenamefont {Mandal},\ and\ \citenamefont
  {Pal}}]{s41699-024-00463-y}%
  \BibitemOpen
  \bibfield  {author} {\bibinfo {author} {\bibfnamefont {R.}~\bibnamefont
  {Pal}}, \bibinfo {author} {\bibfnamefont {B.}~\bibnamefont {Pal}}, \bibinfo
  {author} {\bibfnamefont {S.}~\bibnamefont {Mondal}}, \bibinfo {author}
  {\bibfnamefont {R.~O.}\ \bibnamefont {Sharma}}, \bibinfo {author}
  {\bibfnamefont {T.}~\bibnamefont {Das}}, \bibinfo {author} {\bibfnamefont
  {P.}~\bibnamefont {Mandal}},\ and\ \bibinfo {author} {\bibfnamefont {A.~N.}\
  \bibnamefont {Pal}},\ }\bibfield  {title} {\bibinfo {title}
  {Spin-reorientation driven emergent phases and unconventional
  magnetotransport in quasi-{2D} {vdW} ferromagnet \ce{Fe4GeTe2}},\ }\href
  {https://doi.org/10.1038/s41699-024-00463-y} {\bibfield  {journal} {\bibinfo
  {journal} {npj 2D Mater. Appl.}\ }\textbf {\bibinfo {volume} {8}},\ \bibinfo
  {pages} {30} (\bibinfo {year} {2024})}\BibitemShut {NoStop}%
\bibitem [{\citenamefont {Kong}\ \emph {et~al.}(2025)\citenamefont {Kong},
  \citenamefont {Kov{\'a}cs}, \citenamefont {Charilaou}, \citenamefont
  {Altthaler}, \citenamefont {Prodan}, \citenamefont {Tsurkan}, \citenamefont
  {Meier}, \citenamefont {Han}, \citenamefont {K{\'e}zsm{\'a}rki},\ and\
  \citenamefont {Dunin-Borkowski}}]{acsnano.4c16603}%
  \BibitemOpen
  \bibfield  {author} {\bibinfo {author} {\bibfnamefont {D.}~\bibnamefont
  {Kong}}, \bibinfo {author} {\bibfnamefont {A.}~\bibnamefont {Kov{\'a}cs}},
  \bibinfo {author} {\bibfnamefont {M.}~\bibnamefont {Charilaou}}, \bibinfo
  {author} {\bibfnamefont {M.}~\bibnamefont {Altthaler}}, \bibinfo {author}
  {\bibfnamefont {L.}~\bibnamefont {Prodan}}, \bibinfo {author} {\bibfnamefont
  {V.}~\bibnamefont {Tsurkan}}, \bibinfo {author} {\bibfnamefont
  {D.}~\bibnamefont {Meier}}, \bibinfo {author} {\bibfnamefont
  {X.}~\bibnamefont {Han}}, \bibinfo {author} {\bibfnamefont {I.}~\bibnamefont
  {K{\'e}zsm{\'a}rki}},\ and\ \bibinfo {author} {\bibfnamefont {R.~E.}\
  \bibnamefont {Dunin-Borkowski}},\ }\bibfield  {title} {\bibinfo {title}
  {Strain engineering of magnetic anisotropy in the kagome magnet
  \ce{Fe3Sn2}},\ }\href {https://doi.org/10.1021/acsnano.4c16603} {\bibfield
  {journal} {\bibinfo  {journal} {ACS Nano}\ }\textbf {\bibinfo {volume}
  {19}},\ \bibinfo {pages} {8142} (\bibinfo {year} {2025})}\BibitemShut
  {NoStop}%
\bibitem [{\citenamefont {Cenker}\ \emph {et~al.}(2022)\citenamefont {Cenker},
  \citenamefont {Sivakumar}, \citenamefont {Xie}, \citenamefont {Miller},
  \citenamefont {Thijssen}, \citenamefont {Liu}, \citenamefont {Dismukes},
  \citenamefont {Fonseca}, \citenamefont {Anderson}, \citenamefont {Zhu},
  \citenamefont {Roy}, \citenamefont {Xiao}, \citenamefont {Chu}, \citenamefont
  {Cao},\ and\ \citenamefont {Xu}}]{s41565-021-01052-6}%
  \BibitemOpen
  \bibfield  {author} {\bibinfo {author} {\bibfnamefont {J.}~\bibnamefont
  {Cenker}}, \bibinfo {author} {\bibfnamefont {S.}~\bibnamefont {Sivakumar}},
  \bibinfo {author} {\bibfnamefont {K.}~\bibnamefont {Xie}}, \bibinfo {author}
  {\bibfnamefont {A.}~\bibnamefont {Miller}}, \bibinfo {author} {\bibfnamefont
  {P.}~\bibnamefont {Thijssen}}, \bibinfo {author} {\bibfnamefont
  {Z.}~\bibnamefont {Liu}}, \bibinfo {author} {\bibfnamefont {A.}~\bibnamefont
  {Dismukes}}, \bibinfo {author} {\bibfnamefont {J.}~\bibnamefont {Fonseca}},
  \bibinfo {author} {\bibfnamefont {E.}~\bibnamefont {Anderson}}, \bibinfo
  {author} {\bibfnamefont {X.}~\bibnamefont {Zhu}}, \bibinfo {author}
  {\bibfnamefont {X.}~\bibnamefont {Roy}}, \bibinfo {author} {\bibfnamefont
  {D.}~\bibnamefont {Xiao}}, \bibinfo {author} {\bibfnamefont {J.-H.}\
  \bibnamefont {Chu}}, \bibinfo {author} {\bibfnamefont {T.}~\bibnamefont
  {Cao}},\ and\ \bibinfo {author} {\bibfnamefont {X.}~\bibnamefont {Xu}},\
  }\bibfield  {title} {\bibinfo {title} {Reversible strain-induced magnetic
  phase transition in a van der {Waals} magnet},\ }\href
  {https://doi.org/10.1038/s41565-021-01052-6} {\bibfield  {journal} {\bibinfo
  {journal} {Nat. Nanotechnol.}\ }\textbf {\bibinfo {volume} {17}},\ \bibinfo
  {pages} {256} (\bibinfo {year} {2022})}\BibitemShut {NoStop}%
\bibitem [{\citenamefont {Tian}\ \emph {et~al.}(2016)\citenamefont {Tian},
  \citenamefont {Gray}, \citenamefont {Ji}, \citenamefont {Cava},\ and\
  \citenamefont {Burch}}]{tian2016magneto}%
  \BibitemOpen
  \bibfield  {author} {\bibinfo {author} {\bibfnamefont {Y.}~\bibnamefont
  {Tian}}, \bibinfo {author} {\bibfnamefont {M.~J.}\ \bibnamefont {Gray}},
  \bibinfo {author} {\bibfnamefont {H.}~\bibnamefont {Ji}}, \bibinfo {author}
  {\bibfnamefont {R.~J.}\ \bibnamefont {Cava}},\ and\ \bibinfo {author}
  {\bibfnamefont {K.~S.}\ \bibnamefont {Burch}},\ }\bibfield  {title} {\bibinfo
  {title} {Magneto-elastic coupling in a potential ferromagnetic \rm{2D} atomic
  crystal},\ }\href
  {https://iopscience.iop.org/article/10.1088/2053-1583/3/2/025035} {\bibfield
  {journal} {\bibinfo  {journal} {2D Mater.}\ }\textbf {\bibinfo {volume}
  {3}},\ \bibinfo {pages} {025035} (\bibinfo {year} {2016})}\BibitemShut
  {NoStop}%
\bibitem [{\citenamefont {Lockwood}\ and\ \citenamefont
  {Cottam}(1988)}]{1.342186}%
  \BibitemOpen
  \bibfield  {author} {\bibinfo {author} {\bibfnamefont {D.}~\bibnamefont
  {Lockwood}}\ and\ \bibinfo {author} {\bibfnamefont {M.}~\bibnamefont
  {Cottam}},\ }\bibfield  {title} {\bibinfo {title} {The spin-phonon
  interaction in \ce{FeF2} and \ce{MnF2} studied by {Raman} spectroscopy},\
  }\href
  {https://pubs.aip.org/aip/jap/article/64/10/5876/16837/The-spin-phonon-interaction-in-FeF2-and-MnF2}
  {\bibfield  {journal} {\bibinfo  {journal} {J. Appl. Phys.}\ }\textbf
  {\bibinfo {volume} {64}},\ \bibinfo {pages} {5876} (\bibinfo {year}
  {1988})}\BibitemShut {NoStop}%
\bibitem [{\citenamefont {Huang}\ \emph {et~al.}(2020)\citenamefont {Huang},
  \citenamefont {Cenker}, \citenamefont {Zhang}, \citenamefont {Ray},
  \citenamefont {Song}, \citenamefont {Taniguchi}, \citenamefont {Watanabe},
  \citenamefont {McGuire}, \citenamefont {Xiao},\ and\ \citenamefont
  {Xu}}]{s41565-019-0598-4}%
  \BibitemOpen
  \bibfield  {author} {\bibinfo {author} {\bibfnamefont {B.}~\bibnamefont
  {Huang}}, \bibinfo {author} {\bibfnamefont {J.}~\bibnamefont {Cenker}},
  \bibinfo {author} {\bibfnamefont {X.}~\bibnamefont {Zhang}}, \bibinfo
  {author} {\bibfnamefont {E.~L.}\ \bibnamefont {Ray}}, \bibinfo {author}
  {\bibfnamefont {T.}~\bibnamefont {Song}}, \bibinfo {author} {\bibfnamefont
  {T.}~\bibnamefont {Taniguchi}}, \bibinfo {author} {\bibfnamefont
  {K.}~\bibnamefont {Watanabe}}, \bibinfo {author} {\bibfnamefont {M.~A.}\
  \bibnamefont {McGuire}}, \bibinfo {author} {\bibfnamefont {D.}~\bibnamefont
  {Xiao}},\ and\ \bibinfo {author} {\bibfnamefont {X.}~\bibnamefont {Xu}},\
  }\bibfield  {title} {\bibinfo {title} {Tuning inelastic light scattering via
  symmetry control in the two-dimensional magnet \ce{CrI_3}},\ }\href
  {https://www.nature.com/articles/s41565-019-0598-4} {\bibfield  {journal}
  {\bibinfo  {journal} {Nat. Nanotechnol.}\ }\textbf {\bibinfo {volume} {15}},\
  \bibinfo {pages} {212} (\bibinfo {year} {2020})}\BibitemShut {NoStop}%
\bibitem [{\citenamefont {Jin}\ \emph {et~al.}(2018)\citenamefont {Jin},
  \citenamefont {Kim}, \citenamefont {Ye}, \citenamefont {Li}, \citenamefont
  {Rezaie}, \citenamefont {Diaz}, \citenamefont {Siddiq}, \citenamefont
  {Wauer}, \citenamefont {Yang}, \citenamefont {Li}, \citenamefont {Tian},
  \citenamefont {Sun}, \citenamefont {Lei}, \citenamefont {Tsen}, \citenamefont
  {Zhao},\ and\ \citenamefont {He}}]{s41467-018-07547-6}%
  \BibitemOpen
  \bibfield  {author} {\bibinfo {author} {\bibfnamefont {W.}~\bibnamefont
  {Jin}}, \bibinfo {author} {\bibfnamefont {H.~H.}\ \bibnamefont {Kim}},
  \bibinfo {author} {\bibfnamefont {Z.}~\bibnamefont {Ye}}, \bibinfo {author}
  {\bibfnamefont {S.}~\bibnamefont {Li}}, \bibinfo {author} {\bibfnamefont
  {P.}~\bibnamefont {Rezaie}}, \bibinfo {author} {\bibfnamefont
  {F.}~\bibnamefont {Diaz}}, \bibinfo {author} {\bibfnamefont {S.}~\bibnamefont
  {Siddiq}}, \bibinfo {author} {\bibfnamefont {E.}~\bibnamefont {Wauer}},
  \bibinfo {author} {\bibfnamefont {B.}~\bibnamefont {Yang}}, \bibinfo {author}
  {\bibfnamefont {C.}~\bibnamefont {Li}}, \bibinfo {author} {\bibfnamefont
  {S.}~\bibnamefont {Tian}}, \bibinfo {author} {\bibfnamefont {K.}~\bibnamefont
  {Sun}}, \bibinfo {author} {\bibfnamefont {H.}~\bibnamefont {Lei}}, \bibinfo
  {author} {\bibfnamefont {A.~W.}\ \bibnamefont {Tsen}}, \bibinfo {author}
  {\bibfnamefont {L.}~\bibnamefont {Zhao}},\ and\ \bibinfo {author}
  {\bibfnamefont {R.}~\bibnamefont {He}},\ }\bibfield  {title} {\bibinfo
  {title} {Raman fingerprint of two terahertz spin wave branches in a
  two-dimensional honeycomb {Ising} ferromagnet},\ }\href
  {https://doi.org/10.1038/s41467-018-07547-6} {\bibfield  {journal} {\bibinfo
  {journal} {Nat. Commun.}\ }\textbf {\bibinfo {volume} {9}},\ \bibinfo {pages}
  {5122} (\bibinfo {year} {2018})}\BibitemShut {NoStop}%
\bibitem [{\citenamefont {Kim}\ \emph {et~al.}(2019{\natexlab{b}})\citenamefont
  {Kim}, \citenamefont {Lim}, \citenamefont {Lee}, \citenamefont {Lee},
  \citenamefont {Kim}, \citenamefont {Park}, \citenamefont {Jeon},
  \citenamefont {Park}, \citenamefont {Park},\ and\ \citenamefont
  {Cheong}}]{s41467-018-08284-6}%
  \BibitemOpen
  \bibfield  {author} {\bibinfo {author} {\bibfnamefont {K.}~\bibnamefont
  {Kim}}, \bibinfo {author} {\bibfnamefont {S.~Y.}\ \bibnamefont {Lim}},
  \bibinfo {author} {\bibfnamefont {J.-U.}\ \bibnamefont {Lee}}, \bibinfo
  {author} {\bibfnamefont {S.}~\bibnamefont {Lee}}, \bibinfo {author}
  {\bibfnamefont {T.~Y.}\ \bibnamefont {Kim}}, \bibinfo {author} {\bibfnamefont
  {K.}~\bibnamefont {Park}}, \bibinfo {author} {\bibfnamefont {G.~S.}\
  \bibnamefont {Jeon}}, \bibinfo {author} {\bibfnamefont {C.-H.}\ \bibnamefont
  {Park}}, \bibinfo {author} {\bibfnamefont {J.-G.}\ \bibnamefont {Park}},\
  and\ \bibinfo {author} {\bibfnamefont {H.}~\bibnamefont {Cheong}},\
  }\bibfield  {title} {\bibinfo {title} {Suppression of magnetic ordering in
  {XXZ}-type antiferromagnetic monolayer \ce{NiPS_3}},\ }\href
  {https://doi.org/10.1038/s41467-018-08284-6} {\bibfield  {journal} {\bibinfo
  {journal} {Nat. Commun.}\ }\textbf {\bibinfo {volume} {10}},\ \bibinfo
  {pages} {345} (\bibinfo {year} {2019}{\natexlab{b}})}\BibitemShut {NoStop}%
\bibitem [{\citenamefont {Liu}\ \emph {et~al.}(2021)\citenamefont {Liu},
  \citenamefont {Wang}, \citenamefont {Fu}, \citenamefont {Zhang},
  \citenamefont {Huang}, \citenamefont {Su}, \citenamefont {Lin}, \citenamefont
  {Chen}, \citenamefont {Yu}, \citenamefont {Cui}, \citenamefont {Mei},\ and\
  \citenamefont {Dai}}]{PhysRevB.103.235411}%
  \BibitemOpen
  \bibfield  {author} {\bibinfo {author} {\bibfnamefont {Q.}~\bibnamefont
  {Liu}}, \bibinfo {author} {\bibfnamefont {L.}~\bibnamefont {Wang}}, \bibinfo
  {author} {\bibfnamefont {Y.}~\bibnamefont {Fu}}, \bibinfo {author}
  {\bibfnamefont {X.}~\bibnamefont {Zhang}}, \bibinfo {author} {\bibfnamefont
  {L.}~\bibnamefont {Huang}}, \bibinfo {author} {\bibfnamefont
  {H.}~\bibnamefont {Su}}, \bibinfo {author} {\bibfnamefont {J.}~\bibnamefont
  {Lin}}, \bibinfo {author} {\bibfnamefont {X.}~\bibnamefont {Chen}}, \bibinfo
  {author} {\bibfnamefont {D.}~\bibnamefont {Yu}}, \bibinfo {author}
  {\bibfnamefont {X.}~\bibnamefont {Cui}}, \bibinfo {author} {\bibfnamefont
  {J.-W.}\ \bibnamefont {Mei}},\ and\ \bibinfo {author} {\bibfnamefont {J.-F.}\
  \bibnamefont {Dai}},\ }\bibfield  {title} {\bibinfo {title} {Magnetic order
  in {XY}-type antiferromagnetic monolayer \ce{CoPS3} revealed by {Raman}
  spectroscopy},\ }\href {https://link.aps.org/doi/10.1103/PhysRevB.103.235411}
  {\bibfield  {journal} {\bibinfo  {journal} {Phys. Rev. B}\ }\textbf {\bibinfo
  {volume} {103}},\ \bibinfo {pages} {235411} (\bibinfo {year}
  {2021})}\BibitemShut {NoStop}%
\bibitem [{\citenamefont {Pawbake}\ \emph {et~al.}(2023)\citenamefont
  {Pawbake}, \citenamefont {Pelini}, \citenamefont {Wilson}, \citenamefont
  {Mosina}, \citenamefont {Sofer}, \citenamefont {Heid},\ and\ \citenamefont
  {Faugeras}}]{PhysRevB.107.075421}%
  \BibitemOpen
  \bibfield  {author} {\bibinfo {author} {\bibfnamefont {A.}~\bibnamefont
  {Pawbake}}, \bibinfo {author} {\bibfnamefont {T.}~\bibnamefont {Pelini}},
  \bibinfo {author} {\bibfnamefont {N.~P.}\ \bibnamefont {Wilson}}, \bibinfo
  {author} {\bibfnamefont {K.}~\bibnamefont {Mosina}}, \bibinfo {author}
  {\bibfnamefont {Z.}~\bibnamefont {Sofer}}, \bibinfo {author} {\bibfnamefont
  {R.}~\bibnamefont {Heid}},\ and\ \bibinfo {author} {\bibfnamefont
  {C.}~\bibnamefont {Faugeras}},\ }\bibfield  {title} {\bibinfo {title} {Raman
  scattering signatures of strong spin-phonon coupling in the bulk magnetic van
  der {Waals} material \ce{CrSBr}},\ }\href
  {https://journals.aps.org/prb/abstract/10.1103/PhysRevB.107.075421}
  {\bibfield  {journal} {\bibinfo  {journal} {Phys. Rev. B}\ }\textbf {\bibinfo
  {volume} {107}},\ \bibinfo {pages} {075421} (\bibinfo {year}
  {2023})}\BibitemShut {NoStop}%
\bibitem [{\citenamefont {Baltensperger}\ and\ \citenamefont
  {Helman}(1968)}]{seals-113910}%
  \BibitemOpen
  \bibfield  {author} {\bibinfo {author} {\bibfnamefont {W.}~\bibnamefont
  {Baltensperger}}\ and\ \bibinfo {author} {\bibfnamefont {J.}~\bibnamefont
  {Helman}},\ }\bibfield  {title} {\bibinfo {title} {Influence of magnetic
  order in insulators on the optical phonon frequency},\ }\href
  {https://doi.org/10.5169/seals-113910} {\bibfield  {journal} {\bibinfo
  {journal} {Helv. Phys. Acta}\ }\textbf {\bibinfo {volume} {41}},\ \bibinfo
  {pages} {668} (\bibinfo {year} {1968})}\BibitemShut {NoStop}%
\bibitem [{\citenamefont {Massidda}\ \emph {et~al.}(1999)\citenamefont
  {Massidda}, \citenamefont {Posternak}, \citenamefont {Baldereschi},\ and\
  \citenamefont {Resta}}]{PhysRevLett.82.430}%
  \BibitemOpen
  \bibfield  {author} {\bibinfo {author} {\bibfnamefont {S.}~\bibnamefont
  {Massidda}}, \bibinfo {author} {\bibfnamefont {M.}~\bibnamefont {Posternak}},
  \bibinfo {author} {\bibfnamefont {A.}~\bibnamefont {Baldereschi}},\ and\
  \bibinfo {author} {\bibfnamefont {R.}~\bibnamefont {Resta}},\ }\bibfield
  {title} {\bibinfo {title} {Noncubic behavior of antiferromagnetic
  transition-metal monoxides with the rocksalt structure},\ }\href
  {https://link.aps.org/doi/10.1103/PhysRevLett.82.430} {\bibfield  {journal}
  {\bibinfo  {journal} {Phys. Rev. Lett.}\ }\textbf {\bibinfo {volume} {82}},\
  \bibinfo {pages} {430} (\bibinfo {year} {1999})}\BibitemShut {NoStop}%
\bibitem [{\citenamefont {Fennie}\ and\ \citenamefont
  {Rabe}(2006)}]{PhysRevLett.96.205505}%
  \BibitemOpen
  \bibfield  {author} {\bibinfo {author} {\bibfnamefont {C.~J.}\ \bibnamefont
  {Fennie}}\ and\ \bibinfo {author} {\bibfnamefont {K.~M.}\ \bibnamefont
  {Rabe}},\ }\bibfield  {title} {\bibinfo {title} {Magnetically induced phonon
  anisotropy in \ce{ZnCr2O4} from first principles},\ }\href
  {https://link.aps.org/doi/10.1103/PhysRevLett.96.205505} {\bibfield
  {journal} {\bibinfo  {journal} {Phys. Rev. Lett.}\ }\textbf {\bibinfo
  {volume} {96}},\ \bibinfo {pages} {205505} (\bibinfo {year}
  {2006})}\BibitemShut {NoStop}%
\bibitem [{\citenamefont {Purbawati}\ \emph {et~al.}(2023)\citenamefont
  {Purbawati}, \citenamefont {Sarkar}, \citenamefont {Pairis}, \citenamefont
  {Kostka}, \citenamefont {Hadj-Azzem}, \citenamefont {Dufeu}, \citenamefont
  {Singh}, \citenamefont {Bourgault}, \citenamefont {Nu{\~n}ez-Regueiro},
  \citenamefont {Vogel}, \citenamefont {Renard}, \citenamefont {Marty},
  \citenamefont {Fabre}, \citenamefont {Finco}, \citenamefont {Jacques},
  \citenamefont {Ren}, \citenamefont {Tiwari}, \citenamefont {Robert},
  \citenamefont {Marie}, \citenamefont {Bendiab}, \citenamefont {Rougemaille},\
  and\ \citenamefont {Coraux}}]{acsaelm.2c01256}%
  \BibitemOpen
  \bibfield  {author} {\bibinfo {author} {\bibfnamefont {A.}~\bibnamefont
  {Purbawati}}, \bibinfo {author} {\bibfnamefont {S.}~\bibnamefont {Sarkar}},
  \bibinfo {author} {\bibfnamefont {S.}~\bibnamefont {Pairis}}, \bibinfo
  {author} {\bibfnamefont {M.}~\bibnamefont {Kostka}}, \bibinfo {author}
  {\bibfnamefont {A.}~\bibnamefont {Hadj-Azzem}}, \bibinfo {author}
  {\bibfnamefont {D.}~\bibnamefont {Dufeu}}, \bibinfo {author} {\bibfnamefont
  {P.}~\bibnamefont {Singh}}, \bibinfo {author} {\bibfnamefont
  {D.}~\bibnamefont {Bourgault}}, \bibinfo {author} {\bibfnamefont
  {M.}~\bibnamefont {Nu{\~n}ez-Regueiro}}, \bibinfo {author} {\bibfnamefont
  {J.}~\bibnamefont {Vogel}}, \bibinfo {author} {\bibfnamefont
  {J.}~\bibnamefont {Renard}}, \bibinfo {author} {\bibfnamefont
  {L.}~\bibnamefont {Marty}}, \bibinfo {author} {\bibfnamefont
  {F.}~\bibnamefont {Fabre}}, \bibinfo {author} {\bibfnamefont
  {A.}~\bibnamefont {Finco}}, \bibinfo {author} {\bibfnamefont
  {V.}~\bibnamefont {Jacques}}, \bibinfo {author} {\bibfnamefont
  {L.}~\bibnamefont {Ren}}, \bibinfo {author} {\bibfnamefont {V.}~\bibnamefont
  {Tiwari}}, \bibinfo {author} {\bibfnamefont {C.}~\bibnamefont {Robert}},
  \bibinfo {author} {\bibfnamefont {X.}~\bibnamefont {Marie}}, \bibinfo
  {author} {\bibfnamefont {N.}~\bibnamefont {Bendiab}}, \bibinfo {author}
  {\bibfnamefont {N.}~\bibnamefont {Rougemaille}},\ and\ \bibinfo {author}
  {\bibfnamefont {J.}~\bibnamefont {Coraux}},\ }\bibfield  {title} {\bibinfo
  {title} {Stability of the in-plane room temperature van der {Waals}
  ferromagnet chromium ditelluride and its conversion to chromium-interleaved
  \ce{CrTe2} compounds},\ }\href {https://doi.org/10.1021/acsaelm.2c01256}
  {\bibfield  {journal} {\bibinfo  {journal} {ACS Appl. Electron. Mater.}\
  }\textbf {\bibinfo {volume} {5}},\ \bibinfo {pages} {764} (\bibinfo {year}
  {2023})}\BibitemShut {NoStop}%
\bibitem [{\citenamefont {Bigi}\ \emph {et~al.}(2025)\citenamefont {Bigi},
  \citenamefont {Jego}, \citenamefont {Polewczyk}, \citenamefont {De~Vita},
  \citenamefont {Jaouen}, \citenamefont {Tchouekem}, \citenamefont {Bertran},
  \citenamefont {Le~F{\`e}vre}, \citenamefont {Turban}, \citenamefont
  {Jacquot}, \citenamefont {Miwa}, \citenamefont {Clark}, \citenamefont {Jana},
  \citenamefont {Chaluvadi}, \citenamefont {Orgiani}, \citenamefont {Cuoco},
  \citenamefont {Leandersson}, \citenamefont {Balasubramanian}, \citenamefont
  {Olsen}, \citenamefont {Hwang}, \citenamefont {Jamet},\ and\ \citenamefont
  {Mazzola}}]{s41467-025-59266-4}%
  \BibitemOpen
  \bibfield  {author} {\bibinfo {author} {\bibfnamefont {C.}~\bibnamefont
  {Bigi}}, \bibinfo {author} {\bibfnamefont {C.}~\bibnamefont {Jego}}, \bibinfo
  {author} {\bibfnamefont {V.}~\bibnamefont {Polewczyk}}, \bibinfo {author}
  {\bibfnamefont {A.}~\bibnamefont {De~Vita}}, \bibinfo {author} {\bibfnamefont
  {T.}~\bibnamefont {Jaouen}}, \bibinfo {author} {\bibfnamefont {H.~C.}\
  \bibnamefont {Tchouekem}}, \bibinfo {author} {\bibfnamefont {F.}~\bibnamefont
  {Bertran}}, \bibinfo {author} {\bibfnamefont {P.}~\bibnamefont
  {Le~F{\`e}vre}}, \bibinfo {author} {\bibfnamefont {P.}~\bibnamefont
  {Turban}}, \bibinfo {author} {\bibfnamefont {J.-F.}\ \bibnamefont {Jacquot}},
  \bibinfo {author} {\bibfnamefont {J.~A.}\ \bibnamefont {Miwa}}, \bibinfo
  {author} {\bibfnamefont {O.~J.}\ \bibnamefont {Clark}}, \bibinfo {author}
  {\bibfnamefont {A.}~\bibnamefont {Jana}}, \bibinfo {author} {\bibfnamefont
  {S.~K.}\ \bibnamefont {Chaluvadi}}, \bibinfo {author} {\bibfnamefont
  {P.}~\bibnamefont {Orgiani}}, \bibinfo {author} {\bibfnamefont
  {M.}~\bibnamefont {Cuoco}}, \bibinfo {author} {\bibfnamefont
  {M.}~\bibnamefont {Leandersson}}, \bibinfo {author} {\bibfnamefont
  {T.}~\bibnamefont {Balasubramanian}}, \bibinfo {author} {\bibfnamefont
  {T.}~\bibnamefont {Olsen}}, \bibinfo {author} {\bibfnamefont
  {Y.}~\bibnamefont {Hwang}}, \bibinfo {author} {\bibfnamefont
  {M.}~\bibnamefont {Jamet}},\ and\ \bibinfo {author} {\bibfnamefont
  {F.}~\bibnamefont {Mazzola}},\ }\bibfield  {title} {\bibinfo {title} {Bilayer
  orthogonal ferromagnetism in \ce{CrTe2}-based van der {Waals} system},\
  }\href {https://doi.org/10.1038/s41467-025-59266-4} {\bibfield  {journal}
  {\bibinfo  {journal} {Nat. Commun.}\ }\textbf {\bibinfo {volume} {16}},\
  \bibinfo {pages} {4495} (\bibinfo {year} {2025})}\BibitemShut {NoStop}%
\bibitem [{\citenamefont {Lan}\ \emph {et~al.}(2025)\citenamefont {Lan},
  \citenamefont {Luo}, \citenamefont {Zhou}, \citenamefont {Wang},
  \citenamefont {Cheng}, \citenamefont {Liu}, \citenamefont {Pan},
  \citenamefont {Zhang}, \citenamefont {Li}, \citenamefont {Hou}, \citenamefont
  {Song}, \citenamefont {Lu},\ and\ \citenamefont {Sun}}]{qppq-qsx7}%
  \BibitemOpen
  \bibfield  {author} {\bibinfo {author} {\bibfnamefont {R.}~\bibnamefont
  {Lan}}, \bibinfo {author} {\bibfnamefont {X.}~\bibnamefont {Luo}}, \bibinfo
  {author} {\bibfnamefont {N.}~\bibnamefont {Zhou}}, \bibinfo {author}
  {\bibfnamefont {A.}~\bibnamefont {Wang}}, \bibinfo {author} {\bibfnamefont
  {M.}~\bibnamefont {Cheng}}, \bibinfo {author} {\bibfnamefont
  {L.}~\bibnamefont {Liu}}, \bibinfo {author} {\bibfnamefont {Y.}~\bibnamefont
  {Pan}}, \bibinfo {author} {\bibfnamefont {R.}~\bibnamefont {Zhang}}, \bibinfo
  {author} {\bibfnamefont {J.}~\bibnamefont {Li}}, \bibinfo {author}
  {\bibfnamefont {Y.}~\bibnamefont {Hou}}, \bibinfo {author} {\bibfnamefont
  {W.}~\bibnamefont {Song}}, \bibinfo {author} {\bibfnamefont {Q.}~\bibnamefont
  {Lu}},\ and\ \bibinfo {author} {\bibfnamefont {Y.}~\bibnamefont {Sun}},\
  }\bibfield  {title} {\bibinfo {title} {Thermomagnetic irreversibility in a
  \ce{Cr_{1.45}Te2} crystal: Role of spin-phonon coupling},\ }\href
  {https://link.aps.org/doi/10.1103/qppq-qsx7} {\bibfield  {journal} {\bibinfo
  {journal} {Phys. Rev. B}\ }\textbf {\bibinfo {volume} {112}},\ \bibinfo
  {pages} {104414} (\bibinfo {year} {2025})}\BibitemShut {NoStop}%
\bibitem [{\citenamefont {Staros}\ \emph {et~al.}(2022)\citenamefont {Staros},
  \citenamefont {Hu}, \citenamefont {Tiihonen}, \citenamefont {Nanguneri},
  \citenamefont {Krogel}, \citenamefont {Bennett}, \citenamefont {Heinonen},
  \citenamefont {Ganesh},\ and\ \citenamefont
  {Rubenstein}}]{10.1063/5.0074848}%
  \BibitemOpen
  \bibfield  {author} {\bibinfo {author} {\bibfnamefont {D.}~\bibnamefont
  {Staros}}, \bibinfo {author} {\bibfnamefont {G.}~\bibnamefont {Hu}}, \bibinfo
  {author} {\bibfnamefont {J.}~\bibnamefont {Tiihonen}}, \bibinfo {author}
  {\bibfnamefont {R.}~\bibnamefont {Nanguneri}}, \bibinfo {author}
  {\bibfnamefont {J.}~\bibnamefont {Krogel}}, \bibinfo {author} {\bibfnamefont
  {M.~C.}\ \bibnamefont {Bennett}}, \bibinfo {author} {\bibfnamefont
  {O.}~\bibnamefont {Heinonen}}, \bibinfo {author} {\bibfnamefont
  {P.}~\bibnamefont {Ganesh}},\ and\ \bibinfo {author} {\bibfnamefont
  {B.}~\bibnamefont {Rubenstein}},\ }\bibfield  {title} {\bibinfo {title} {A
  combined first principles study of the structural, magnetic, and phonon
  properties of monolayer \ce{CrI3}},\ }\href
  {https://doi.org/10.1063/5.0074848} {\bibfield  {journal} {\bibinfo
  {journal} {J. Chem. Phys.}\ }\textbf {\bibinfo {volume} {156}},\ \bibinfo
  {pages} {014707} (\bibinfo {year} {2022})}\BibitemShut {NoStop}%
\bibitem [{\citenamefont {Calder}\ \emph {et~al.}(2015)\citenamefont {Calder},
  \citenamefont {Lee}, \citenamefont {Stone}, \citenamefont {Lumsden},
  \citenamefont {Lang}, \citenamefont {Feygenson}, \citenamefont {Zhao},
  \citenamefont {Yan}, \citenamefont {Shi}, \citenamefont {Sun}, \citenamefont
  {Tsujimoto}, \citenamefont {Yamaura},\ and\ \citenamefont
  {Christianson}}]{ncomms9916}%
  \BibitemOpen
  \bibfield  {author} {\bibinfo {author} {\bibfnamefont {S.}~\bibnamefont
  {Calder}}, \bibinfo {author} {\bibfnamefont {J.~H.}\ \bibnamefont {Lee}},
  \bibinfo {author} {\bibfnamefont {M.~B.}\ \bibnamefont {Stone}}, \bibinfo
  {author} {\bibfnamefont {M.~D.}\ \bibnamefont {Lumsden}}, \bibinfo {author}
  {\bibfnamefont {J.~C.}\ \bibnamefont {Lang}}, \bibinfo {author}
  {\bibfnamefont {M.}~\bibnamefont {Feygenson}}, \bibinfo {author}
  {\bibfnamefont {Z.}~\bibnamefont {Zhao}}, \bibinfo {author} {\bibfnamefont
  {J.~Q.}\ \bibnamefont {Yan}}, \bibinfo {author} {\bibfnamefont {Y.~G.}\
  \bibnamefont {Shi}}, \bibinfo {author} {\bibfnamefont {Y.~S.}\ \bibnamefont
  {Sun}}, \bibinfo {author} {\bibfnamefont {Y.}~\bibnamefont {Tsujimoto}},
  \bibinfo {author} {\bibfnamefont {K.}~\bibnamefont {Yamaura}},\ and\ \bibinfo
  {author} {\bibfnamefont {A.~D.}\ \bibnamefont {Christianson}},\ }\bibfield
  {title} {\bibinfo {title} {Enhanced spin-phonon-electronic coupling in a 5d
  oxide},\ }\href {https://doi.org/10.1038/ncomms9916} {\bibfield  {journal}
  {\bibinfo  {journal} {Nat. Commun.}\ }\textbf {\bibinfo {volume} {6}},\
  \bibinfo {pages} {8916} (\bibinfo {year} {2015})}\BibitemShut {NoStop}%
\bibitem [{\citenamefont {Garc\'{\i}a-Flores}\ \emph
  {et~al.}(2012)\citenamefont {Garc\'{\i}a-Flores}, \citenamefont {Moreira},
  \citenamefont {Kaneko}, \citenamefont {Ardito}, \citenamefont {Terashita},
  \citenamefont {Orlando}, \citenamefont {Gopalakrishnan}, \citenamefont
  {Ramesha},\ and\ \citenamefont {Granado}}]{PhysRevLett.108.177202}%
  \BibitemOpen
  \bibfield  {author} {\bibinfo {author} {\bibfnamefont {A.~F.}\ \bibnamefont
  {Garc\'{\i}a-Flores}}, \bibinfo {author} {\bibfnamefont {A.~F.~L.}\
  \bibnamefont {Moreira}}, \bibinfo {author} {\bibfnamefont {U.~F.}\
  \bibnamefont {Kaneko}}, \bibinfo {author} {\bibfnamefont {F.~M.}\
  \bibnamefont {Ardito}}, \bibinfo {author} {\bibfnamefont {H.}~\bibnamefont
  {Terashita}}, \bibinfo {author} {\bibfnamefont {M.~T.~D.}\ \bibnamefont
  {Orlando}}, \bibinfo {author} {\bibfnamefont {J.}~\bibnamefont
  {Gopalakrishnan}}, \bibinfo {author} {\bibfnamefont {K.}~\bibnamefont
  {Ramesha}},\ and\ \bibinfo {author} {\bibfnamefont {E.}~\bibnamefont
  {Granado}},\ }\bibfield  {title} {\bibinfo {title} {Spin-electron-phonon
  excitation in \ce{Re}-based half-metallic double perovskites},\ }\href
  {https://link.aps.org/doi/10.1103/PhysRevLett.108.177202} {\bibfield
  {journal} {\bibinfo  {journal} {Phys. Rev. Lett.}\ }\textbf {\bibinfo
  {volume} {108}},\ \bibinfo {pages} {177202} (\bibinfo {year}
  {2012})}\BibitemShut {NoStop}%
\bibitem [{\citenamefont {Sadhukhan}\ \emph {et~al.}(2022)\citenamefont
  {Sadhukhan}, \citenamefont {Bergman}, \citenamefont {Kvashnin}, \citenamefont
  {Hellsvik},\ and\ \citenamefont {Delin}}]{PhysRevB.105.104418}%
  \BibitemOpen
  \bibfield  {author} {\bibinfo {author} {\bibfnamefont {B.}~\bibnamefont
  {Sadhukhan}}, \bibinfo {author} {\bibfnamefont {A.}~\bibnamefont {Bergman}},
  \bibinfo {author} {\bibfnamefont {Y.~O.}\ \bibnamefont {Kvashnin}}, \bibinfo
  {author} {\bibfnamefont {J.}~\bibnamefont {Hellsvik}},\ and\ \bibinfo
  {author} {\bibfnamefont {A.}~\bibnamefont {Delin}},\ }\bibfield  {title}
  {\bibinfo {title} {Spin-lattice couplings in two-dimensional \ce{CrI3} from
  first-principles computations},\ }\href
  {https://link.aps.org/doi/10.1103/PhysRevB.105.104418} {\bibfield  {journal}
  {\bibinfo  {journal} {Phys. Rev. B}\ }\textbf {\bibinfo {volume} {105}},\
  \bibinfo {pages} {104418} (\bibinfo {year} {2022})}\BibitemShut {NoStop}%
\bibitem [{\citenamefont {Zhang}\ \emph
  {et~al.}(2019{\natexlab{b}})\citenamefont {Zhang}, \citenamefont {Hou},
  \citenamefont {Wang},\ and\ \citenamefont {Wu}}]{PhysRevB.100.224427}%
  \BibitemOpen
  \bibfield  {author} {\bibinfo {author} {\bibfnamefont {B.~H.}\ \bibnamefont
  {Zhang}}, \bibinfo {author} {\bibfnamefont {Y.~S.}\ \bibnamefont {Hou}},
  \bibinfo {author} {\bibfnamefont {Z.}~\bibnamefont {Wang}},\ and\ \bibinfo
  {author} {\bibfnamefont {R.~Q.}\ \bibnamefont {Wu}},\ }\bibfield  {title}
  {\bibinfo {title} {First-principles studies of spin-phonon coupling in
  monolayer $\ce{Cr2Ge2Te6}$},\ }\href
  {https://link.aps.org/doi/10.1103/PhysRevB.100.224427} {\bibfield  {journal}
  {\bibinfo  {journal} {Phys. Rev. B}\ }\textbf {\bibinfo {volume} {100}},\
  \bibinfo {pages} {224427} (\bibinfo {year} {2019}{\natexlab{b}})}\BibitemShut
  {NoStop}%
\bibitem [{\citenamefont {Granado}\ \emph {et~al.}(1999)\citenamefont
  {Granado}, \citenamefont {Garc\'{\i}a}, \citenamefont {Sanjurjo},
  \citenamefont {Rettori}, \citenamefont {Torriani}, \citenamefont {Prado},
  \citenamefont {S\'anchez}, \citenamefont {Caneiro},\ and\ \citenamefont
  {Oseroff}}]{PhysRevB.60.11879}%
  \BibitemOpen
  \bibfield  {author} {\bibinfo {author} {\bibfnamefont {E.}~\bibnamefont
  {Granado}}, \bibinfo {author} {\bibfnamefont {A.}~\bibnamefont
  {Garc\'{\i}a}}, \bibinfo {author} {\bibfnamefont {J.~A.}\ \bibnamefont
  {Sanjurjo}}, \bibinfo {author} {\bibfnamefont {C.}~\bibnamefont {Rettori}},
  \bibinfo {author} {\bibfnamefont {I.}~\bibnamefont {Torriani}}, \bibinfo
  {author} {\bibfnamefont {F.}~\bibnamefont {Prado}}, \bibinfo {author}
  {\bibfnamefont {R.~D.}\ \bibnamefont {S\'anchez}}, \bibinfo {author}
  {\bibfnamefont {A.}~\bibnamefont {Caneiro}},\ and\ \bibinfo {author}
  {\bibfnamefont {S.~B.}\ \bibnamefont {Oseroff}},\ }\bibfield  {title}
  {\bibinfo {title} {Magnetic ordering effects in the {Raman} spectra of
  \ce{La_{1-x}Mn_{1-x}O3}},\ }\href
  {https://link.aps.org/doi/10.1103/PhysRevB.60.11879} {\bibfield  {journal}
  {\bibinfo  {journal} {Phys. Rev. B}\ }\textbf {\bibinfo {volume} {60}},\
  \bibinfo {pages} {11879} (\bibinfo {year} {1999})}\BibitemShut {NoStop}%
\bibitem [{\citenamefont {Chen}\ \emph {et~al.}(2011)\citenamefont {Chen},
  \citenamefont {Jia}, \citenamefont {Kemper}, \citenamefont {Singh},\ and\
  \citenamefont {Devereaux}}]{PhysRevLett.106.067002}%
  \BibitemOpen
  \bibfield  {author} {\bibinfo {author} {\bibfnamefont {C.-C.}\ \bibnamefont
  {Chen}}, \bibinfo {author} {\bibfnamefont {C.~J.}\ \bibnamefont {Jia}},
  \bibinfo {author} {\bibfnamefont {A.~F.}\ \bibnamefont {Kemper}}, \bibinfo
  {author} {\bibfnamefont {R.~R.~P.}\ \bibnamefont {Singh}},\ and\ \bibinfo
  {author} {\bibfnamefont {T.~P.}\ \bibnamefont {Devereaux}},\ }\bibfield
  {title} {\bibinfo {title} {Theory of two-magnon {Raman} scattering in iron
  pnictides and chalcogenides},\ }\href
  {https://link.aps.org/doi/10.1103/PhysRevLett.106.067002} {\bibfield
  {journal} {\bibinfo  {journal} {Phys. Rev. Lett.}\ }\textbf {\bibinfo
  {volume} {106}},\ \bibinfo {pages} {067002} (\bibinfo {year}
  {2011})}\BibitemShut {NoStop}%
\end{thebibliography}%

\onecolumngrid
\clearpage

\newcounter{pdfpagecount}
\setcounter{pdfpagecount}{1}

\whiledo{\value{pdfpagecount} < 18}{
    \thispagestyle{empty} 
    \noindent
    \centerline{
        \includegraphics[page=\value{pdfpagecount}, width=0.95\paperwidth, height=0.95\paperheight, keepaspectratio]{Supplemental_Material.pdf}
    }
    \clearpage
    
    \ifnum\value{pdfpagecount}=17
        \makeatletter
        \let\ps@plain\ps@empty
        \let\ps@headings\ps@empty
        \makeatother
        \thispagestyle{empty}
    \fi
    
    \stepcounter{pdfpagecount}
}

\pagestyle{empty}
\twocolumngrid

\end{document}